\documentclass[9pt,twocolumn,twoside]{pnas-new}

\usepackage{subcaption}

\templatetype{pnasresearcharticle} 

\begin{document}

\title{Measuring Global Migration Flows using Online Data}

\author[a,1]{Guanghua Chi}
\author[b,c,1]{Guy J. Abel}
\author[d]{Drew Johnston}
\author[a]{Eugenia Giraudy}
\author[a]{Mike Bailey}

\affil[a]{Meta, Menlo Park, CA 94025}
\affil[b]{Department of Sociology, Faculty of Social Sciences, University of Hong Kong, Hong Kong SAR, China}
\affil[c]{International Institute for Applied Systems Analysis (IIASA), 2361 Laxenburg, Austria}
\affil[d]{Department of Economics, Harvard University, Cambridge, MA 02138}

\leadauthor{Chi}

\significancestatement{This paper introduces a novel approach to measuring international migration, improving on existing efforts in both breadth and resolution. Despite the importance of these figures to researchers, policymakers, and the public, most countries do not report migration flows, and estimates where they exist are often outdated. The International Organization for Migration argues that improved migration data would be highly beneficial, with a value of \$35 billion to governments alone. Our estimates improve substantially on existing figures, allowing us to estimate monthly migration flows between 181 countries by drawing on data from more than 3 billion individuals. We are collaborating with NGOs on humanitarian applications of this data, and release it publicly to aid researchers and policymakers.}

\authorcontributions{Author contributions: G.C. designed research; G.C., G.J.A., D.J., E.G., and M.B. performed research; G.C., G.J.A., D.J., E.G., and M.B. analyzed data; and G.C., G.J.A., D.J., and M.B. wrote the paper.}
\authordeclaration{Competing interest statement: G.C., D.J., E.G., and M.B. are employees at Meta. Meta contributed server time to create the analyses in the project.}
\correspondingauthor{\textsuperscript{1}To whom correspondence should be addressed. E-mail: gchi@meta.com, guyabel@hku.hk}


\begin{abstract}
Existing estimates of human migration are limited in their scope, reliability, and timeliness, prompting the United Nations and the Global Compact on Migration to call for improved data collection. Using privacy protected records from three billion Facebook users, we estimate country-to-country migration flows at monthly granularity for 181 countries, accounting for selection into Facebook usage. Our estimates closely match high-quality measures of migration where available but can be produced nearly worldwide and with less delay than alternative methods. We estimate that 39.1 million people migrated internationally in 2022 (0.63\% of the population of the countries in our sample). Migration flows significantly changed during the COVID-19 pandemic, decreasing by 64\% before rebounding in 2022 to a pace 24\% above the pre-crisis rate. We also find that migration from Ukraine increased tenfold in the wake of the Russian invasion. To support research and policy interventions, we will release these estimates publicly through the Humanitarian Data Exchange.
\end{abstract}

\dates{This manuscript was compiled on \today}
\doi{\url{www.pnas.org/cgi/doi/10.1073/pnas.XXXXXXXXXX}}

\maketitle
\thispagestyle{firststyle}
\ifthenelse{\boolean{shortarticle}}{\ifthenelse{\boolean{singlecolumn}}{\abscontentformatted}{\abscontent}}{}

\firstpage[12]{3}

\dropcap{E}stimates of migration flows are widely used in evidence-based policymaking, informing efforts to address domestic labor shortages \cite{oecd_getting_2016}, mitigate the negative effects of emigration \cite{docquier_brain_2007}, and increase immigrants' employment rates \cite{bevelander_employment_1999}. Despite the value of these estimates, the International Organization for Migration (IOM) noted in its 2022 World Migration Report that only 45 governments provide data on migration flows, in part because the collection of accurate figures is ``extremely difficult'' \cite{mcauliffe2022migration, willekens_international_2016}. The report also noted that these figures use inconsistent methodologies and definitions of migration, and are often out of date: the latest estimates catalogued by the United Nations Population Division date to 2015, while figures for a smaller group of wealthy countries are published only after a three-year delay \cite{undesa_flow,oecd_flow}. As a result, the collection of accurate migration statistics was listed as the top objective of the 2018 Global Compact for Migration, with the IOM estimating that better estimates could be worth as much as \$35 billion to governments and migrants \cite{kraly_data_2020, iom2018morethan}.

To address the limitations of the existing data, researchers have used alternative methods to estimate the rate of migration between countries. One such approach estimates migration flows using time-series data on migrant stocks from surveys or censuses. Though most countries produce data on migrant stocks, such data is usually collected infrequently, making it impossible to calculate monthly migration flows \cite{abel_quantifying_2014,azose_estimation_2019,abel_bilateral_2019, mig_stock}. Recently, researchers have used new data sources, such as cell phone records and geotagged Tweets, to measure migration in a more timely fashion, but these approaches have been limited in their geographic scope or representativeness \cite{sirbu_human_2021,UNECE,zagheni_leveraging_2017,hawelka_geo-located_2014,lu_unveiling_2016,wesolowski_quantifying_2015,zagheni_inferring_2014}.

In this paper, we estimate migration flows between 181 countries\footnote{A complete list of included countries can be found in Table S9. We exclude very small countries and countries in which Facebook is banned or where other constraints restrict our estimation. We produce experimental estimates for China in SI Estimates for China. These estimates use a separate methodology and are not included in most of the figures in this paper.} for each month from January 2019 through December 2022 using privacy protected data from 3 billion active users of Facebook, the world's largest social network. To do this, we identify changes in the predicted home location of each user over time. We define a migration event as an instance in which an individual who has resided in a country for more than a year moves to another country, which they then reside in for more than one year. This approach matches the United Nations' recommended definition of migration\footnote{See the definition provided by the Statistics Division of the United Nations Department of Economic and Social Affairs \href{https://www.iom.int/key-migration-terms}{here}.}. If individuals or families who leave a country and go from place to place in transit (staying less than 12 months) until they reach the destination, they are not considered migrants to those countries in transit. They are only counted in the migration estimates when they remain in the destination for at least 12 months. We then aggregate these individual migration events to country-by-country-by-month counts and apply weights to make our estimates representative of the actual migration flows at the population level, accounting for different Facebook usage and levels of economic development along each migration corridor. We validate our estimates against administrative data, which are currently the most reliable sources of information about migration patterns. We find that our figures are strongly correlated with these traditional data sources, where they exist. We will release the migration estimates publicly through the Humanitarian Data Exchange (\href{https://data.humdata.org/dataset/international-migration-flows}{https://data.humdata.org/dataset/international-migration-flows}) as a resource for researchers and policymakers. It contains the monthly estimates for each country pair from 2019 to 2022. The list of countries is in Table S9. For each country in the dataset we release the flow for both directions.

\section*{Global migration patterns}

\begin{figure*}
\centering
\includegraphics[width=17.8cm]{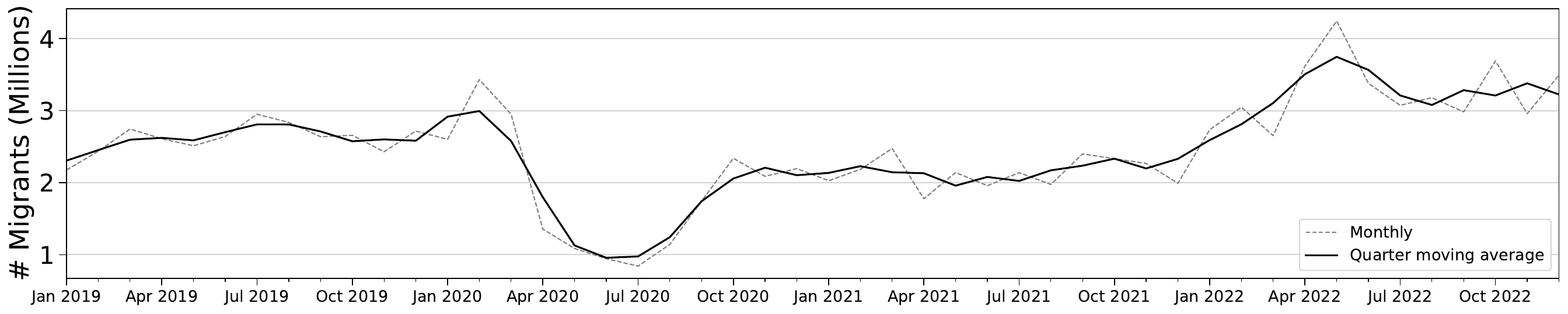}
\caption{Global international migration over time. The dashed line depicts the estimated level of migration in each month, following the procedure described in SI Algorithm to detect migrants using the selection rate weights described in SI Weighting. The solid line smooths the data over a three-month window with one month on each side of the current month.}
\label{migration_trend}
\end{figure*}

\begin{figure*}[!h]
    \begin{center}
    \begin{subfigure}[t]{0.56\textwidth}
        \caption{}
        \includegraphics[width=1\linewidth]{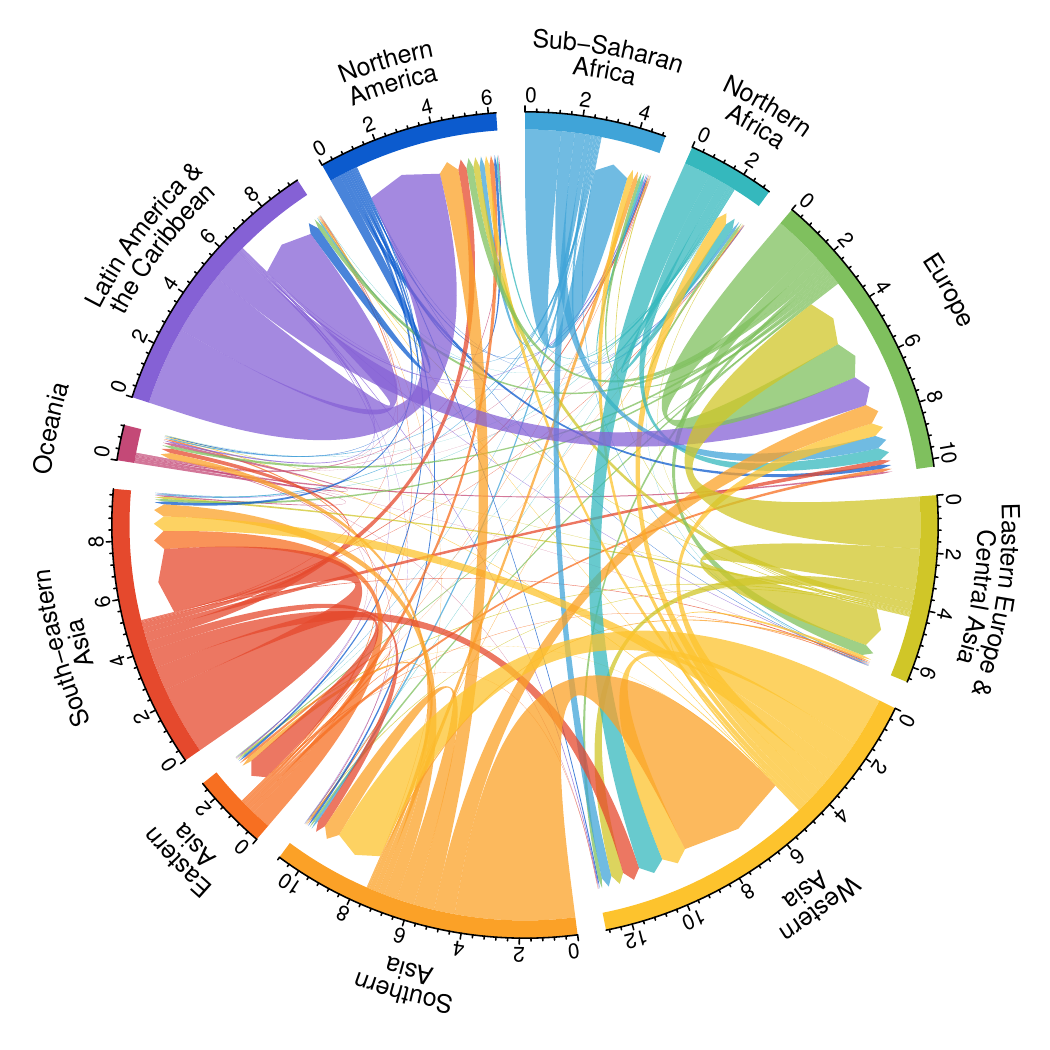} 
    \end{subfigure}
    \vspace{0.5cm}
    \begin{subfigure}[t]{0.42\textwidth}
        \caption{}
        \vspace{1cm}
        \includegraphics[width=1\linewidth]{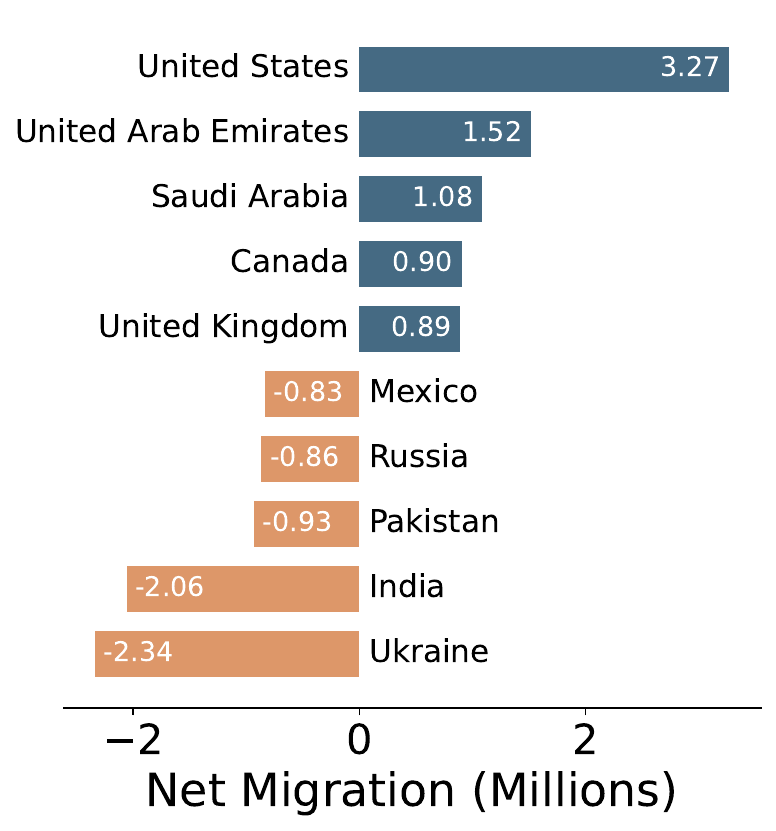}
    \end{subfigure} 
    \begin{subfigure}[t]{0.3\textwidth}
        \caption{}
        \includegraphics[width=1\linewidth]{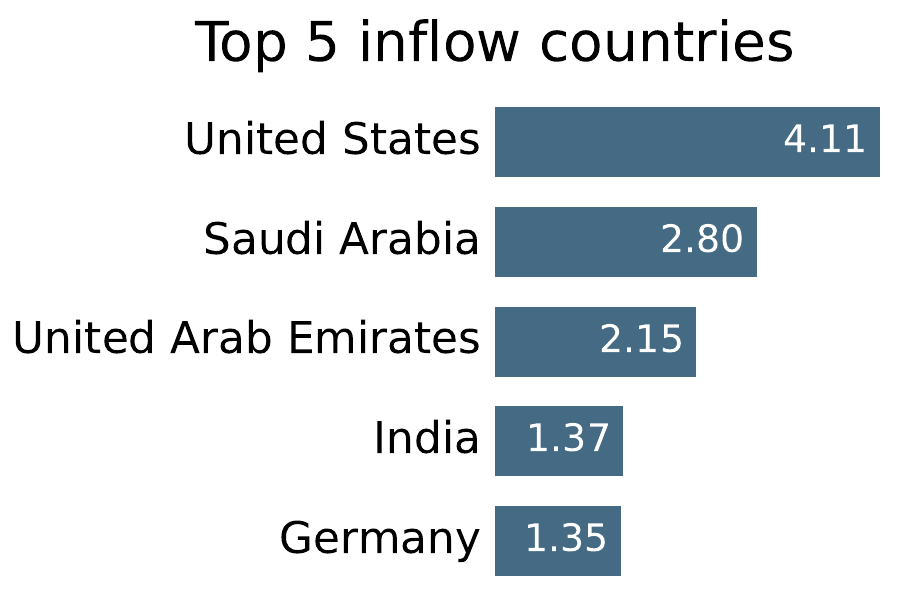} 
    \end{subfigure}
    \vspace{0.5cm}
    \begin{subfigure}[t]{0.3\textwidth}
        \caption{}
        \includegraphics[width=1\linewidth]{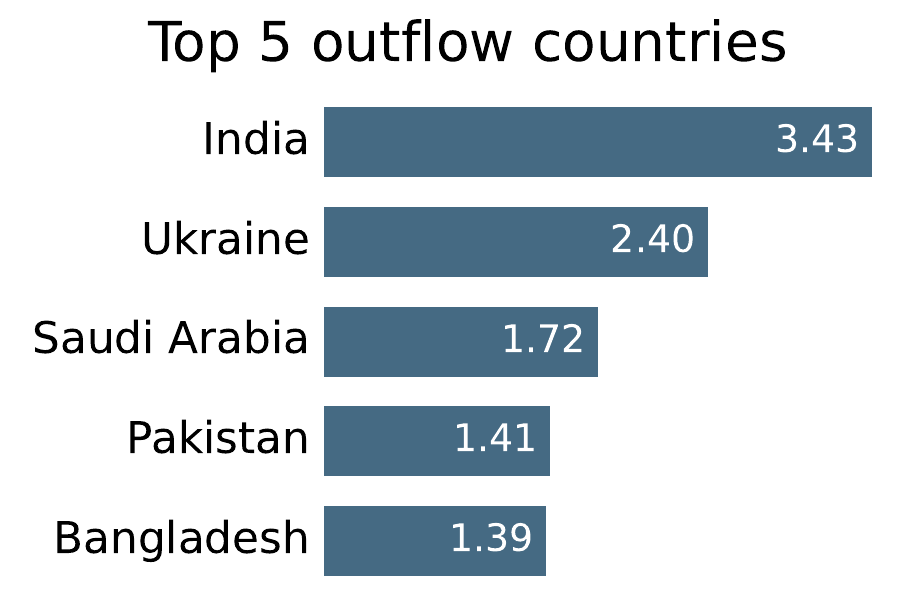} 
    \end{subfigure} 
    \begin{subfigure}[t]{0.3\textwidth}
        \caption{}
        \includegraphics[width=1\linewidth]{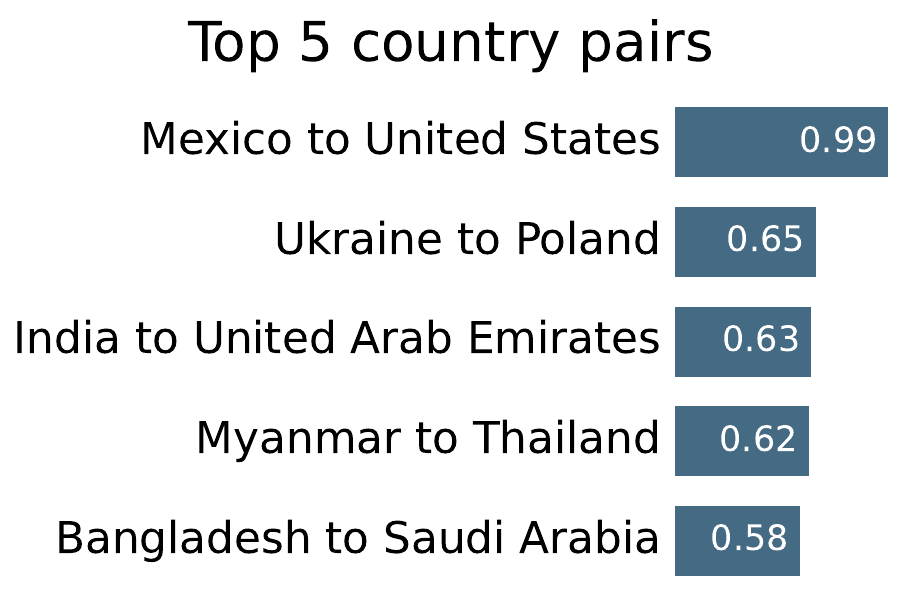} 
    \end{subfigure} 
    \end{center}
    \vspace{-0.6cm}
\caption{Estimated international migration flows in 2022, in millions of people. \textbf{(A)} International migration flows between and within regions in 2022. Lines measure aggregate flows between all country pairs in each region. \textbf{(B)} Top and bottom five countries by annual net migration in 2022. Orange bars depict countries with net emigration, blue bars depict those with net immigration. \textbf{(C)} Top five countries by annual gross migration inflows in 2022. \textbf{(D)} Top five countries by annual gross migration outflows in 2022. \textbf{(E)} Top five (directed) country pairs for annual migration in 2022.}
\label{migration_regional}
\end{figure*}

We estimate that around 3.3 million people\footnote{The United Nation estimates a global population of 281 million migrants in 2020 \cite{mcauliffe2021world}. Note that this figure is referring to the migration stock population, not the migration flow. The migration stock is a count of persons who are currently not in their country of birth, regardless of when they arrived at their destination country, while we count the number of people that moved between countries in a given year.} migrated each month in 2022 between the 181 countries\footnote{The 181 countries in our study account for 79.2\% of the global population. In SI Estimates for China, we demonstrate how we can adapt our methodology to estimate migration to and from China, the largest country not included in our main results. When we include China in our estimates, the countries in our sample account for 97.4\% of the world population. Estimates for China are not included in our public data release.} in our study (Fig. \ref{migration_trend}). In total, we estimate that 39.1 million people migrated internationally that year, approximately 0.63\% of the population of the countries in our sample\footnote{In SI Estimates for China, we estimate that in 2022 around 81 thousand people migrated to China and 1.1 million migrated away, boosting our global migration figures to 40.2 million and 0.53\% of the population of the larger 182-country sample.}. The United States had the highest positive net migration in 2022, with over 3.27 million more people migrating to the country than leaving it (Fig. \ref{migration_regional}B). Ukraine saw the largest net outflow, losing over 2.34 million people in 2022 after it was invaded by Russia\footnote{Our estimate is lower than the number of Ukrainians who sought Temporary Protected Status after leaving the country. We believe this is due to our more conservative definition of migration; see SI Crisis-induced migration for more details.}. The United States, Saudi Arabia, and United Arab Emirates are the top three countries in terms of the migration inflows (Fig. \ref{migration_regional}C), while for the migration outflows, the top three countries are India, Ukraine, and Saudi Arabia (Fig. \ref{migration_regional}D).

In Fig. \ref{migration_regional}A, we plot the geographic distribution of these flows at the regional level. This plot displays migration in 2022, though we find broadly similar patterns of migration corridors in the other years we study\footnote{A regression of the natural logarithm of the number of migrants in each year on year and country-pair fixed effects explains 88.9\% of the variation in the data.}. At the country pair level, Fig. \ref{migration_regional}E shows that close to one million people migrated from Mexico to the United States in 2022. In addition to country pairs with close proximity where migration corridors are prevalent, we also observe the existence of persistent migration between a number of far-away country pairs. To explain this persistence, we next discuss several factors which drive migration along these established corridors.

Distance is a strong predictor of the rate of migration between countries: over 20.9\% of migration in 2022 occurred between bordering countries, despite fewer than 1.7\% of country pairs sharing a border. Even among countries that do not share a border, we observe substantial clustering of migration at the regional level, with particularly large flows of migrants within Europe, in part due to the lack of internal migration restrictions within the European Union. We also observe notably large flows within Latin America and the Caribbean, driven by migrants from Venezuela to nearby countries.

We also observe several long-distance migration corridors that see persistently high levels of migration. For instance, we observe large flows between West Asia and South Asia, driven largely by labor migration from India, Pakistan, and Bangladesh to Saudi Arabia, Qatar, and the United Arab Emirates. Unusually, this corridor sees largely symmetric migration patterns, since migrants usually return to their origin at the conclusion of their contract \cite{mcauliffe2022migration}. Other destinations, particularly Europe and North America, see inflows from a more diffuse set of regions due to the differing set of institutions governing migration to these regions. In SI Social networks and migration, we show that social ties between the origin and destination are strongly predictive of these flows, even between geographically distant country pairs. 

We also observe that both origin- and destination-country development are strongly predictive of migration rates, with countries classified as high-income by the World Bank attracting 67\% of global migrants \cite{world2022world}. Additionally, high-income countries send 33\% of the global migrants despite comprising just 19\% of the world population, aligning with prior research that migrants tend to be wealthier \cite{clemens2014does}. These patterns are visualized in SI National income levels and migration trends. 

Although migration corridors are generally stable, our monthly data allows us to observe how migration patterns shift in response to crises or policy changes. Such shifts, despite their importance in policymaking, have historically been hard to observe due to the limited resolution of existing migration estimates. We observe that crises can lead to dramatic changes in migration; 2.3 million people emigrated from Ukraine following Russia's invasion of the country in February 2022 and settled elsewhere between February and December 2022 for at least a year, a tenfold increase over the pre-war emigration rate. We find that Poland, Germany, Czech Republic, the United States, and United Kingdom have received the highest number of migrants, which are closely aligned with estimates from the UNHCR\footnote{See \href{https://data.unhcr.org/en/situations/ukraine}{here} for more details. Note that the website shows the cumulative estimates of Ukrainian refugees since 2022, while our estimates show the number of migrants from Ukraine in 2022. We assume that most of the migrants from Ukraine in 2022 are caused by the Ukraine war as shown in Fig. S20.} \cite{gonzalez-leonardo_where_2024}, though Czech Republic and Estonia have received the largest share relative to their population. We estimate that flows from Hong Kong to the United Kingdom increased more than fifteenfold in the wake of Hong Kong's passage of a contentious security law in June 2020, while migration from Myanmar to Singapore increased more than fivefold in the wake of a coup in the former country in February 2021. We discuss these patterns in more detail in SI Crisis-induced migration.

In recent years, global factors such as the COVID-19 pandemic and its associated migration policy responses have also played an important role in shaping migration. Specifically, after the onset of the COVID-19 pandemic, the global flow of international migrants fell 64\%, driven in part by the imposition of border controls by many countries \cite{benton_covid-19_2020,gonzalez-leonardo_assessing_2024}\footnote{In Fig. S18, we present time series estimates for China, and find a largely similar pattern. Migration from China fell from an average of 170 thousand per month in 2019 to 93 thousand in June 2020. Migration to the country fell by a similar amount, from an average of 21 thousand per month to 6 thousand.}. International migration flows started to rebound after July 2020 as countries began to lift controls \cite{benton_covid-19_2020}. Migration rates remained low in 2021, with an average of 2.14 million people migrating each month, 18\% below the 2019 average. Migration reached pre-pandemic levels for the first time in January 2022 and remained high throughout the year, with 3.3 million people migrating on average each month, 24\% above the rate in 2019.

\begin{figure*}
\centering
\includegraphics[width=17.8cm]{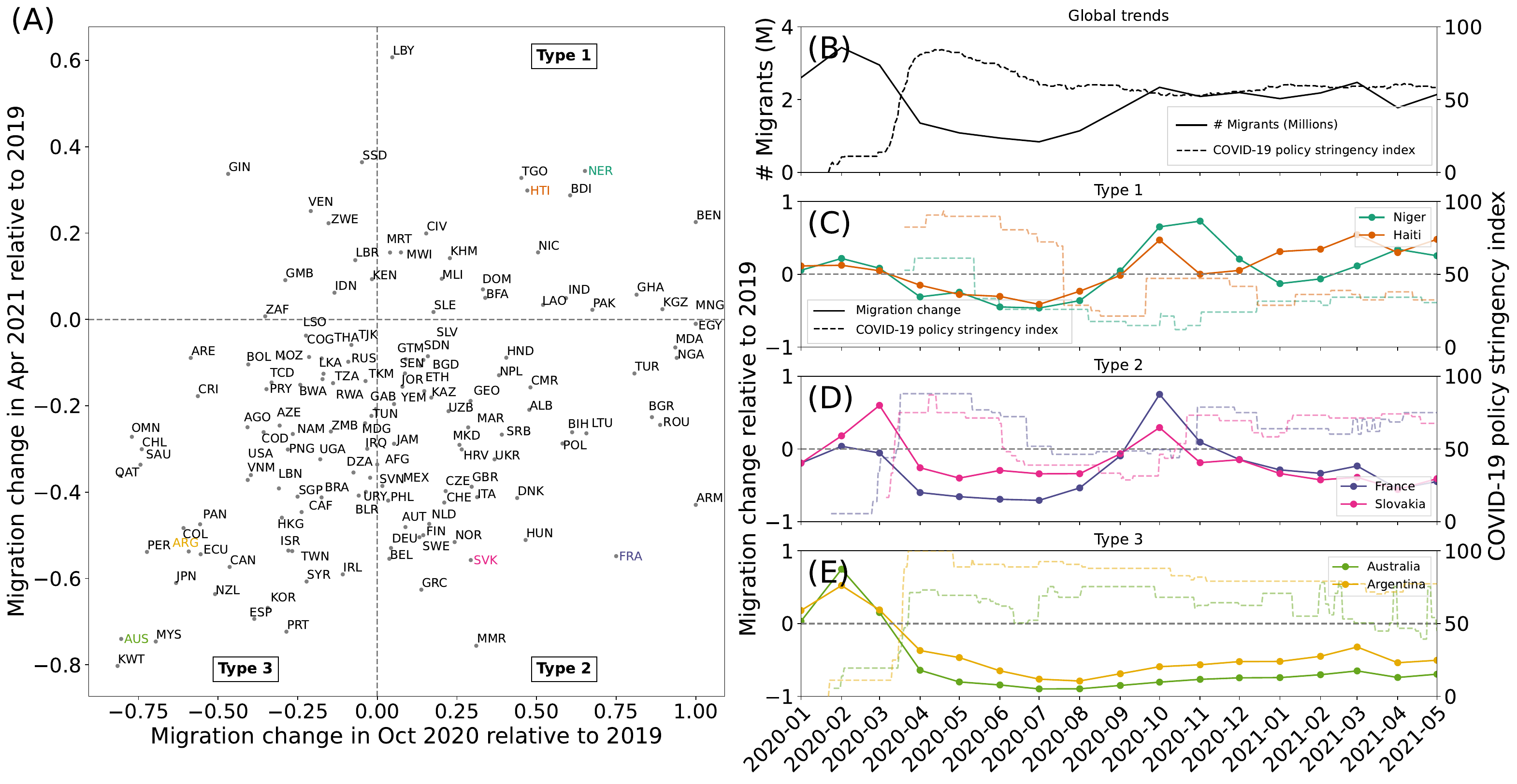}
\caption{Migration Flows and COVID-19 policy stringency. \textbf{(A)} Changes in immigration flows during October 2020 and April 2021, relative to each country's 2019 average. Countries are divided into three types based on their quadrant in this plot. \textbf{(B)} Global migration and the mean COVID-19 policy stringency index over time. \textbf{(C)} Migration inflows (solid lines) and COVID-19 stringency (dashed line) for Niger and Haiti (two ``Type 1" countries where restrictions were relatively low throughout the sample period). \textbf{(D)} Migration inflows and COVID-19 stringency for France and Slovakia (two ``Type 2" countries where restrictions were relaxed and then tightened). \textbf{(E)} Migration inflows and COVID-19 stringency for Australia and Argentina (two ``Type 3" countries where restrictions were relatively high throughout the sample period).}
\label{migration_covid}
\end{figure*}

Our data also allows us to observe the relationship between COVID-related policies and migration at high frequency. There was a large degree of heterogeneity in national approaches to COVID-19 control; in some countries, the strict border controls enacted in early 2020 remained through 2021, while in other countries controls were never enacted or quickly reversed (Fig. \ref{migration_covid}). We find that immigration is negatively correlated with the COVID-19 policy stringency index\footnote{This index by the Oxford Coronavirus Government Response Tracker summarizes a number of COVID-19 policies, including international travel controls.}: a 10 point increase on the 100 point scale is associated with a 5.1 percent decline in monthly immigrants the following month \cite{hale_global_2021}. The COVID-induced drop in migration is more than 50\% across all the regions. Migration from South Asia saw the largest decrease in 2020 (dropping 81.5 percent in 2020 compared to a 2019 baseline), while in North America migration decreased by 55.6 percent in 2020 relative to 2019. 

\section*{Data and methods}

Our first step in estimating migration between countries is detecting a change in a person's country of residence\footnote{This research was conducted at Meta and not reviewed by an Institutional Review Board, but has been approved via the privacy review process described at \href{https://about.meta.com/privacy-progress/}{https://about.meta.com/privacy-progress/}. The Privacy Review section in the link provided outlines the internal privacy expectations, such as purpose limitation, data limitation, and fairness. All analysis in this study was conducted on privacy protected data.}. We determine a predicted country location for each user based on a combination of signals, including the self-reported location on Facebook profiles and the IP addresses used to connect to Facebook. We use a segment-based method to turn these estimates into a time frame of each user’s country location \cite{chi_general_2020}. We define migration as living in one country for the majority of a 12-month period before moving to another country for the majority of the following 12 months, matching the definition of migration recommended by the United Nations \cite{uns}\footnote{This algorithm is explained in more detail in SI Algorithm.}
\footnote{Although we follow this recommended definition in our main analyses, it is easy to adapt our approach to use alternative definitions of migration. For instance, in Fig. S3, we produce estimates using a 6-month threshold for migration, allowing us to produce timelier estimates and to capture more temporary labor migration.}. 

We next aggregate these individual-level migration events to the country-pair level for each month, which provides us with a measure of the number of Facebook users migrating across each corridor. To account for the fact that Facebook usage varies across countries, we weight our data to make it representative at the population level. After experimenting with alternatives, the weight we apply depends on the country-wide Facebook usage rate and country income level. A single worldwide offset, which is calculated based on the administrative data in New Zealand, is added to control the degree to which the bias of migrants on Facebook varies with development. This method (called selection rate) allows us to account for the fact that in poorer countries, wealthier individuals are more likely to both use Facebook \cite{poushter2018social} and to migrate\footnote{A recent study of selection into emigration found that wealthier individuals were more likely to emigrate in 93 of 99 countries studied \cite{clemens2024migration}.}. We describe the details of this weighting method and compare it with a variety of alternative weighting methodologies in SI Weighting. In Tables S2 and S3, we benchmark our selection model against a naïve approach, in which we weight migration flows using the inverse of the country-wide Facebook usage rate. We find that the selection model reduces the difference between our estimates and those of the Swedish government by about 53.3\% relative to the naïve approach\footnote{We consider a variety of alternative weighting methodologies in SI Weighting.}. In Fig. S5, we present a country-by-country analysis of the weighted flows and show that the model allows us to account for the fact that selection into both Facebook usage and migration is higher in poor countries. This improves the accuracy of our estimates in the developing world considerably without harming our performance in wealthier countries. We describe our methodology and present further benchmarks in SI Weighting.

\begin{figure*}
\centering
\includegraphics[width=13cm]{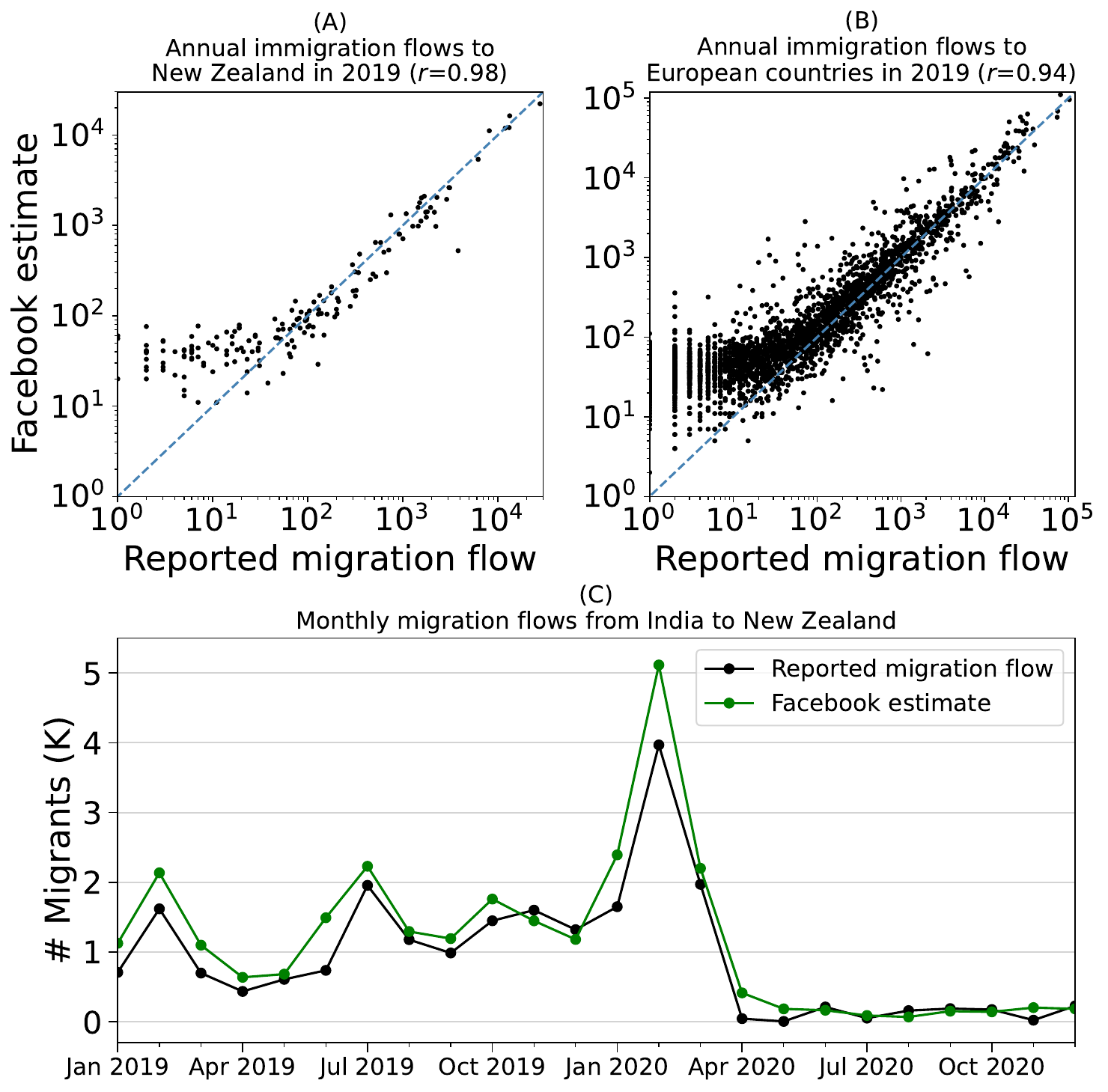}
\caption{Reported country-to-country migration flows vs. Facebook estimates in thousands. \textbf{(A)} Validation of Facebook estimates against 2019 annual data from the New Zealand Statistical Office. Each point stands for the number of immigrants from a country of origin to New Zealand in 2019. \textbf{(B)} Validation of Facebook estimates against 2019 annual data from Eurostat, the statistical office of the European Union. Each point stands for the number of immigrants from a country of origin to one European country in 2019. \textbf{(C)} Validation of Facebook estimates of monthly migration from India to New Zealand using data from the New Zealand Statistical Office.}
\label{validation}
\end{figure*}

After applying the weights, we add noise to our estimates using techniques from the differential privacy literature to preserve the privacy of individual-level data. This generally results in a small amount of noise, with 95\% of monthly estimates changing by fewer than 7 people. We describe our methodology in more detail in SI Differential privacy.

\section*{Validation}

We next benchmark our estimates against administrative datasets drawn from statistical offices of New Zealand and the European Union. These two datasets are thought to be of unusually high quality, but only capture migration flows to a subset of countries \cite{wisniowski_utilising_2013,hugo2016migration}. In SI Validation, we benchmark our data against other administrative records, as well as against low-frequency estimates of global migration.

We first benchmark our data against the immigration statistics reported by the government of New Zealand \cite{newzealand}. New Zealand has a strong statistical infrastructure and is one of the only countries to release monthly estimates of migration, allowing us to validate both the magnitude and temporal pattern of migration in our data\footnote{The New Zealand government uses a non-standard definition of migration in its statistics, requiring an individual to reside in the country for 12 months in a 16 month period to be a migrant, rather than the 6 months in a 12 month period used by most nations. In the validation exercises we present in this section, we produced figures using the New Zealand government's definition of migration, though we use the United Nations definition of migration in our public data release and in all other figures in this paper which include data from New Zealand.}.

In Fig. \ref{validation}A, we compare our 2019 estimates of migration from each country to New Zealand against the government's official figures. The two series are closely aligned, with similar magnitudes and a correlation of 0.98\footnote{Note that our estimates are systematically larger for smaller estimated flows compared to the corresponding reported flows. This is in part because in the differential privacy step, if the number of migrants between a pair of countries becomes negative after adding this noise for each month, we censor the data at 0. After adding those noise annually, our estimates will be consistently greater than our raw estimates.}. In Fig. \ref{validation}C, we consider the time series dimension of the data, focusing on migration flows to New Zealand from India. Our estimates of monthly migration align closely with the official figures, picking up the spikes in Indian migration in February 2019 and February 2020, at the beginning of the academic year for tertiary education in New Zealand\footnote{In 2019, around 13,295 Indian students were pursuing government-funded tertiary education in New Zealand. This was the second-largest group of international students in the country \cite{eel2022}.}\footnote{Time series plots of migration from other countries to New Zealand are available in Fig. S8.}. We also observe that our data captures the dramatic decrease in migration that followed the government's almost-total closure of the border in March 2020, in response to the onset of COVID-19 \cite{nz2022}.

We next benchmark our data against the 2019 migration estimates provided by Eurostat, the statistical office of the European Union. Relative to the data from New Zealand, the Eurostat data is broad in scope, with information on inflows into 22 countries in Europe. The Eurostat data does have several limitations; it is only available at an annual granularity and is collected inconsistently across countries, making use of different methodologies and definitions of migration. We discuss these considerations in more detail in SI Validation. Nevertheless, we show in Fig. \ref{validation}B that our estimates are in general aligned with the figures provided by Eurostat, with a correlation of 0.94 and generally similar levels of migration. 

In SI Validation, we explore these benchmarks in greater detail, highlighting several instances in which our estimates diverge from the official statistics. In some cases, these discrepancies are driven by outliers in our model of the relationship between Facebook usage, income, and migration. For instance, we underestimate migration between Samoa and New Zealand, which is driven by a unique policy that allows for an unusually wide swath of Samoans to migrate\footnote{This policy, called the \href{https://www.immigration.govt.nz/about-us/media-centre/common-topics/samoan-quota-scheme}{Samoan Quota}, allows around 0.5\% of the Samoan population (and their dependents) to migrate to New Zealand each year. This allocation is unusually large and conducted by lottery, so migrants admitted under this scheme have a different pattern of selection than those admitted under other provisions of New Zealand's immigration system, which normally favors relatively educated individuals from sending countries \cite{nzpc2021}. The selection of Samoans migrating to New Zealand also contrasts sharply with the global pattern of migration to rich countries, which favors wealthier migrants and which is used to fit our model of selection.}. On other migration corridors, we find that our data seems to capture migration excluded from administrative statistics. For instance, the figures reported by Slovakia to Eurostat include only migrants who receive a permanent residency permit, rather than all those who meet the United Nations' definition of migration. In general, however, we find that our estimates are aligned with administrative figures.

On the whole, these validation exercises highlight the contribution of our estimates relative to existing measures of migration. Our figures generally match existing data where available, but cover a wider fraction of the world and provide finer time granularity at the monthly level. This data provides a strong basis for future research and policymaking related to the causes and consequences of migration.

\acknow{We are grateful for support on project management from Lada Adamic, Alex Dow, Laura McGorman, Alex Pompe, Itamar Rosenn, Omari Sefu; thoughtful research feedback from Jakub Bijak, Joshua Blumenstock, Joel Cohen, Eric Dunford, Harshil Sahai, Andrew Tatem, Yunus Berndt, Robert Beyer, Damien Jusselme, Marie McAuliffe, Robert Trigwell, Jean-christophe Dumont, Sara Mouhoud, Julia Gelatt, Stefano Iacus, Umberto Minora, Nadwa Mossaad, Jason Schachter, Dean Eckles, Martin Koenen; differential privacy from Brian Karrer, Yue Wang, Arjun Wilkins; and data validation from Betsy Bell, Pedram Jahangiry, Rafael Labrador, Wesley Smith, Troy Wissenbach.}

\showacknow{} 

\bibliography{main}

\end{document}



\maketitle

\SItext

\section*{Materials and methods}

\subsection*{Data}

In this paper, we use a variety of data sets to produce our migration estimates for Facebook users, to weight our estimates to capture population-level migration, and to validate our figures against external data. In this section, we describe the data we use in each of these processes.

\subsubsection*{Location assignment}

Our estimates of global migration flows are based on aggregated, weighted observations of changes in individual countries of residence for 3 billion monthly active users on the Facebook platform\footnote{See \href{https://investor.fb.com/investor-news/press-release-details/2022/Meta-Reports-Third-Quarter-2022-Results/default.aspx}{Meta Reports Third Quarter 2022 Results}}. We currently use data on migration between the start of 2018 to the end of 2023. Since the United Nations definition of migration requires continued residence in the origin and destination for a year, this allows us to produce migration figures for all months between January 2019 and December 2022. We determine the country of residence for each individual using an algorithm that takes into account signals such as the IP addresses they use, their self-reported home country, and their activity on the platform. The same predictive model for home country location has been used in other studies based on active users of the Facebook platform \cite{sahai_social_2022,cherifi_who_2020,herdagdelen_social_2016}. Due to a one-off change in the location prediction algorithm, many users were assigned to a new location in October 2021 and subsequent months. As these changes reflect a shift in methodology, rather than a true change in migration patterns, we impute the number of migrants in October 2021 for each country pair by averaging the level of migration in September 2021 and November 2021. The influence of the home country prediction algorithm change is typically realized within less than a month, which does not have a long-term effect. This means that the change in the algorithm represented a one-time adjustment to certain users' location assignments, rather than an ongoing impact on the process through which moves are imputed. Besides, the location prediction algorithm change resulted in a minimal proportion of users experiencing a change in their predicted home country because the existing algorithm already has a high degree of precision. We drop a small number of other data points, which include 47 country pair-by-day cells, from our dataset when changes in the home prediction algorithm affect individual country pairs before aggregating our data to the country pair-by-month level\footnote{This represents around 0.0001\% of the total country pair-by-day cells in our sample (47 cells / ((181 * 180) country pairs * 1460 days)).}. 
We additionally drop a small number of country pair-by-month cells due to data irregularities\footnote{We omit, in total, 101 country pair by month cells, which comprises 0.006\% of total cells (101 / (48 months * (181 * 180) country pairs)). These cells are: Uganda to Syria between May 2019 and September 2019, Syria to Uganda between May 2020 and February 2021, Myanmar to Cambodia between November 2020 and December 2022,  Senegal to Guinea-Bissau between January 2019 and July 2020,  Timor-Leste to Indonesia between September 2019 and December 2022 and  Afghanistan, Argentina, Chad, Turkey, and Uganda to Sudan between January 2019 and February 2019.}.
We also omit certain country pairs due to sensitive geographies or countries with disputed borders where we are not sure our estimates would be capturing migrants as opposed to large volumes of other cross border movement.

\subsubsection*{Population data}

As part of the migration estimation process, we use population counts when calculating the weights in each country, as described in SI Weighting section. We use annual population estimates for 2019-2022 from the World Bank\footnote{See \href{https://data.worldbank.org/indicator/SP.POP.TOTL}{here} for the data source}, which compiles data from the United Nations Population Division, national statistical publications, Eurostat, the United Nations Statistical Division's Population and Vital Statistics Report, and the Secretariat of the Pacific Community. We also use the population data as part of our validation exercises to normalize the immigration counts by the population of the receiving country.

\subsubsection*{Validation data}

We use a variety of alternative migration estimates to validate our data, which are described in more detail in SI Validation. In all the analyses below, we focus on using data on reported immigration, since this is generally thought to be more accurate than data on emigration \cite{willekens_monitoring_1994,willekens_evidence-based_2019}. 

In the Validation section and in SI Validation, we make extensive use of migration data from the National Statistical Office (NSO) of New Zealand, which provides monthly migration statistics on its website \cite{newzealand}. This data has been widely explored in prior research and is thought to be of comparatively high quality \cite{hugo2016migration}. In this context, individuals are counted as migrants if they are ``an overseas resident who arrives in New Zealand and cumulatively spends 12 out of the next 16 months in New Zealand." This contrasts with the recommendation of the United Nations, which requires migrants to live in their destination for the majority of 12 consecutive months to be considered residents. To account for this difference, we adjust the parameters in our model (see SI Algorithm) to match those used by the New Zealand government when benchmarking our data against theirs. This modification only applies to the benchmarking step; when measuring migration to New Zealand in the body of the paper, we use the United Nations' definition of migration. 

In SI Validation, we also make use of data from Sweden, which publishes migration statistics based on the national population register\footnote{See \href{https://www.statistikdatabasen.scb.se/pxweb/en/ssd/START__BE__BE0101__BE0101J/ImmiEmiFlyttN/}{Immigrations and emigrations data} from Statistics Sweden}. Migrants from abroad who intend to reside in Sweden are required to register in order to obtain a personal identification number that is widely used by the government. During the registration process, the previous country of residence is recorded, alongside other details, which allows for  the enumeration of annual immigration flow data by country of origin\footnote{See \href{https://unece.org/sites/default/files/2022-10/G2_Presentation_Forsberg_ENG.pdf}{this report} by M. Forsberg for further information.}. 

In the Validation section and in SI Validation, we also employ data from Eurostat\footnote{This data is available on the Eurostat \href{https://ec.europa.eu/eurostat/databrowser/product/view/migr_imm5prv?lang=en&category=migr.migr_cit.migr_immi}{website} as series migr\_im5prv.}, the statistical office of the European Union, which collates data on immigration statistics from various NSOs. Each member nation has developed different systems for collecting and defining migration. Some member countries, such as Sweden and Denmark, base their figures on population registers, while others, such as Ireland and Portugal, employ estimates based on labor market surveys. Consequently, both data quality and the exact definition of migration vary across countries. In recent decades the European Union has introduced legal frameworks to harmonize the definitions used by countries in their required reports to Eurostat, though the definitions and methodologies are not yet perfectly compatible\footnote{A full list of methodologies can be found in Section 18 of the description of Eurostat series ``migr\_immi'', at \href{https://ec.europa.eu/eurostat/cache/metadata/en/migr_immi_esms.htm}{this} link.}. The data covers reported immigration flows to the 27 current members of the EU, along with Iceland, Norway, North Macedonia, Montenegro, Switzerland, and the United Kingdom. Most countries are present in each year, but data are missing in some countries, such as the United Kingdom, which is only present in 2019. For most countries, the immigration data is broken down by the country of origin (which can be a non-Eurostat nation). Some countries only report pairwise data to Eurostat for a subset of origin countries. When validating the total rate of migration to or from a country, we also include migrants whose exact origin is not specified\footnote{For more information, see \href{https://eur-lex.europa.eu/legal-content/EN/TXT/?uri=CELEX\%3A02007R0862-20210701}{here}. Article 3(1)(A)(iii) describes the mandate to collect this data, and Article 9(1) discusses allowable sources and methodologies. Table 18 of \href{https://ec.europa.eu/eurostat/cache/metadata/en/migr_immi_esms.htm}{this link} describes the processes used by each country to produce their immigration estimates.}.

We also validate our estimates using data from Germany\footnote{See \href{https://www-genesis.destatis.de/genesis//online?operation=table&code=12711-0011&bypass=true&levelindex=0&levelid=1689630546458\#abreadcrumb}{here} for data source and \href{https://www-genesis.destatis.de/genesis/online?operation=ergebnistabelleInfo&levelindex=2&levelid=1689630604870\#abreadcrumb}{here} for data details.} in a secondary validation exercise discussed later. The international migration flows to Germany are collected by the registration offices in Germany and are available at a monthly level, which makes them an appealing source of validation for our high-frequency estimates of migration. However, the time frame used by the German government to define migration events is much shorter than that recommended by the United Nations. We discuss these differences in more detail in SI Validation.

Finally, we compare our data against global estimates of migration that are imputed using changes in migrant stocks. This data is only available at a five-year granularity so in our comparisons, we compare our estimated level of migration in 2019 against the annual average in the 2015-2020 period \cite{abel_bilateral_2019}.

\subsection*{Algorithm for detecting migration events}

To determine if users migrate internationally, we search within each individual sequence of home country locations to determine if their home country changes for at least 12 months. This definition aligns closely with that used by the United Nations Statistics Division, which defines a person's usual residence as “the place at which the person has lived continuously for most of the last 12 months (that is, for at least six months and one day), not including temporary absences for holidays or work assignments, or intends to live for at least six months” (UN Statistics Division, 2008, p102).

We assume that all changes of 12 months or longer in the user’s home country correspond to changes in their country of residence (i.e., migration events), rather than extended absences for holidays or work assignments, as it is not possible to determine the purpose of any user’s change in their home country from signals on Facebook.

Our algorithm for detecting migrant events for each individual is based on Chi et al. \cite{chi_general_2020}. The method first detects each user's segments in time when they live in the same location (allowing for small periods in other places due to travel). The country of each segment is a person’s country of residence over that period of time. The method then designates them as a migrant if two adjacent segments are in different countries. Assume we detect two segments for a person: a segment in country A from Apr. 1, 2019 to May 1, 2020 and a segment in country B from May 2, 2020 to Dec. 1, 2024. Then this person is a migrant who migrated in May 2020 from country A to country B. This person is a resident of country A from Apr. 1, 2019 to May 1, 2020, and a resident of country B from May 2, 2020 to Dec. 1, 2024.

\begin{figure}
\centering
\includegraphics[width=\linewidth]{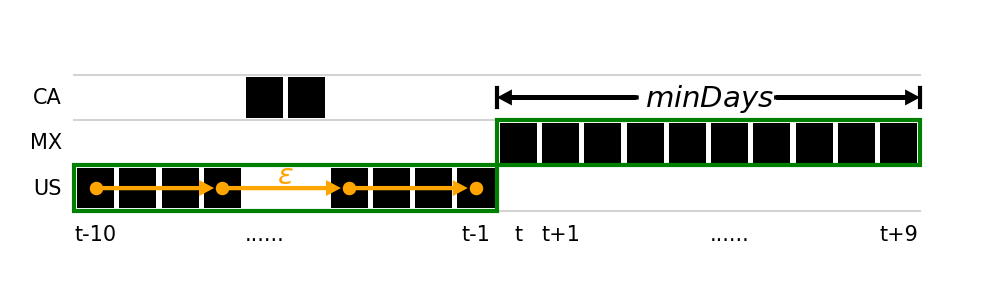}
\caption{Diagram of the segment-based algorithm, portraying the location history of one individual. Each row is one country, while each column is one day. The black square means this person lives in that country on that day. The green rectangles are two segments. $\epsilon$ defines the maximum gap between consecutive days. $minDays$ defines the length of each segment. $propDays$ defines the proportion of days in each segment. In this case, it is 80\% in the segment of Mexico, and 100\% in the segment of the US. \label{algorithm}}
\end{figure}

To detect segments, we define the maximum gap between consecutive days $\epsilon$. A segment of time in a given country will be considered continuous if there are no periods of more than $\epsilon$ days in that period in which the user is seen only outside the country. Detected segments might have different lengths and might have different proportions of days when a person lives in the country of residence. To ensure that the detected segments meet the length of residency and that a person lives there for most of their time, two additional parameters are used by the algorithm for 1) the minimum length of the segment $minDays$ and 2) the proportion of days in each segment $propDays$ (Fig. \ref{algorithm}). We set each of these parameters to match the definition of long-term migration recommended by the United Nations. We set the segment length to at least 12 months and the proportion of days in each segment to 50\%. As the United Nations does not provide a recommendation of the maximum gap between consecutive days in the destination country (referred to as the radius), we set this parameter to 60 days (see SI Alternative radii for details). In some cases, gaps between the last day of a segment and the start of the subsequent segment are visible. We drop individuals from the estimation if there is a gap of longer than 60 days due to uncertainty about the migrant's residence during this time. See Chi et al. \cite{chi_general_2020} for a full explanation of the segment-based algorithm to detect migration events. 

We explore variations in the parameter values for the algorithm elsewhere in the paper. For instance, when validating our data against the migration data provided by the New Zealand NSO, we set $minDays$ to be 16 months (487 days), and $propDays$ to be 75\% (12 months), to match the definition used in their administrative data. We also present results after setting $minDays$ to be 6 months (182 days), and $propDays$ to be 50\% (3 months), which allows us to measure more recent migration trends (see SI Alternative minimum segment lengths).

\subsubsection*{Alternative radii}

We considered several factors to inform our decision to set the radius $\epsilon$ to be 60 days in our segment-based algorithm. As the value of $\epsilon$ increases, it allows users to spend more consecutive days outside of their country of residence during a segment. When $\epsilon$ is large, we can better account for users who temporarily travel outside their country of residence, but segments are then more likely to overlap, if, for instance, a user alternates stints in two countries. A smaller radius lessens the number of overlapping segments but imposes stricter requirements on the length of trips that users can take abroad during their residence in a given area. 

To justify our choice of $\epsilon$, we consider three metrics. Our first metric is based on analysing the dominance of the modal country within each segment. Concretely, we measure the percentage of segments in which the modal country within the segment accounts for more than 90\% of days in the segment. We also consider an alternative metric which measures the number of segments in which the most common country and second-most common country are almost as frequent as one another, differing by fewer than 20\% of the total number of days in the segment. 

Our third metric combines the number of transitions and the longitudinal entropy, following the procedure outlined in Gabadinho, Ritschard, Studer, and Müller \cite{gabadinho_indice_2010}. Specifically, we calculate for any sequence of $s$:
\begin{equation*}
    C(s)= \sqrt{\frac{\ell_d(s) - 1}{\ell_{s} - 1} \,\frac{h(s)}{h_{max}}}
\end{equation*}

\noindent where $h_{max}$ is the theoretical maximum entropy given the cardinality of our set of countries. $h_s$ is the entropy of a sequence $s$. $\ell_d(s) - 1$ is the number of transitions in a sequence $s$. $\ell(s) - 1$ is the maximum number of transitions in a sequence $s$. The complexity index attains its minimum value of 0 when there are no transitions in a sequence and its maximum value when each state in a sequence is different. In other words, if a sequence consists of a single state that is repeated throughout, the complexity index will be 0, whereas if each state in the sequence is unique and different from all the others, the complexity index will be at its maximum.

\begin{figure}
\centering
\includegraphics[width=\linewidth]{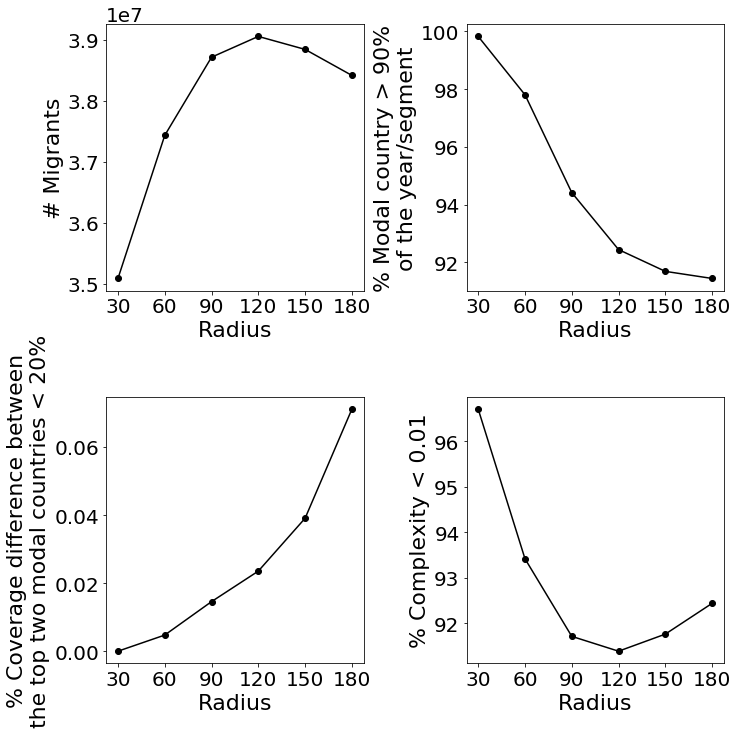}
\caption{Effect of different radii on migration results. The first plot shows the number of migrants detected using different radii. The three other plots show the effect of different radii on different metrics. \label{segment_different_radius}}
\end{figure}

As shown in Fig. \ref{segment_different_radius}, we find that, when we set a lower value of epsilon, migrants' location histories exhibit lower complexity, and there are fewer cases where users split their time nearly equally between two countries during a segment. At the same time, we find that the overall number of detected migrants falls substantially when we set $\epsilon$ to lower values, particularly when we set it equal to 30 days. In this case, we drop over 6\% of our total number of migrants, as we ignore segments that might include short visits to family, tourist excursions, or business trips. When $\epsilon$ is small, the detected segments need to have a very continuous record in a country because the gap between two days when a person lives in the country must be smaller or equal to $\epsilon$. In the extreme, when $\epsilon = 1$, a person must live in the same country every day to form a valid location segment. Consequently, using a smaller $\epsilon$ in the algorithm will lead to fewer migration events detected. 
When $\epsilon$ is large two potential issues might arise: (1) the proportion of days ($propDays$) a user is identified in the segment can become smaller as the larger $\epsilon$ allows for more gaps between the days a person is located in their country of residence and (2) neighboring segments are more likely to overlap, which will be dropped in our algorithm. The remaining segments after dropping the overlaps might be shorter than 12 months, and hence will no longer be valid for detecting migration events based on the UN definition. These two issues result in fewer migrant events detected when using larger $\epsilon$, and hence we choose to set $\epsilon$ to equal 60 days to balance these considerations.

\subsubsection*{Alternative minimum segment lengths}

Estimates of international migration are typically available only with a considerable time lag for publication. Our estimates rely on digital trace data that is gathered in near real-time. As we adopt the United Nations' recommended definition of migration, which requires migrants to reside in their destination for the majority of a year, we need to wait until the completion of 12 months to detect migration events for the past year. In this section, we explore an alternative definition of migration which could allow us to measure global migration with a shorter time lag. 

\begin{figure}
\centering
\includegraphics[width=.5\linewidth]{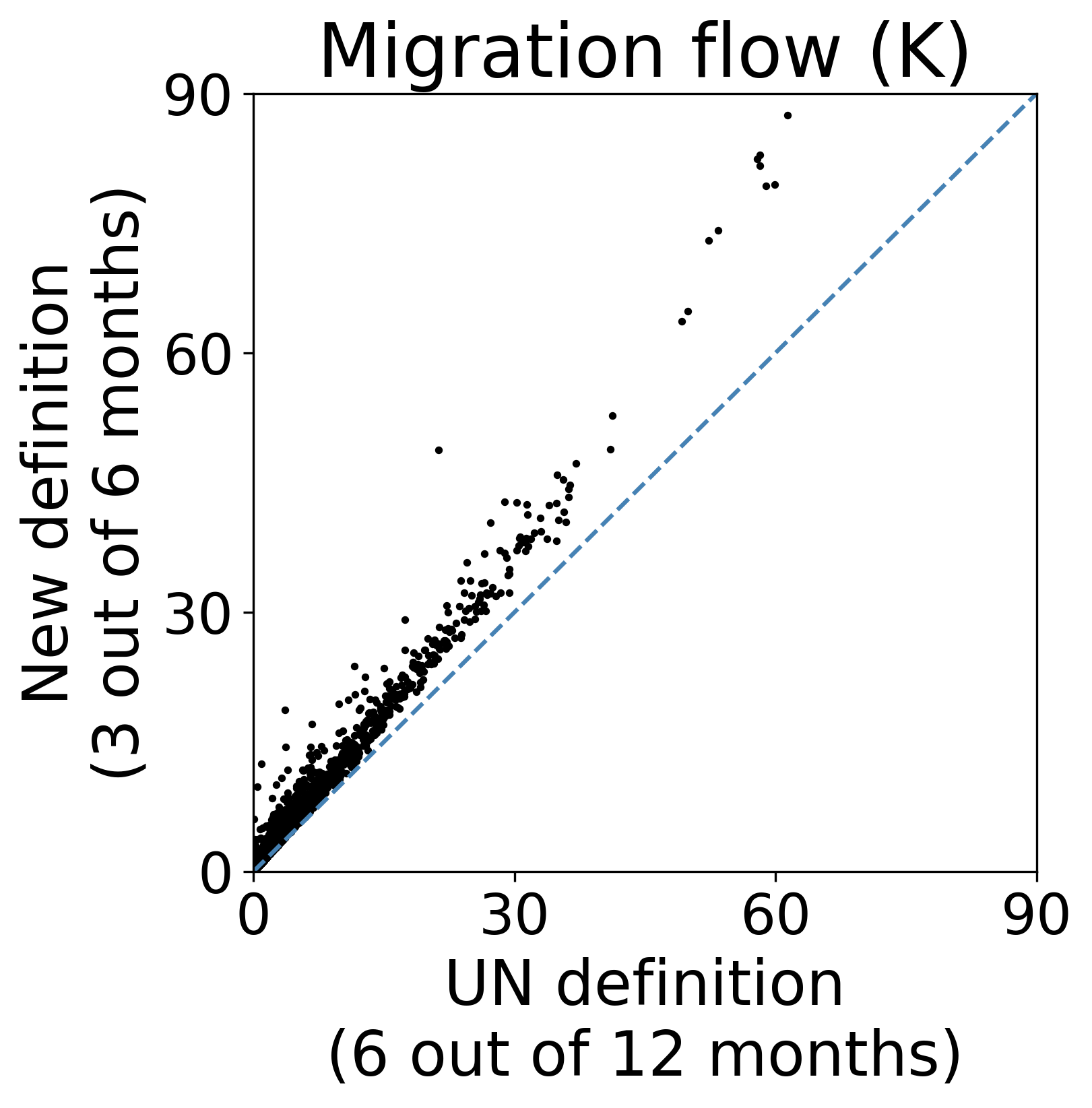}
\caption{Relationship between the number of migration flows based on the United Nation's recommendation (which requires residence in the destination for at least 6 out of 12 months) and based on the new definition (which requires residence in the destination for at least 3 out of 6 months), for each country pair and month in 2019.\label{new_definition_vs_un_definition}}
\end{figure}

To create timelier estimates, we consider a different definition for residence, in which a migrant still needs to be a resident of their origin region for more than six months in a 12-month period, but the required duration in their destination is shorter--only 3 months in a 6-month span. This change allows us to capture migration with only a 6-month lag, at the cost of capturing some migration that does not match the United Nations’ definition. As shown in Fig. \ref{new_definition_vs_un_definition}, using this shorter definition, we detect slightly more migrants than we do when applying the United Nations' definition due to our inclusion of shorter-term migrants, which is consistent with the findings of Nowok and Willekens \cite{nowok_probabilistic_2011}.

\begin{figure}
\centering
\includegraphics[width=.8\linewidth]{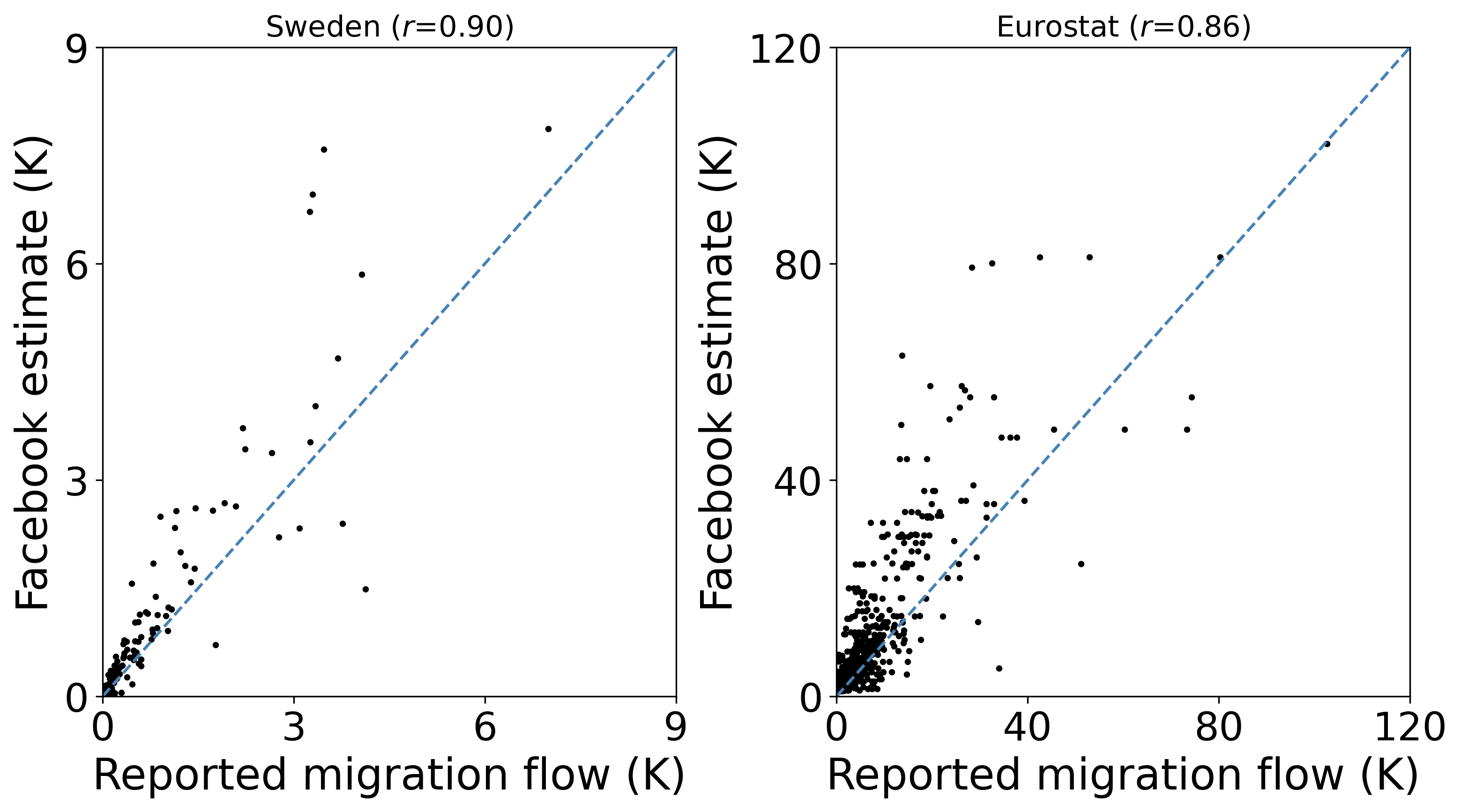}
\caption{Validation with Sweden and Eurostat migration datasets based on the new definition (a person living in a new place for at least 3 out of 6 months). \label{new_definition_validation}}
\end{figure}

In Fig. \ref{new_definition_validation}, we correlate our estimated migration levels in 2019 using this method against data from Sweden and Eurostat. We can see that, although we tend to slightly overestimate migration with this new definition, we still see a strong relationship between the reported migration statistics, with a Pearson correlation of 0.90 with the Swedish data and 0.86 with the Eurostat data. We do not benchmark the short-term results against New Zealand's data because its definition of migration implies even longer minimum stays than the United Nations definition.

\subsubsection*{Comparison with frequency-based method}

The frequency-based approach, in which one's residence is considered to be the location where a user spends the majority of days in a given calendar year, is a common method to detect migration \cite{lu_unveiling_2016,zagheni_inferring_2014,blumenstock_migration_2019}. In this methodology, an individual is considered a migrant if their modal location changes between calendar years. Prior work has found that the segment-based method has a better performance than the frequency-based method based on human-labelled data \cite{chi_general_2020}, but these two methods have not been previously compared in the context of the digital trace data that we use in this paper. In this section, we benchmark our preferred segment-based method against a frequency-based method and highlight how the two methods map onto the United Nations' preferred definition of migration. 

Although the modal country is measured annually with the frequency-based methodology, this approach does not ensure that a migrant lives primarily in a country for an entire year. Correspondingly, the frequency-based definition classifies a number of short-term moves as migration events. In Table \ref{segment_vs_frequency}, we show that 18.7\% of moves under the frequency-based method have a duration of fewer than 300 days, which are not classified as migration events in the segment-based definition of migration or the UN definition. The frequency-based approach also diverges from the United Nations’ definition of migration and from the segment-based approach as it does not impose any constraints on the minimum number of days that an individual must be present in their destination during the residence period.  As a result, we find that the frequency-based definition detects many migration events for individuals who frequently move between locations, even though these short-term moves are not captured by the segment-based approach or by the United Nations' definition of migration.

In Table \ref{segment_vs_frequency} we demonstrate that using the segment-based definition, 97.8\% of the individuals spend 90\% or more of their residence period in the country assigned to them. In the frequency-based approach, individuals only spend 70.5\% of their time in their assigned country of residence. We also find that the frequency-based method leads to a far higher number of cases in which individuals split their time evenly between two countries, but are still recorded as migrants. 
As a result of these factors, we find that the frequency-based method assigns users more complex and less continuous migration histories.

The frequency-based approach also has another additional drawback relative to the segment-based approach, namely, it does not neatly capture the date on which a migrant moved. For example, for a person whose modal home country changed from one country in 2019 to another in 2020, it is unclear if the person migrated in 2019 or 2020. This issue makes it harder to directly compare the migration estimates from the frequency-based technique with those provided by administrative sources. Nevertheless, we can still get a sense of how our migration estimates using this method compare to administrative data by comparing them to both of the neighboring years. For instance, we can compare our frequency-based migration estimates for 2018-2019 and 2019-2020 against the Eurostat migration estimates from 2019 at the country-pair level, finding Pearson correlations of 0.79 and 0.86, respectively. These results compare unfavourably with those from our segment-based method in 2019, which had a correlation of 0.93 with the Eurostat estimates, see SI Validation for full details on our validation exercises. 

In summary, we find the segment-based approach performs better than traditional frequency-based methods for detecting international migration events in four aspects: (1) individuals spend a higher proportion of their time living in their detected county of residence, (2) individuals have fewer transitions across countries, (3) the precise timing of migration events are known, and (4) comparisons with validation data are facilitated by the use of more-similar definitions of migration.

\subsection*{Weighting}

We observe migration decisions only for individuals who actively use Facebook. In order to provide estimates for the total population level of migration, including migration among non-users, we adjust our initial estimates using a variety of weighting mechanisms. An ideal weighting mechanism would allow us to account for a variety of (potentially unobserved) factors that affect both Facebook usage rates and migration propensity. In this section, we discuss a range of weighting methods used in migration estimation. To benchmark these methods, we use each one to weight the level of migration we observe among Facebook users, and benchmark these series against administrative records documenting the number of individuals arriving in Sweden during each year between 2019 and 2022. Sweden's migration data is derived from the country's mandatory population register and is considered to be of high quality, as it is available at the country-pair level and captures essentially all long-term stays in the country, regardless of visa type \cite{wisniowski_utilising_2013}.

\subsubsection*{Raw estimates}

We first benchmark our initial estimates against the Swedish administrative data, calculating both the Pearson correlation between the two series and the sum of the absolute errors. As shown in the third column of Table \ref{reweight_methods_result}, the raw estimates are highly correlated with the Swedish data ($r$ = 0.84, 0.93, 0.95, 0.96 from 2019 to 2022)\footnote{We do not include the correlation plot here because it could be used to infer the penetration rate of Facebook users in each country, which is not public information.}. Though the two series are highly correlated, we can also see that the sum of the distance between the initial estimates and reported migration statistics is relatively high. In Table \ref{reweighting_methods_migrants_num}, we show that this is largely because we underestimate the level of immigration from almost all countries since we do not account for the migrants to Sweden that do not use Facebook.

\subsubsection*{Selection rate}

\begin{figure}
\centering
\includegraphics[width=\linewidth]{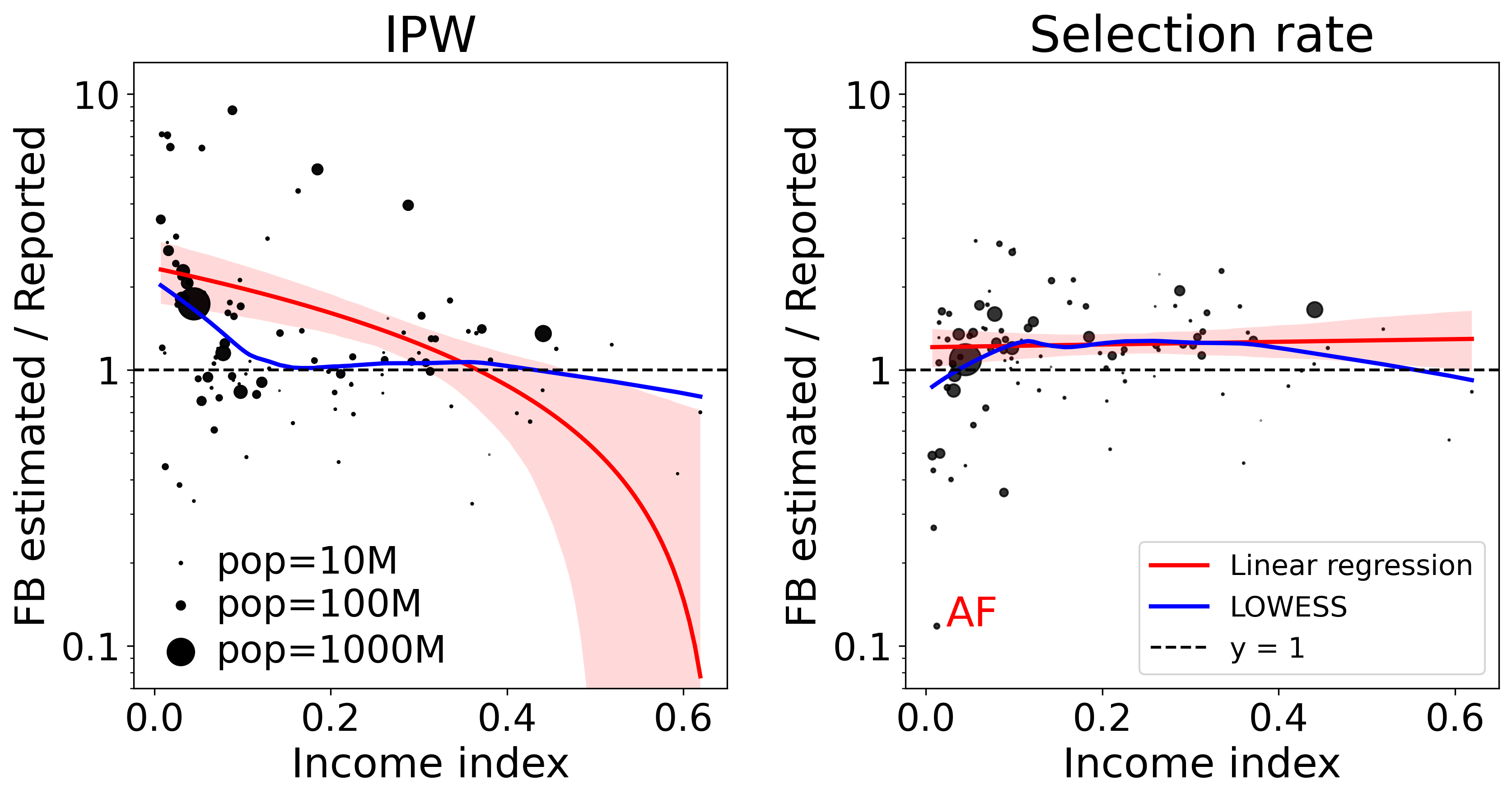}
\caption{Ratio of estimated flows to administrative records in 2019, using the inverse penetration rate weights (left panel) and the selection rate weights (right panel). The income index is described in Equation 2. Countries with under 100 migrants in the administrative data in 2019 are excluded from the plot.\label{reweighting_selection_rate_vs_ipw}}
\end{figure}

In the main results discussed in the paper, we use a hybrid approach that harnesses several features of the joint relationship between socioeconomic status, migration, and Facebook usage. In Fig. \ref{reweighting_selection_rate_vs_ipw}, we show that when we weight our raw data using the inverse penetration rates, we obtain relatively accurate estimates for high-income countries, but we tend to dramatically overestimate the rate of migration from lower-income countries. These results indicated that there is variation in the degree of selection into migration and Facebook usage across countries, with less-developed countries tending to have a less representative sample on the platform. This finding aligns with a large literature in economics highlighting the barriers that poor individuals face both in migrating and in accessing communications technology \cite{clemens2014does, poushter2018social}. In developed countries, these constraints bind fewer people, and the degree of selection into migration and Facebook usage is likely to be less severe.

With this intuition in mind, we can describe the construction of the weights used in our main specification. In high-income origin countries, we would like our weights to resemble the inverse of the penetration rate in origin country $o$ ($\frac{FBUsers_{o,t}}{Population_{o,t}}$), given that Facebook users in such countries are broadly representative of the broader population in their migration propensity (following the analysis in the first panel of Fig. \ref{reweighting_selection_rate_vs_ipw}). In lower income origin countries, we have observed that Facebook users tend to be more likely to migrate than the population at large, so we would like our weights to assume that the degree of non-random selection into our sample increases as we consider origin countries with lower incomes. Correspondingly, we would like to place relatively less importance on the penetration rate when weighting the data for these countries. To parameterize this relationship, we define the denominator of our weights as a linear combination of the penetration rate in origin country $o$ and a fitted constant $r_t$ (whose construction we will describe shortly), which captures the degree to which selection into our sample is more severe in low-income countries. The contribution of each term to the overall weights is determined by the per-capita income of each country, scaled such that the highest-income country has a value of 1. Concretely, we can define the weight placed on each migrant from origin $o$ in year $t$ as:

\begin{equation}
\label{selection_weights}
W_{o,t}= \frac{1}{\text{Income}_{o} \times \frac{FBUsers_{o,t}}{Population_{o,t}} + (1 - \text{Income}_o) \times r_t}
\end{equation}

Here, $\text{Income}_o$ is calculated using the 2019 Gross National Income per capita (GNIpc)\footnote{See \href{https://hdr.undp.org/data-center/documentation-and-downloads}{here} for data source} of country $o$, using the following formula:
\begin{equation}
    \text{Income}_o = \frac{GNIpc_o}{\max{(GNIpc_o)}}
\end{equation}

\begin{figure}
\centering
\includegraphics[width=.7\linewidth]{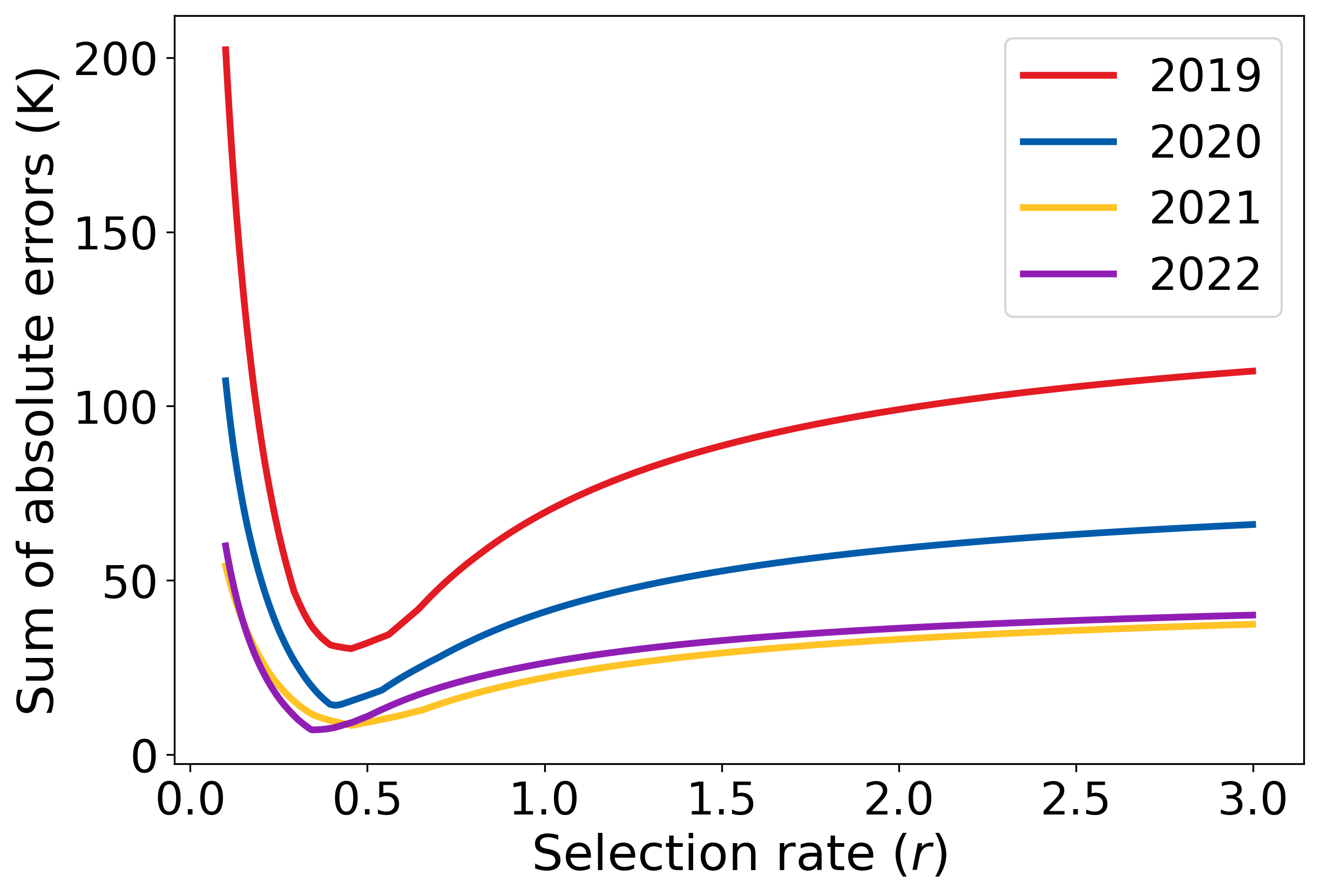}
\caption{The effect of selection rates on the sum of absolute errors based on the NSO dataset in Sweden. We exclude migrants from Ukraine to Sweden from these analyses since Sweden did not include those protected under the Temporary Protection Directive from its immigration figures; see SI Validation Sweden for more details.\label{reweighting_selection_rate_optimal}}
\end{figure}

To apply these weights in practice, we need to tune a single parameter for each year, $r_t$, which controls the degree to which selection into the sample varies with development in year $t$. To choose the parameter used in the paper,  we test a variety of possible values for $r_t$ and benchmark the resulting weighted migration figures against figures from New Zealand's NSO at the country pair level. For each year, we test candidate values of $r_t$ between 0 and 3 using the New Zealand data in steps of 0.01 and select the parameter that minimizes the sum of the absolute errors across all origin countries. We find that the optimal value of $r_t$ is 0.46 in 2019, 0.41 in 2020, 0.46 in 2021, and 0.35 in 2022. We then use these values of $r_t$ to weight our data for all other country pairs in that year. Moreover, while the migration patterns of New Zealand may not be representative for most countries, we observe that a small variation in the $r$ measure results in a minor change in the absolute errors, which further strengthens the validity for calibrating $r$ using New Zealand data. In Fig. \ref{reweighting_selection_rate_optimal}, we illustrate how the sum of the absolute errors in the Swedish data varies with $r_t$ in each year. The optima we find here are quite similar to the optima estimated with the New Zealand data, hinting at the generalizability of the parameters we estimate. We show in columns 11 and 12 of Table \ref{reweight_methods_result} that this weighting method results in the lowest total error rate on the Swedish data for 2019, 2021 and 2022, and on the Eurostat data for all the years in our sample.

In principle,  this method could be expanded to allow for $r_t$ to differ according to the level of development of the destination country. This could capture the fact that the degree of selection on socioeconomic characteristics into migration could vary according to the wealth of the destination, even holding fixed one's origin country. Our ability to investigate the sensitivity of $r_t$ to these destination characteristics is limited by the scarcity of high-quality immigration data for less-developed countries. With that said, in Table \ref{reweight_methods_result_eurostat}, we highlight the strong performance of this methodology using a second validation set from Eurostat. Though this is not a globally representative dataset, the figures in Table \ref{reweight_methods_result_eurostat} show that the selection rate methodology outlined here performs well across a variety of destination countries whose GDP per capita varies by more than a factor of 10 (from a high of \$113,196 in Luxembourg to \$9,518 in Bulgaria)\footnote{The Eurostat figures are reported individually by each country and vary somewhat in their collection methods and definitions, so some degree of caution is needed in interpreting these results.}. We present more analyses in SI Validation below.

\subsubsection*{Inverse penetration rate weights}

The simplest weighting approach is to weight our initial migration estimate from origin country $o$ to destination $d$ in year $t$ by the inverse of the Facebook penetration rate in the origin country in that year: $W_{o,t}= \frac{Population_{o,t}}{FBUsers_{o,t}}$.  Under this approach, we assume that no factors affect both an individual's migration decision and their rate of Facebook usage. In columns five and six of Table \ref{reweight_methods_result}, we demonstrate the implausibility of this assumption, showing that applying the inverse penetration rate weights increases the absolute level of error in our estimates and decreases the correlation of our figures with the Swedish data. 

In Column 5 of Table \ref{reweighting_methods_migrants_num}, we present the estimates of the total level of migration to Sweden after applying the inverse penetration rate weights. We see that we now overstate the aggregate level of migration to Sweden, which suggests that individuals we observe on Facebook are more likely than an average person in their country to migrate to Sweden. We show in Fig. \ref{reweighting_selection_rate_vs_ipw} that the degree of selection is non-uniform with respect to the income of one's origin country. Among wealthy sending countries, it seems that there is little selection into the sample on the basis of migration likelihood, with the inverse penetration weights causing our estimated flows to closely match the administrative data. However, in less developed countries, we see that Facebook usage appears to be non-random, with the inverse propensity weights causing us to overstate migration several-fold. For this reason, we will now turn our attention to weighting methods that can account for these differential rates of selection.

\subsubsection*{Raking ratio estimation}

In addition to a user's location, we are also able to identify users' age, sex and region within their country. We can use these demographic characteristics to further adjust for the non-random nature of Facebook usage and migration. In particular, we can use these additional details to calculate user weights that account for uneven Facebook usage within a country. We use an iterative proportional fitting method (also known as raking) to create weights for each age by sex by region group in our sample, such that the (weighted) prevalence of each group in our data matches the true distribution. To do this, we use WorldPop estimates of the population of each demographic group in each GADM-1 region\footnote{GADM-1 regions are a global set of top-level administrative divisions. In the United States, these correspond to states. See \href{https://hub.worldpop.org/doi/10.5258/SOTON/WP00698}{here} for the data source.}. The raking process was carried out using the survey package in R \cite{surveyr}. 

We are only able to calculate weights for subgroups that are observed in both Facebook users and possible global data sets on factors that might impact the variation in Facebook usage patterns. Consequently, we are restricted to considering a relatively narrow set of characteristics in our weighting model, where we are unable to directly target other potentially relevant socioeconomic characteristics. We see in columns 7 and 8 of Table \ref{reweight_methods_result} that the raking methodology we employ does not improve the correlation of our data with the Swedish administrative figures. Instead, our correlation decreases and the sum of the absolute errors increases, particularly in low-income countries. 

\subsubsection*{Coefficient}

The methods discussed above are appealing because the weights used can be calculated directly from country-level demographic data, which is available globally. However, as we have seen, these methods tend to perform poorly because there is likely to be selection into our sample based on characteristics that are not observed across all countries in both Facebook and administrative datasets. In this section, we demonstrate how we can use additional migration data to help us estimate the parameters of a simple model of selection into our sample.

We use a second set of administrative data from the NSO of New Zealand to estimate this ratio, regressing the government's estimates of the number of arrivals from each country on our raw estimates. We use data from New Zealand since its migration figures are also thought to be of exceptionally high quality \cite{hugo2016migration}. We then use the estimated coefficient we find as our weighting factor, which is common across all country pairs. In this model, we assume that we observe a fraction of the true number of migrants between each pair of countries, where this fraction is constant across country pairs and over time. 

We present the results on the Swedish arrival data in columns 9 and 10 of Table \ref{reweight_methods_result}. Since we are solely multiplying all of our estimates by a common factor, the correlation is unchanged, although this methodology slightly improves the overall error rate. This implies that the overall rate of selection into migration is relatively similar across the two destinations when we consider all origin countries. That said, the rate of selection into migration seems to vary according to the characteristics of the sending county (see Fig. \ref{reweighting_selection_rate_vs_ipw}), which we cannot capture using this single-parameter approach \cite{clemens2014does}.

\subsection*{Differential privacy}

We use techniques from the literature on differential privacy to protect information about individuals in the aggregated data we make public. Dwork and Roth (2014) \cite{dwork_algorithmic_2014} proposed the definition of differential privacy and provided typical mechanisms to achieve differential privacy. A randomized algorithm $M$ is $(\epsilon, \delta)$-DP if for any pair of data points, $X$ and $Y$ differing at most one row (e.g., an individual in a dataset), and any event $E$ (i.e. all potential output of $M$),

$\Pr[M(X)\in E]\leq e^\epsilon\Pr[M(Y)\in E]+\delta$

Balle and Wang (2018) \cite{balle_improving_2018}, and Dong, Roth, and Su (2020) \cite{dong_gaussian_2022} recently proposed the approach to inject random noise from the Gaussian distribution to the output, to achieve $(\epsilon, \delta)$-DP. A randomized algorithm $M(x) = f(x) + Z$ with $Z \sim \mathcal{N}(0,\,\sigma^{2})$ is $(\epsilon, \delta)$-DP if and only if

$\Phi(\frac{\Delta}{2\sigma} - \frac{\epsilon\sigma}{\Delta}) - e^\epsilon\Phi(-\frac{\Delta}{2\sigma} - \frac{\epsilon\sigma}{\Delta})\leq\delta$

\noindent where $\Delta$ is L2 sensitivity, which is defined as the L2 norm ($\sqrt{\sum_{i=1}^{k} v_{i}^{2}}$) of the maximum possible difference in all the output aggregates caused by adding or removing one individual. We adopt such an approach as it is easy to implement and the noise magnitude is relatively small. Since an individual can only migrate once per year under the definition we use in this paper, the impact of adding or deleting an individual from our data can only affect the value in any specific cell by at most +1 or -1. Under the assumption that we will collect and release yearly data for a total of 10 years, and we will share the data of the total number of migrants, the number of migrants by age, and the number of migrants by sex, the L2 sensitivity will be $\sqrt{30}$. 

We use $\epsilon$ = 10, $\delta$ = $10^{-9}$, $\Delta$ = $\sqrt{30}$, which implies that we should inject random noise drawn from a Gaussian distribution with parameters (0, $\sigma$ = 3.56). Accordingly, in 95\% of the cases, we will add or subtract fewer than 7 people to each country-by-country-by-month cell. If the number of migrants between a pair of countries becomes negative after adding this noise, we censor the data at 0.

\section*{Validation}

\subsection*{New Zealand}

\begin{figure}
\centering
\includegraphics[width=1\linewidth]{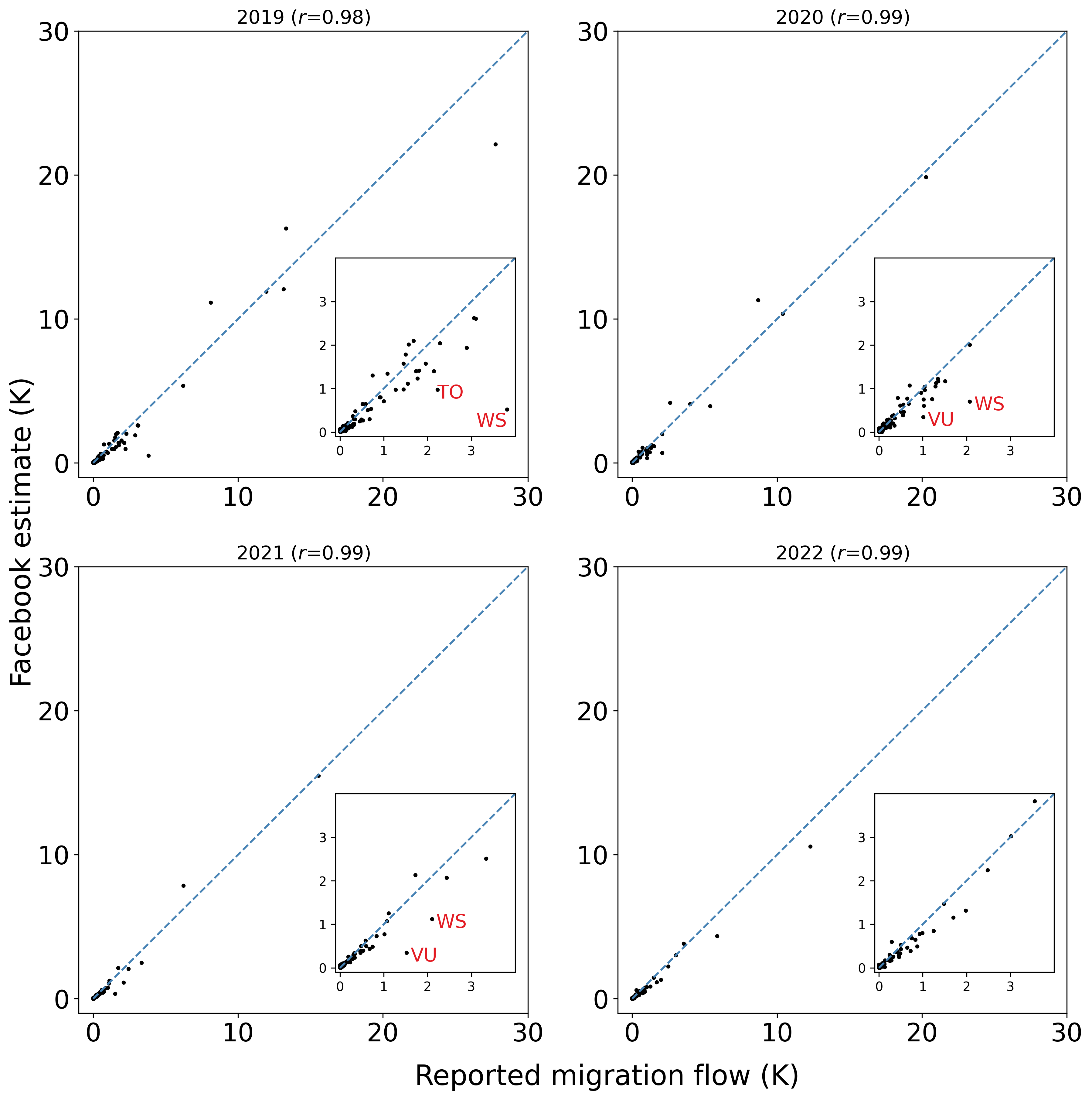}
\caption{Validation in New Zealand from 2019 to 2022. Each dot represents annual migration from an origin country to New Zealand. All axes are in thousands of people. \label{validation_nz_2019_to_2022}}
\end{figure}

In Fig. \ref{validation_nz_2019_to_2022}, we compare our estimated rates of migration into New Zealand with the figures reported by the country's NSO. On the X axis, we present the official estimate of the migration rate into the country from each origin. We benchmark this against our estimated level of migration, which is weighted using the selection weighting approach described in SI Weighting\footnote{The data from New Zealand is based on a slightly different definition of migration than is used elsewhere in the paper. In general, we follow the United Nations definition, in which migrants are defined to be individuals who move to a new region for 12 months, having resided elsewhere for 12 months. In the case of New Zealand, migrants are defined as those who arrive in the country and live there for 12 of the next 16 months (perhaps discontinuously) and who lived in a different country for 12 of the previous 16 months. In all of the analyses presented in this section in which we compare our estimates to the figures from the New Zealand government, we use a set of Facebook estimates that are produced using parameters that match the New Zealand government's definition of migration. We use our baseline definition for migration when comparing our data against the figures produced by other NSOs.}. We find that our data matches the New Zealand government figures extremely well, with a Pearson correlation above 0.98 each year. These figures remain robust in 2020, 2021 and 2022, despite the onset of the COVID-19 pandemic, which dramatically changed the pattern of migration into New Zealand.

Nevertheless, we do highlight in Fig. \ref{validation_nz_2019_to_2022} several sending countries for which our estimates of migration to New Zealand are less aligned with the official figures, namely Tonga (TO), Samoa (WS), and Vanuatu (VU). These three countries are all categorized by three factors: strong cultural and regional ties to New Zealand, relatively low levels of Facebook usage, and preferential visa programs that make migration to New Zealand accessible to a wider swath of society than it would be without government intervention. For instance, New Zealand administered Samoa before the country's independence and signed a treaty in 1962 allowing for preferential immigration of Samoans under the Samoan Quota (SQ), which led to widespread emigration from the country, which now has 50\% of its population living abroad. Similar programs exist in New Zealand for Tongans (under the Pacific Access Category) and Vanuatuans (under a variety of temporary worker programs) \cite{doan2023migration}. These programs highlight a limitation of our selection weight approach---though we are generally able to approximate the rate at which economic selection into Facebook usage and migration coincides, we perform less well for country pairs where policies facilitate or hinder migration for certain groups. Since we fit only a single parameter in the selection weight model to explain migration under facilitated conditions and more traditional conditions (and there are many more cases of the latter), our model performs less well for these unusual cases.

\begin{figure}
\centering
\includegraphics[width=1\linewidth]{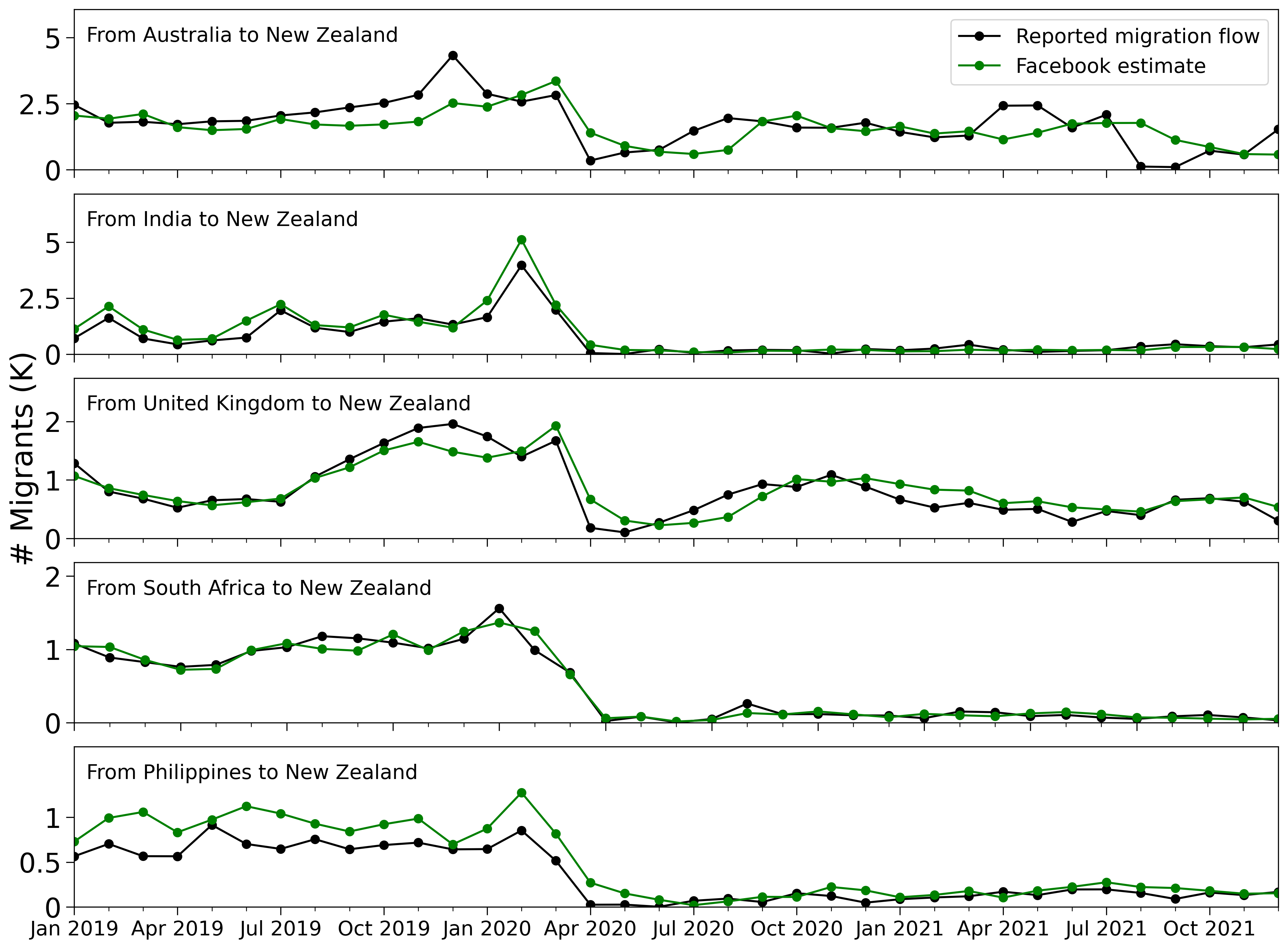}
\caption{Validation in New Zealand at the monthly level in the top 5 countries: Australia, India, United Kingdom, South Africa, and the Philippines. \label{validation_nz_monthly_top5}}
\end{figure}

In Fig. \ref{validation_nz_monthly_top5}, we further analyze trends in monthly migration to New Zealand from the top five sending countries. We demonstrate that our data capture the same monthly migration patterns as the administrative data, including seasonal variations and the impact of COVID-19 on inflows.

\subsection*{Sweden}

\begin{figure}
\centering
\includegraphics[width=1\linewidth]{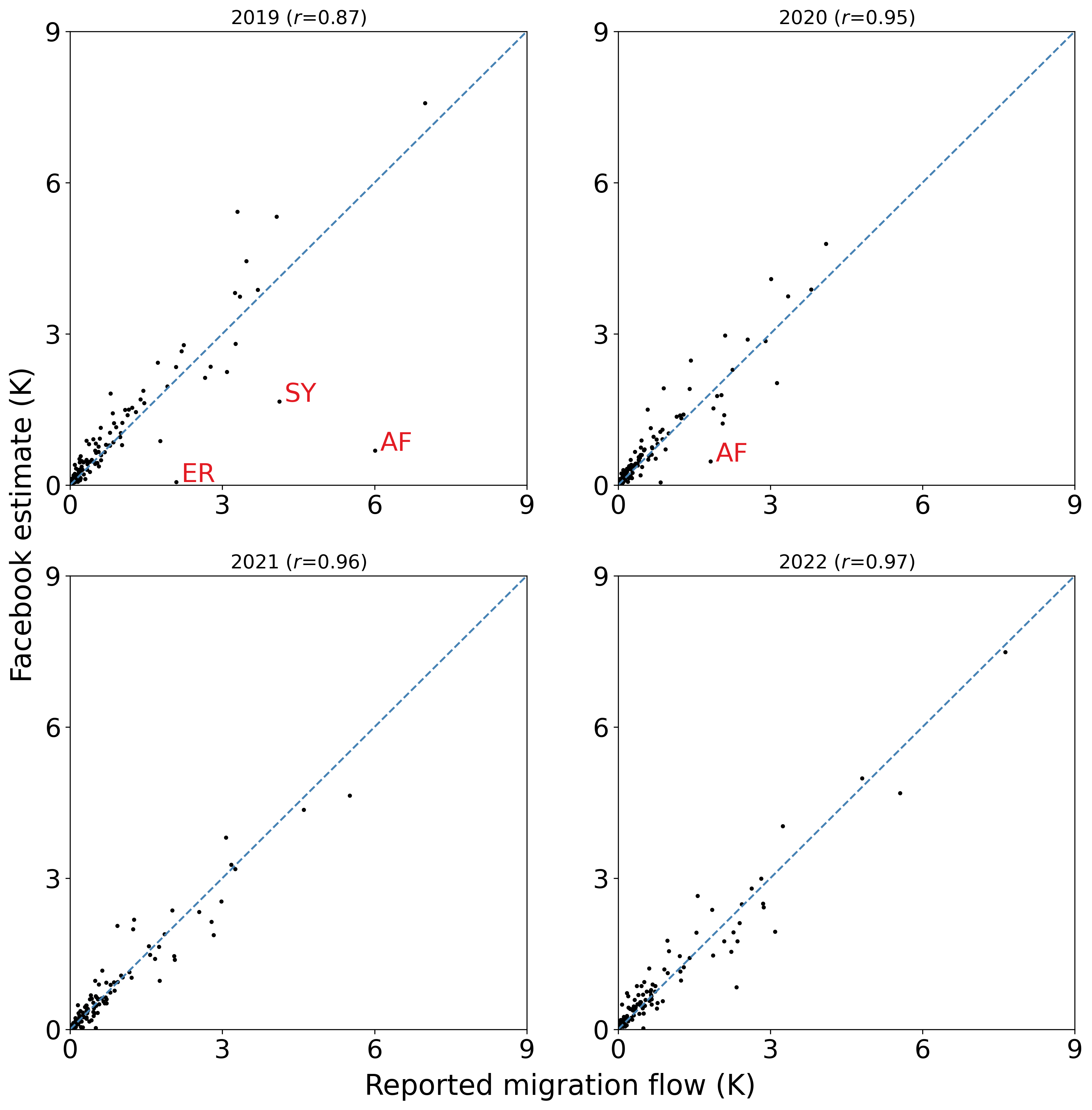}
\caption{Validation in Sweden in 2019, 2020, 2021, and 2022. Note that we do not add the point from Ukraine to Sweden in 2022 because our estimate of 25.6K is much larger than the scale in the y axis. The correlation in 2022 is calculated only using the data points shown in this plot (without Ukraine). \label{validation_se_2019_to_2022}}
\end{figure}

In Fig. \ref{validation_se_2019_to_2022}, we present analogous validation exercises, comparing our migration estimates to the administrative figures from Sweden. We find high Pearson correlations between the two sources of data, ranging from 0.87 in 2019 to 0.97 in 2022. Our estimates also closely align with the Swedish government figures in their magnitudes, indicating we are able to match the absolute as well as relative scale of the migration flow for most sending countries.

As in the case of the New Zealand validation exercises, several outliers stand out in this exercise. In 2019, for instance, our estimates are substantially lower than the Swedish government's figures for flows from Afghanistan and Syria, two countries which saw large outflows of asylum seekers during this period. Here, as in New Zealand, policy decisions made by the receiving government, paired with relatively low rates of Facebook usage in the sending countries (especially in Afghanistan and Eritrea) contributed to an unusual pattern of selection in our sample, one in which migration and Facebook usage are less tightly linked than for most migration corridors. We can see further evidence for this interpretation when we compare the 2019 and 2020 data. In 2020, the Swedish government changed its policy on residence permits, eliminating the presumption of eligibility that most Syrians were able to take advantage of in prior years\footnote{For more information on this policy, see \href{https://web.archive.org/web/20201220200922/https://www.migrationsverket.se/English/About-the-Migration-Agency/For-press/News-archive/News-archive-2019/2019-08-29-New-judicial-position-concerning-Syria.html}{here}.}. The following year, we no longer underestimate Syrian migration so substantially, perhaps reflecting that the pattern of selection into migration and Facebook usage based on unobserved characteristics shifted as a result of this policy change.

In 2022, we estimate that 25.6K people migrated from Ukraine to Sweden, far above the 800 migrants that the Swedish government reported to Eurostat. We believe that this distinction is driven by the fact that the Swedish government had not included most Ukrainians resettled under the Temporary Protection Directive in the population register used to calculate migration figures\footnote{For more information on this topic, see \href{https://www.migrationsverket.se/English/Private-individuals/Protection-under-the-Temporary-Protection-Directive/Frequently-asked-questions-about-the-Temporary-Protection-Directive.html}{here}. Some exceptions apply, such as migrants who are the partner of an EEA citizen.}. In 2022, the Swedish government separately reported that it had granted Temporary Protected Status to 47 thousand people after the Russian invasion of Ukraine\footnote{See \href{https://ec.europa.eu/eurostat/databrowser/view/migr_asytpfq/default/table?lang=en}{Eurostat} for these figures.}, which supports a substantially higher estimate of migration from Ukraine. We do not add these recipients of Temporary Protected Status to our validation figures due to discrepancies in the migration concept considered. Concretely, not all those who are granted Temporary Protected Status meet the standard definition of migrants, as individuals counted in this enumeration may return home or move to a third country before spending a year in Sweden. 

\subsection*{Germany}

To further assess the quality of our data at the monthly level, we compare our data against data from the German government on entries at the monthly level. The German government issues these figures at a monthly level, which makes this an appealing source of high-frequency validation data. With that said, the definition of migration used in constructing this data differs from our own in several key ways, leading to a difference in the level of the two series. Concretely, the German data measures the number of arrivals from abroad who declare their intention to remain in Germany for more than three months, rather than the twelve months required by the United Nations definition of migration\footnote{Additionally, this definition is based on the intended duration of one's stay, not the true duration, as in our implementation.}. The German data also does not require that the migrant be present in their previous location for a year.

\begin{figure}
\centering
\includegraphics[width=1\linewidth]{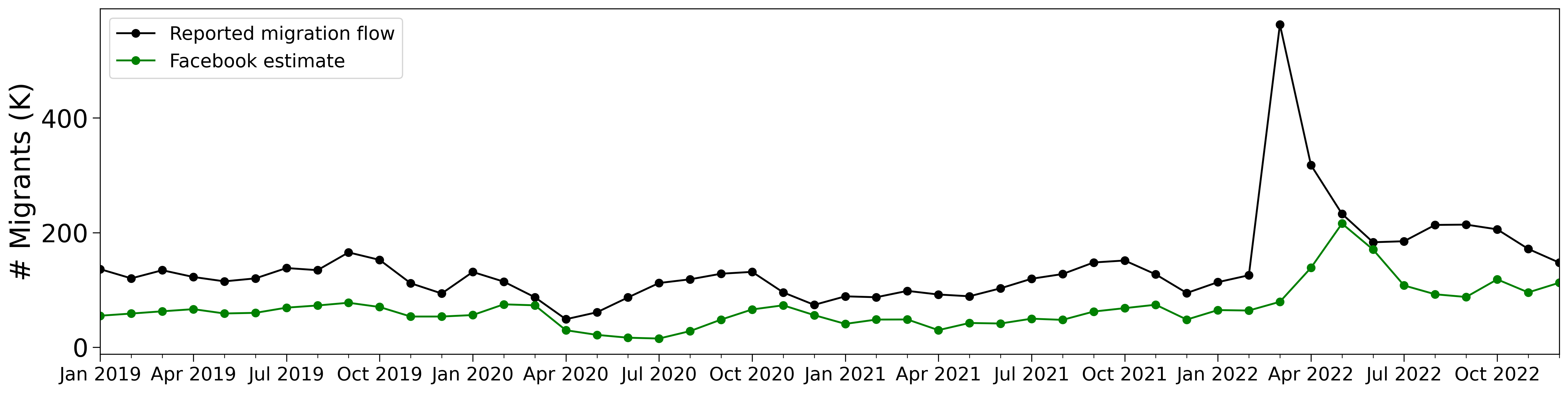}
\caption{Validation in Germany at the monthly level from 2019 to 2022. \label{validation_germany}}
\end{figure}

In Fig. \ref{validation_germany}, we plot the level of migration in the German administrative data against our estimate of the total level of migration to Germany. The two series have very similar patterns in terms of the temporal variation in the rate of migration, though the overall levels differ due to the looser definition of migration used in the German data. We also observe a slight lag in our data, which seems to lag the administrative figures by about one month, a pattern that does not appear in the New Zealand validation exercises presented in Fig. 4. This discrepancy could reflect a difference in administrative procedures or definitions across countries or a factor that affects the speed at which one's Facebook home prediction updates. This distinction is likely to be more muted at higher levels of temporal aggregation.

\subsection*{United States}

U.S. Customs and Border Protection (CBP) shares the number of migrants encountered at the border each month \footnote{More information about this dataset can be found \href{https://www.cbp.gov/newsroom/stats/nationwide-encounters}{here}.}. In recent years, such encounters have accounted for a large share of migration to the United States from several countries, reflecting in part the spike in asylum claims initiated at the southern border, which normally involve migrants coming into contact with CBP agents. Many migrants whose presence is handled under the agency's Title 8 authority are allowed to stay in the country while their claims are pending, which often takes several years due to long backlogs in the immigration court system\footnote{Data compiled by Syracuse University's indicated that the average wait time for a hearing in 2020 was 1552 days, over 4 years. Further statistics on processing times are available \href{https://trac.syr.edu/phptools/immigration/asylumbl/}{here}.}\footnote{CBP also provides information on Title 42 encounters, which refers to provisions under which the government could process asylum claims in an accelerated fashion due for public health reasons. Most migrants processed under this provision do not remain in the United States for a year; as such, we do not compare encounters conducted under Title 42 authority to our data. This provision was widely used to turn back asylum seekers after the onset of the COVID-19 pandemic, before authorization for the provision was revoked in May 2023.}.

Comparing our estimates to the estimated number of individuals processed under Title 8 is difficult for a number of reasons. Importantly, the two measures define migrants' origins differently---we define migrants' origin using the country in which they last lived for 12 months, while CBP uses the migrant's citizenship. The figures presented by CBP also do not include non-asylum migration, such as students and people for work, which is present in our data. As a consequence, in certain cases our estimates of migrant flows diverge from those reported by CBP. For instance, these differences are strong in the case of migrants from Haiti, where CBP reports a considerably larger flow than we measure in our data. In part, we believe that this reflects the fact that many Haitian citizens arriving in the United States actually left their home country after the 2010 earthquake and lived during the interim in Latin America\footnote{See \href{https://www.migrationpolicy.org/article/haitian-migration-through-americas}{this article} for a description of the process of migration that led such migrants to the United States.}. Using our definition of migration, these Haitian citizens arriving in the United States are counted as migrants from the last country in which they spent one year.

Conversely, our estimates of migration from Mexico to the United States are higher than the number of encounters reported by CBP. In part, this reflects the high level of non-asylum migration from Mexico to the US, but our higher levels may also reflect the fact that many migrants from third countries spend a substantial amount of time in Mexico before arriving in the United States. In some cases, these delays can be substantial, as in 2019 when Mexico paused its issuance of exit permits for third-country migrants bound for the United States\footnote{See \href{https://www.migrationpolicy.org/article/haitian-migration-through-americas}{here} and \href{https://www.reuters.com/world/americas/their-prospects-dim-haitian-migrants-strain-mexicos-asylum-system-2021-10-05/}{here} for more information on this policy shift.}. Migrants who remain in Mexico for more than a year as a part of their journey to the United States are classified as migrants from Mexico under our definition, but migrants from their country of citizenship under the definition used by CBP.

\subsection*{Eurostat}

\begin{figure}
\centering
\includegraphics[width=1\linewidth]{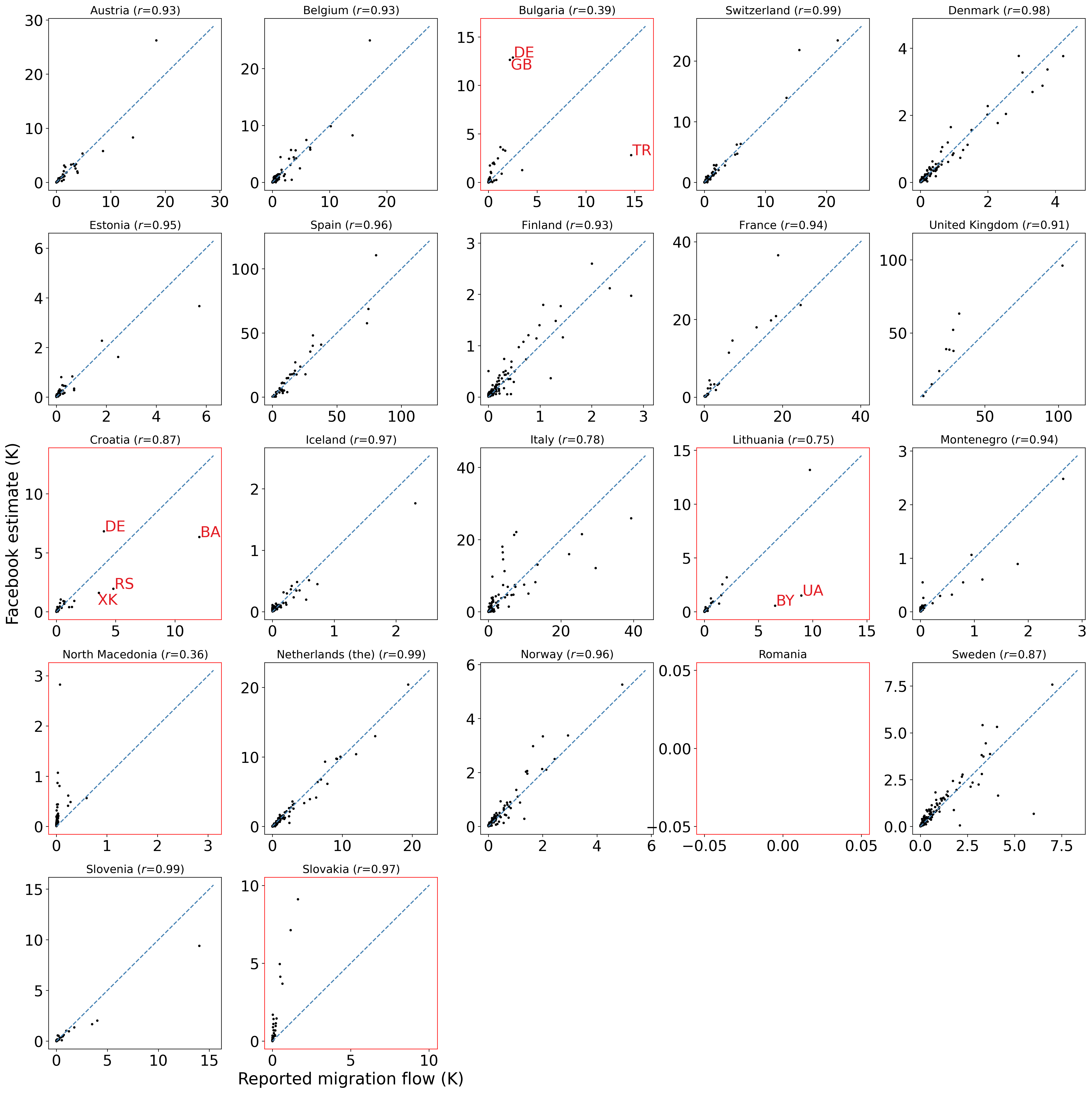}
\caption{Comparison of our estimated immigration figures for 2019 with those compiled by Eurostat. Each destination country is presented in a separate plot.\label{reweighting_selection_rate_eurostat_2019}}
\end{figure}

\begin{figure}
\centering
\includegraphics[width=1\linewidth]{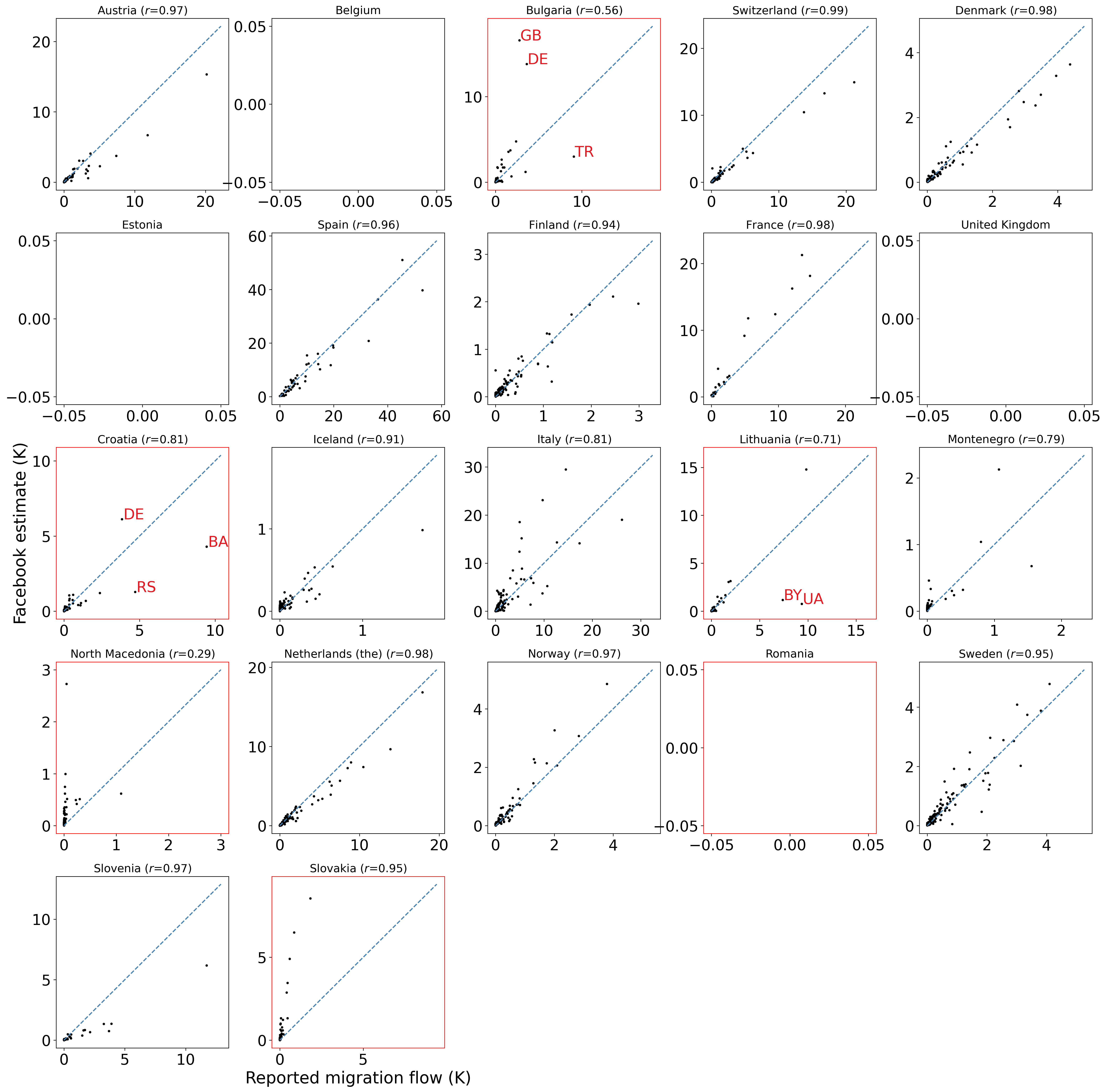}
\caption{Comparison of our estimated immigration figures for 2020 with those compiled by Eurostat. Each destination country is presented in a separate plot.\label{reweighting_selection_rate_eurostat_2020}}
\end{figure}

\begin{figure}
\centering
\includegraphics[width=1\linewidth]{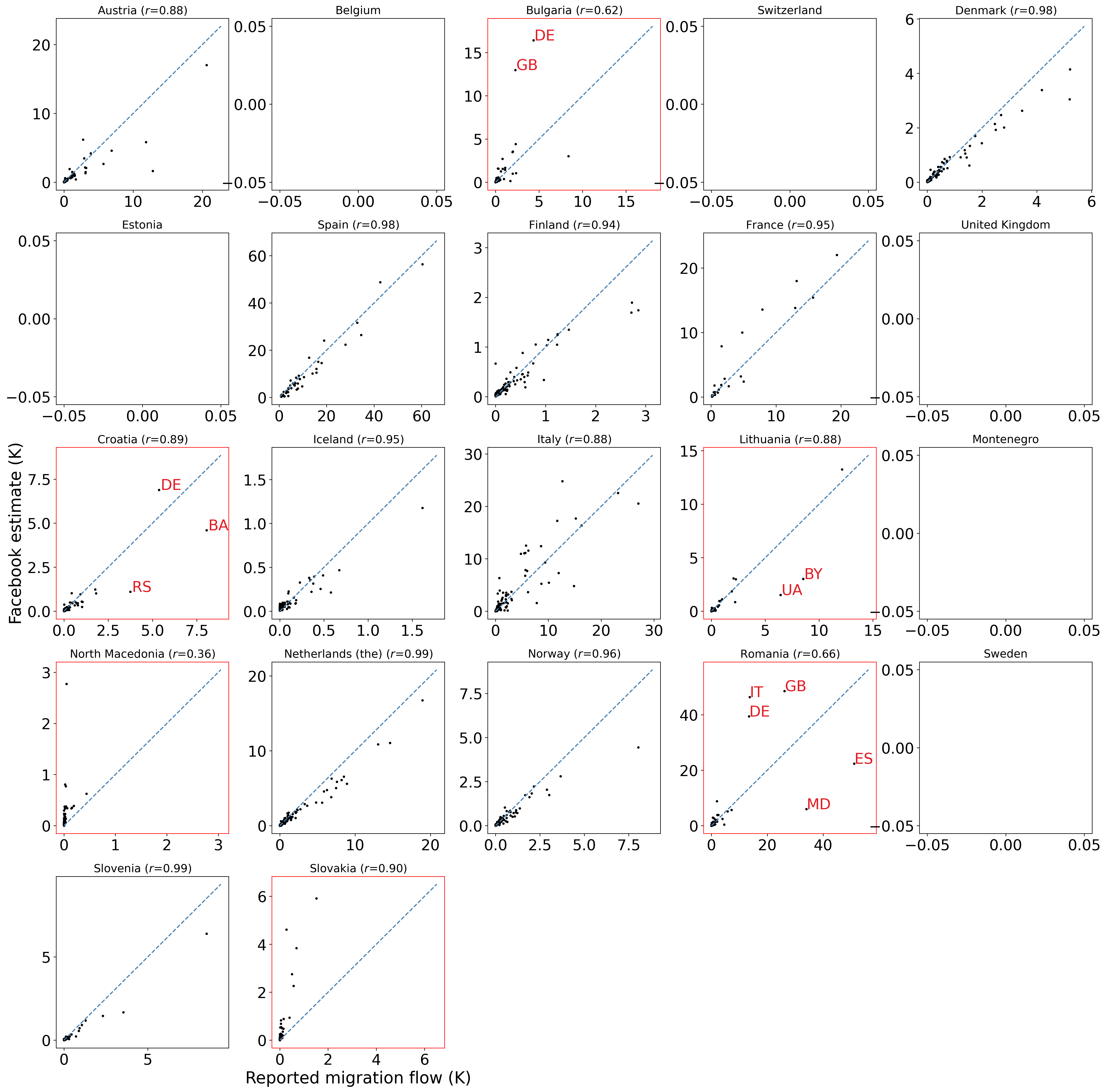}
\caption{Comparison of our estimated immigration figures for 2021 with those compiled by Eurostat. Each destination country is presented in a separate plot.\label{reweighting_selection_rate_eurostat_2021}}
\end{figure}

\begin{figure}
\centering
\includegraphics[width=1\linewidth]{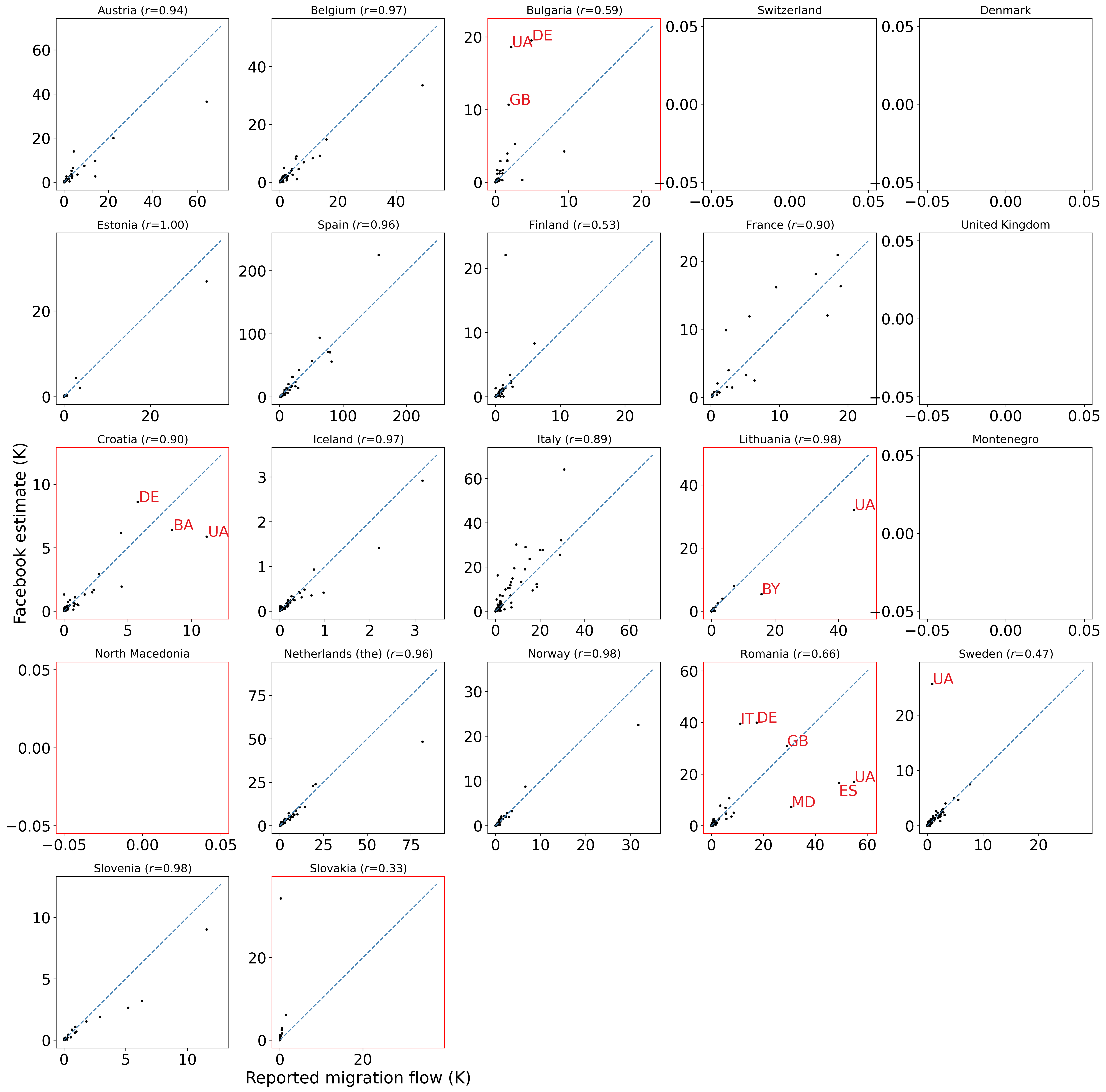}
\caption{Comparison of our estimated immigration figures for 2022 with those compiled by Eurostat. Each destination country is presented in a separate plot.\label{reweighting_selection_rate_eurostat_2022}}
\end{figure}

In Fig. \ref{reweighting_selection_rate_eurostat_2019}, we present analogous validation exercises using the 2019 Eurostat estimates for a wide array of countries. Although Eurostat publishes the estimates in a standardized format, the collection procedures underlying this data vary by country. Some nations, such as Sweden, derive their estimates from high-quality population register data, while others use sample surveys, healthcare records, or a number of other techniques\footnote{More information about the procedures used can be found \href{https://ec.europa.eu/eurostat/statistics-explained/index.php?title=Migration_and_migrant_population_statistics}{here} and \href{https://ec.europa.eu/eurostat/cache/metadata/en/migr_immi_esms.htm}{here}.}. Despite this heterogeneity in the underlying quality of the data, we find that our data is in general highly correlated with the figures from Eurostat, with a correlation above 0.9 in 15 of the 21 countries reporting migration figures in 2019. We see similarly strong performances in Figures \ref{reweighting_selection_rate_eurostat_2020}, \ref{reweighting_selection_rate_eurostat_2021},  and \ref{reweighting_selection_rate_eurostat_2022}, which repeat these validation exercises in 2020, 2021, and 2022.

With that said, we do find several large outliers in the data which could be informative to highlight. In the case of Bulgaria, we estimate that flows into the country from the United Kingdom and Germany greatly exceed the government estimates. We find a similar pattern in Romania, which sees higher inflows from Germany, the United Kingdom, and Italy than the government figures suggest. Although we do not have definitive evidence, it seems plausible that these flows represent Bulgarian and Romanian nationals returning to their native country after working or studying abroad. Both Bulgaria and Romania saw large-scale emigration to the higher-wage countries of Western Europe after labor mobility barriers were lifted in 2014, and have since seen individuals return in the wake of improving economic conditions locally and following the legal uncertainty for EU nationals imposed after Brexit. Official statistics may undercount such return migrants if they do not respond as other migrants do to questions about their migration status \cite{fic_migration_2013}. We find a similar pattern in North Macedonia, where our estimates of the flow into the country from Germany are substantially larger than the level reported in the Eurostat data, which might reflect dual citizens or temporary workers returning from working in the country. The large flow we find to North Macedonia from Germany, which has both a strong labor market and a large stock of Macedonians, is consistent with this interpretation.

In some countries, we find that our estimates have a strong correlation with the figures from Eurostat, but the overall level of migration we detect differs substantially from the administrative figures. We find this most dramatic in the case of Slovakia, where our figures have a Pearson correlation of 0.33-0.97 across years, though our estimates find a substantially higher level of immigration than the government reports. We find that this discrepancy is likely caused by the method used by the Slovak government to report their figures to Eurostat---they include only individuals who have been granted a permanent residency permit\footnote{More information about the Slovak government's definition of migration can be found \href{https://slovak.statistics.sk/wps/portal/ext/themes/demography/population/about/!ut/p/z0/jY1BTsMwFETPwqJL5zu2m-8uE5CqtipSQILgDXJct3VD7DS2Atye5AasZmYx74GCBpTXk7vo5ILXX_P-UMVnjTtZVXlJZfXG6A6fX_JDXW8PTwiv1sMelGv77Nv0Gc1QIuWIUjCxKdacLwh3u99VCcoEn-xPgia0UV9J7IjzZ6K7tKJzCWM_aydvSRxGPf2u6BRt6uY8n1AwKzlZC2OIMJyTFjmSTd6eLGWMF5ouHjYeH48XUINO1wUdoPnXdei27zKWD38kSXQ2/}{here}.}. The majority of migrants enter on temporary permits, which are renewable for several years. Such permits enable these temporary permit holders to meet the United Nations' definition of migration, which only requires migrants to reside in their destination country for one year\footnote{Information about temporary residency permits in Slovakia can be found \href{https://www.mic.iom.sk/en/download/info-cards/itemlist/category/73-temporary-residence.html}{here}.}. These temporary residence permits are far more common than the permanent variety, with the OECD estimating that 20,000 temporary permits valid for more than one year were issued in 2019, relative to the 7,000 total migrants reported in the Eurostat figures that year\footnote{For more information about the temporary permits, see \href{https://www.oecd-ilibrary.org/sites/b3775efb-en/index.html?itemId=/content/component/b3775efb-en}{here}.}. For this reason, we think the estimates we report better capture the United Nations' definition of migration than the figures Slovakia provides to Eurostat.

In the case of Croatia, we find smaller levels of migration from Bosnia and Herzegovina (BA) and Serbia (RS) than are suggested by the administrative data. It is possible that our estimates here are complicated by people who hold dual citizenship across the countries, and who move between them relatively frequently, a problem raised by Dario Pavić et al. \cite{pavic_razlike_2019}. It is possible that the people the official statistics record as moving to Croatia from these origin countries did not reside in their country of origin (or, perhaps, Croatia) for more than 12 months and thus do not meet our criteria for migration.

\subsection*{OECD}

The Organisation for Economic Co-operation and Development (OECD) releases the International Migration Outlook, a comprehensive report that provides the number of permanent migrants to each OECD country, as well as in-depth analyses of current migration trends and policies\footnote{See \href{https://doi.org/10.1787/b0f40584-en}{International Migration Outlook 2022} by OECD for the details}. The definition of permanent migrants in this report is irrespective of the actual duration of stay. Instead, it is based on the type of permit the individual uses to enter the destination country, except in cases of migration within free-circulation areas where no permit is required. This definition also includes individuals who transition from a temporary permit to a permanent one. Other definition differences exist; some countries, such as Australia, Ireland and the United States report their migration statistics using fiscal years instead of calendar years.

\begin{figure}
\centering
\includegraphics[width=1\linewidth]{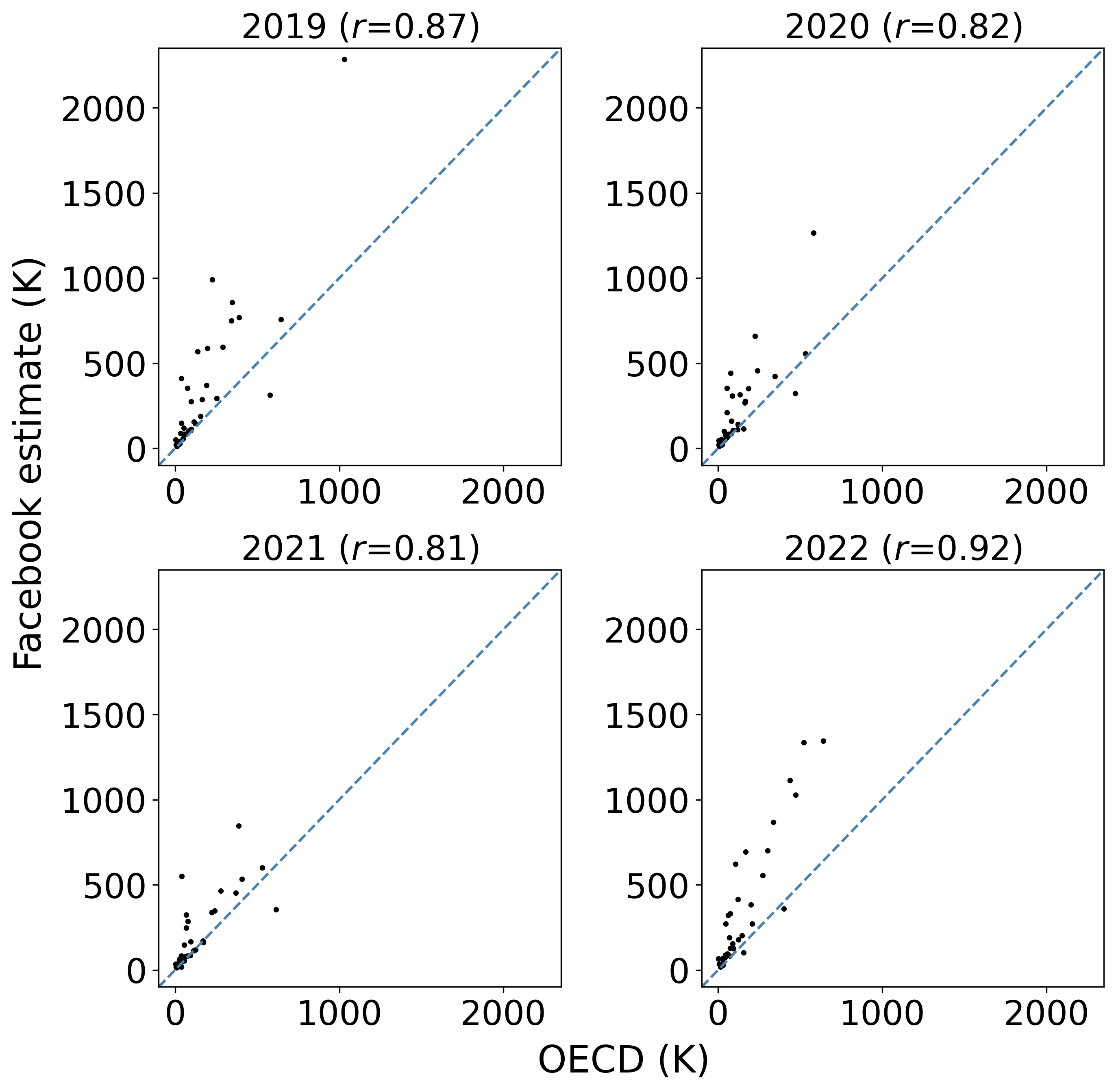}
\caption{Validation of in-flows in OECD countries. \label{validation_oecd}}
\end{figure}

More fundamentally, the OECD's definition of migration differs from our own in two key ways. First, their definition does not account for returned residents. Second, it does not include ``temporary'' migrants who live in a country for more than one year, such as students and workers. These factors can explain why our estimates are higher than OECD; the OECD estimates that 6.1 million people migrated to OECD countries in 2019, while our estimate stands at 12.1 million (Fig. \ref{validation_oecd}). For example, the NSO of New Zealand reported that the nation received 165,741 immigrants in 2019 \cite{newzealand}, while the OECD reported that there were only 38.3K immigrants to New Zealand in that year\footnote{See \href{https://www.oecd-ilibrary.org/sites/b0f40584-en/index.html?itemId=/content/publication/b0f40584-en}{Table 1.1} in the International Migration Outlook 2023 by OECD}. Overall, our estimates exhibit a strong positive correlation with the OECD's figures (Pearson correlations of 0.87 in 2019, 0.82 in 2020, 0.81 in 2021, and 0.92 in 2022) despite the differing migration concepts employed in the two datasets.

\subsection*{Global indirect estimates}

To incorporate a more global perspective for our validation we compare our estimates against indirect estimates of global migration. In recent years, the United Nations and World Bank have both published data on migrant stocks, broken down primarily by their country of birth in five-year intervals. Several methods have been developed to indirectly estimate the migration flows to match the changes in migrant stocks over the five-year intervals \cite{abel_bilateral_2019, abel_bilateral_2022}. These methodologies estimate migrant transitions, that is, individuals who resided in country $i$ at the start of the five-year time period and in country $j$ at the end of the time period, regardless of any other moves made in the interim. This transition-based definition of migration undercounts the total number of moves during the five-year time period as return moves within the interval are not captured in the changes in the migrant population data.

\begin{figure}
\centering
\includegraphics[width=1\linewidth]{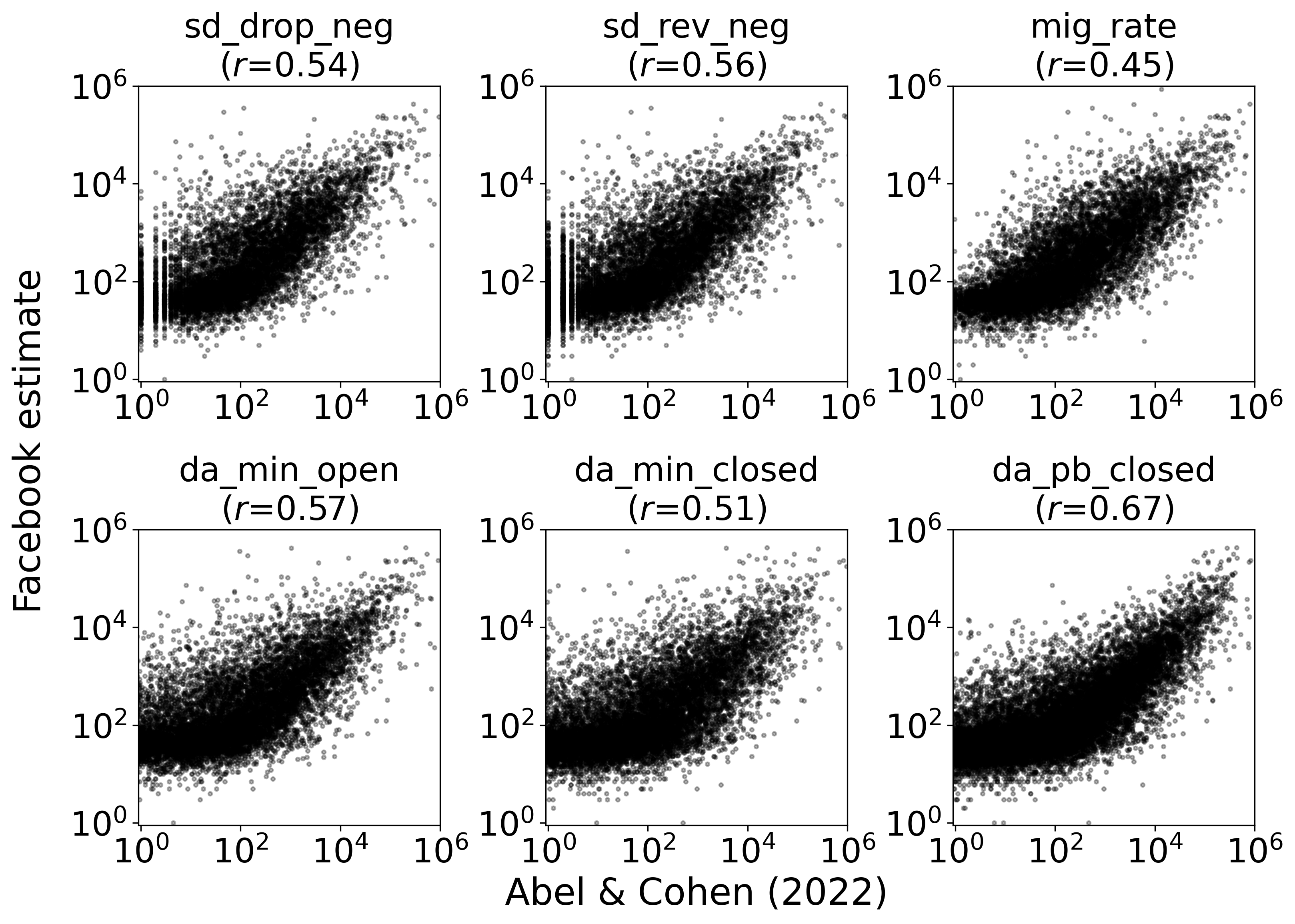}
\caption{Validation of country-to-country estimates flows during 2019 with updated estimates of \cite{abel_bilateral_2019} for the 2015-2020 period. Six methods were used in this paper, including (1) stock differencing drop zeros (sd\_drop\_neg), (2) stock differencing reverse negative flows (sd\_rev\_neg), (3) migration rates (mig\_rate), (4) demographic accounting in an open system with a minimisation approach (da\_min\_open), (5) demographic accounting using a closed system with a minimisation approach (da\_min\_closed) and (6) demographic accounting using a closed system and a Pseudo-Bayesian approach (da\_pb\_closed). \label{validation_abel_cohen}}
\end{figure}

The use of indirect migration estimates as a validation source also presents some challenges. Most notably, the stock figures on which the estimates are based are only available every five years, so we are forced to compare our migration estimates for a single year (2019) to the average annual rate of migration over the 2015-2020 period in the indirect estimates. The definition of migration also differs across the two concepts, as the indirect estimates employ a transition-based, rather than a movement-based, definition of migration. For this reason, we find that our correlations with this source of validation data are lower than our correlations against alternative validation sets (see Fig. \ref{validation_abel_cohen}). Nevertheless, we find our estimates correlate best with indirect estimates from the Pseudo-Bayesian approach within the demographic accounting closed system \cite{azose_estimation_2019}, which themselves have been shown to provide the closest match to reported migration figures of the six methods considered \cite{abel_bilateral_2019}. 

\subsection*{Alternative validations of migration}

In Tables \ref{validation_corr_2019}, \ref{validation_corr_2020}, \ref{validation_corr_2021}, and \ref{validation_corr_2022}, we present validations using the data sources above on a number of alternative migration measures. In these exercises, we transform both our data and the administrative data before correlating the two series. For each metric, we present the number of observations included in the comparison\footnote{This varies across rows for several reasons. In the row correlating the log of each series, we must exclude country pairs in which either of the series records no migrants, as the log of 0 is undefined. For the total outbound, total inbound, and net migration rows, we aggregate the country pair level to the country level.}. When calculating the aggregations, we compute totals solely over the set of country pairs included in both data sets. For instance, though we can observe outbound migration between almost all country pairs, not all countries who participate in the Eurostat dataset report their outbound migration to non-Eurostat countries. As a result, when we calculate the aggregate level of migration from these countries in the Facebook data, we exclude migration that involves a country pair not observed in Eurostat.

In general, our data is strongly correlated with the administrative data across the different measures that we examine. There is a generally lower level of correlation across data sources after we apply the log transformation. This is largely due to the effects of our differential privacy protections, which add relatively large amounts of noise to country pairs with small estimated migration flows. This noise has a larger impact on the overall correlation following the log transformation.

\section*{Estimates for China}

In most of this paper, our migration estimates cover the sample of 181 countries defined in Table S9. These countries account for around 79\% of the world's population. China accounts for the bulk of the remainder of the population, but the low level of Facebook usage in the country makes it difficult to estimate migration involving the nation with our standard methodology. With that said, there are a substantial number of Facebook users in China, despite the low penetration rate. In this section, we use an alternate methodology to estimate the level of migration to and from China at the population level.

\begin{figure}
\centering
\includegraphics[width=.5\linewidth]{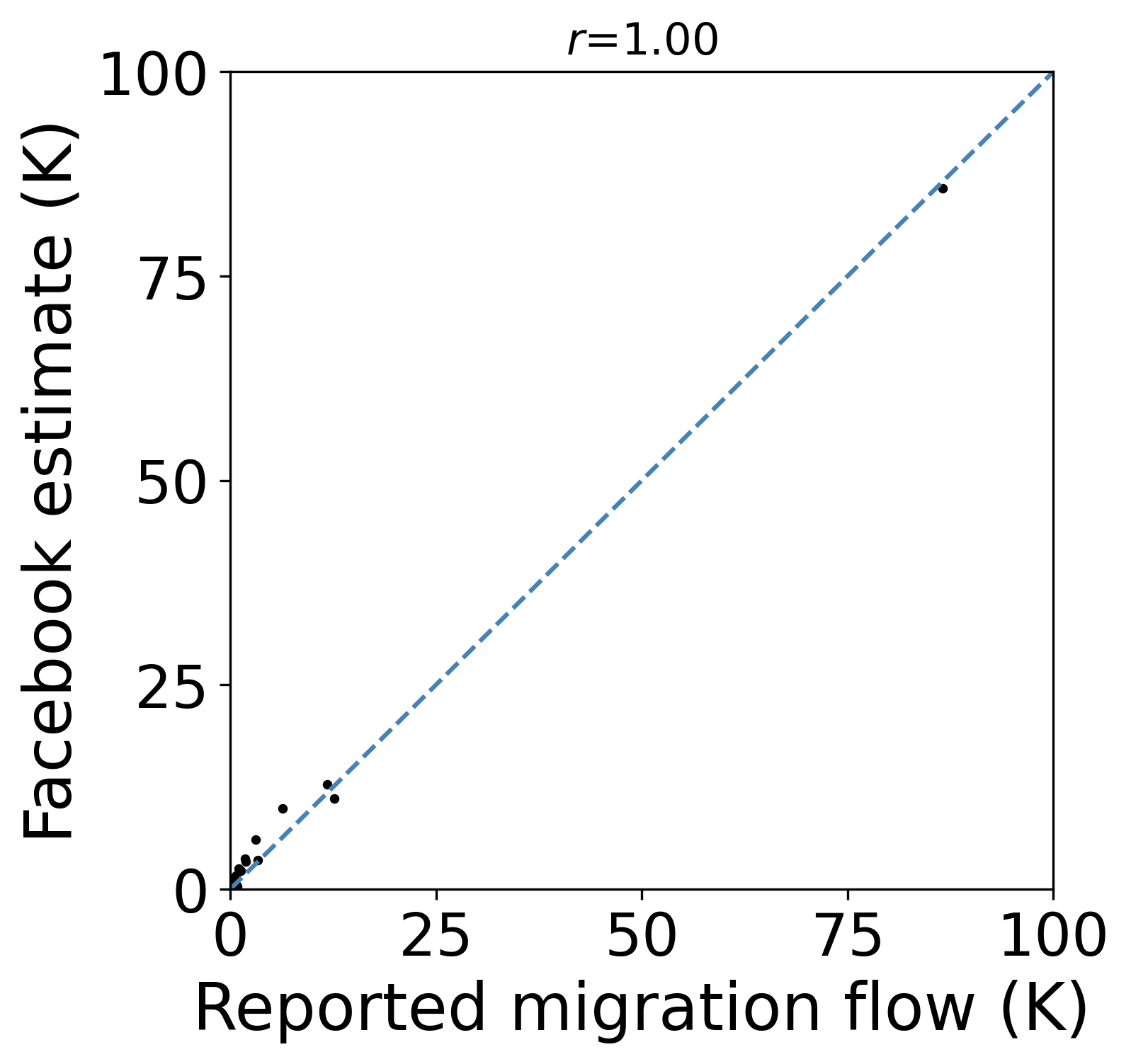}
\caption{Validation of our 2019 estimates of Chinese migration to 21 countries using Eurostat data\label{china_validation}}
\end{figure}

To do this, we make use of data from Eurostat, which reports migration from China to 21 countries in 2019. We fit a simple linear model without an intercept:
\begin{equation*}
    \text{MigEurostat}_{\text{China},d,2019} = \beta \text{MigFB}_{\text{China},d,2019}
\end{equation*}
Here, $\text{MigEurostat}_{\text{China},d,2019}$ is Eurostat's figure for migration from China to country $d$ in 2019 and $\text{MigFB}_{\text{China},d,2019}$ is the number of users in our sample who migrate to country $d$ from China in that year. We find that $\beta=11.36$ is the best fit and use this factor to rescale all of our raw estimates of migration to and from China. In Fig. \ref{china_validation}, we show that the resulting estimates of migration from China have a very high correlation with the estimates from Eurostat\footnote{Migration from China to countries outside of Europe might be driven by a different pattern of selection. Unfortunately, we do not know of any validation data that would allow us to benchmark the performance of our technique in the developing world.}.

\begin{figure}
\centering
\includegraphics[width=\linewidth]{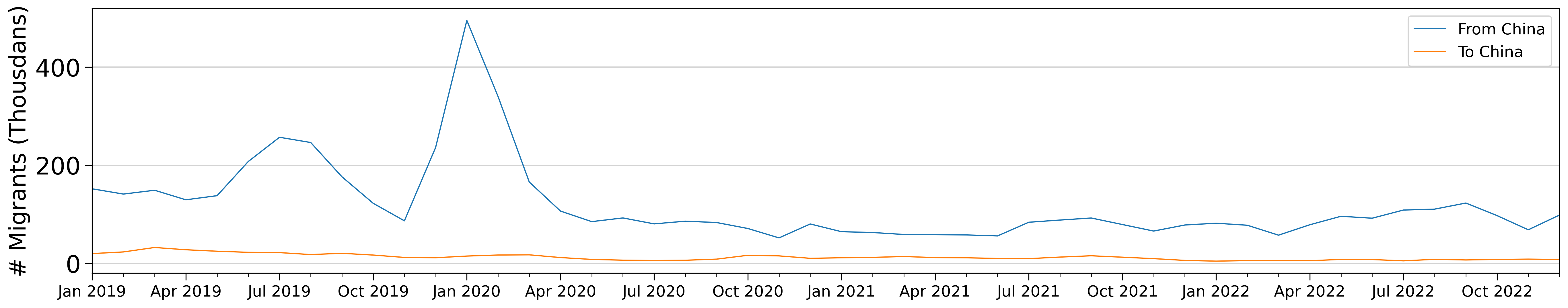}
\caption{Migration to and from China\label{china_time_series}}
\end{figure}

Using this rescaling technique, we estimate that 2.04 million people migrated from China to another country in 2019\footnote{As we do not know the nationality of Facebook users, it is important to note that our estimated number of migrants from China should not be assumed to represent Chinese nationals who have relocated to another country.}. We show in Fig. \ref{china_time_series} that migration fell dramatically following the onset of COVID-19 before recovering during 2021.

\section*{National income levels and migration trends}

The global scale of our dataset allows us to explore how international migration patterns vary with the development level of the origin and the destination. We find that migrants tend to move to relatively wealthy countries: countries classified as high-income by the World Bank attract 67\% of the global migrants. Correspondingly, only 6\% of global migrants choose low-income countries as their destinations. 

\begin{figure}
\centering
\includegraphics[width=1\linewidth]{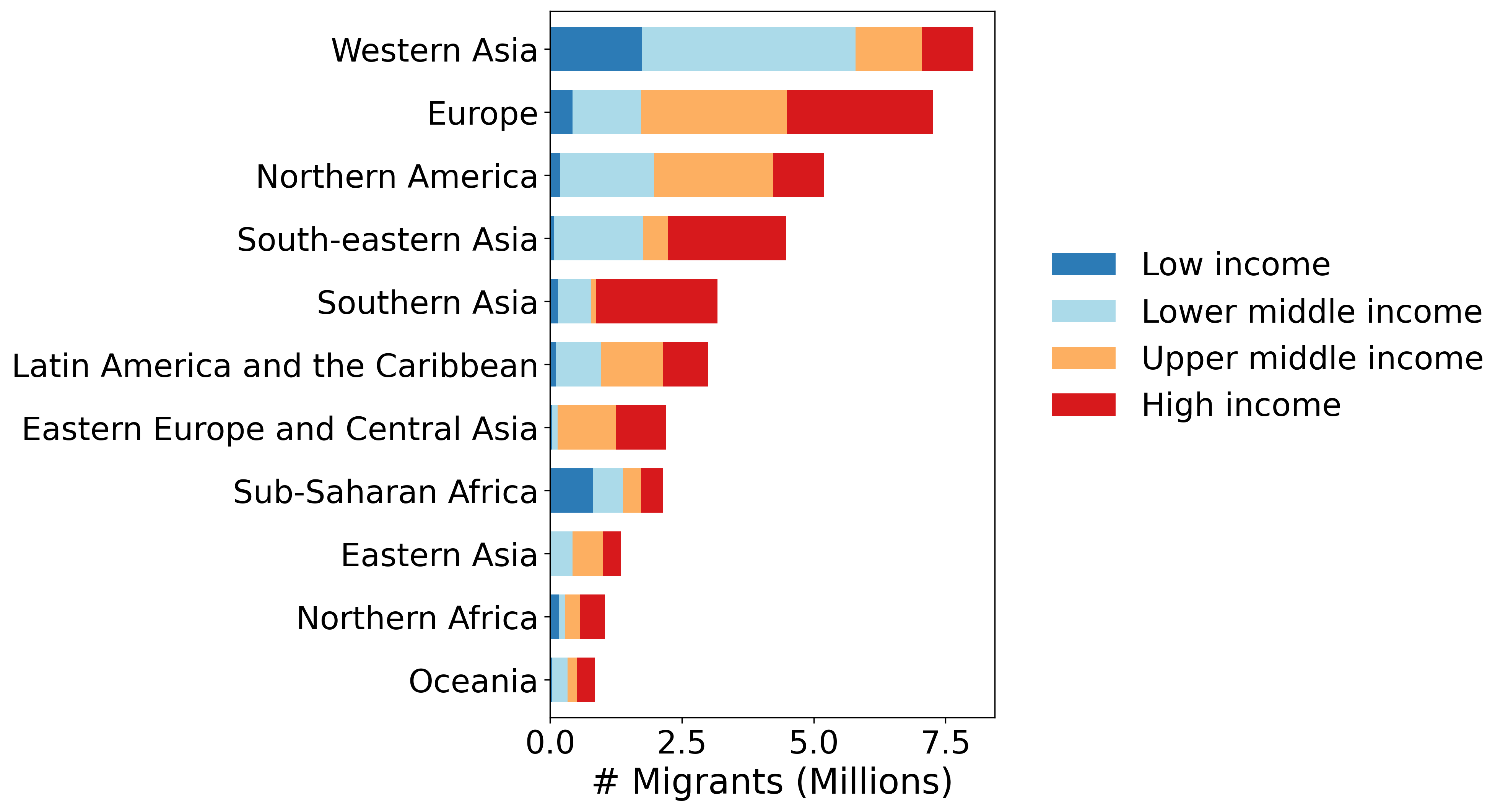}
\caption{2022 regional migration by income level of origin country. \label{region_income}}
\end{figure}

High-income countries are also over-represented as sources of migrants. High-income countries comprise just 19\% of the world population but represent 33\% of global migrants. This pattern is driven by the combination of several factors. Globally, migrants from high-income countries are more able to finance migration and tend to enjoy a more favorable treatment in their destination countries. We explore this pattern in Fig. \ref{region_income}. In Europe a majority of migrants come from high-income countries, probably tied to the freedom of movement in the European Union. South Asia also sees large-scale inflows from high-income countries, which is driven by individuals originally from the region returning from labor contracts in Western Asia.

\section*{Crisis-induced migration}

\begin{figure}
\centering
\includegraphics[width=1\linewidth]{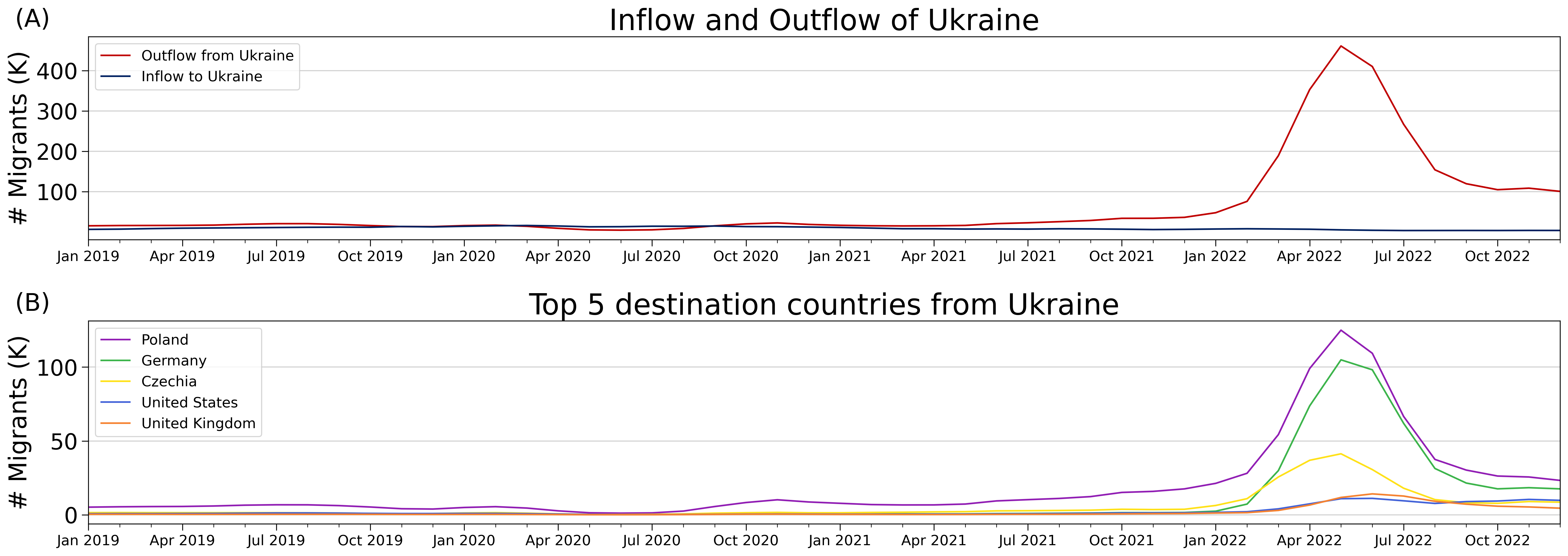}
\caption{Migration flow changes before and after the Ukraine war. Top: Inflow to Ukraine and Outflow from Ukraine from 2019 to 2022; Bottom: Top 5 outflow destinations from Ukraine during the Ukraine war from Feb. 2022 to Dec. 2022.  \label{ukraine_trend}}
\end{figure}

Our estimates can also help to illustrate the effect of conflicts and civil unrest on migration, in a context in which gathering more traditional indicators of migration may be difficult due to logistical or political sensitivities. In Panel A of Fig. \ref{ukraine_trend}, we highlight the particularly large outflows from Ukraine following Russia's invasion of the country in February 2022. In total, we estimate that 2.3 million people emigrated from the country and settled elsewhere for at least a year between February and December 2022, a tenfold increase over the pre-war emigration rate. In Panel B of Fig. \ref{ukraine_trend}, we show the top five countries that received the most migrants population -- Poland, Germany, Czech Republic, the United States, and United Kingdom\footnote{Our measures of the top destination countries are closely aligned with estimates from the UNHCR; see \href{https://data.unhcr.org/en/situations/ukraine}{here} for more details.}. Our estimates are lower than the 3.8 million recipients of Temporary Protected Status (TPS) reported by Eurostat in January 2023\footnote{This figure is derived from the \href{https://ec.europa.eu/eurostat/databrowser/view/migr_asytpsm__custom_10894513/default/table?lang=en}{migr\_asytpsm series}.}, which we believe is due to our more conservative concept of migration. Concretely, the Eurostat figures do not impose a minimum length of stay in either the origin or the destination---for instance, a migrant who moved to Poland in October 2022 and applied for TPS before returning to Ukraine in August 2023 would be counted in the TPS figures but not in our migration estimates. Most migrants from Ukraine in 2022 planned to return to the country, with a survey by the International Organization on Migration finding that 77\% intended to return\footnote{A summary of the survey's findings can be found \href{https://www.unrefugees.org/news/unhcr-one-year-after-the-russian-invasion-insecurity-clouds-return-intentions-of-displaced-ukrainians/}{here}.}.

\begin{figure}
    \begin{center}
    \begin{subfigure}[t]{0.8\textwidth}
        \caption{}
        \includegraphics[width=1\linewidth]{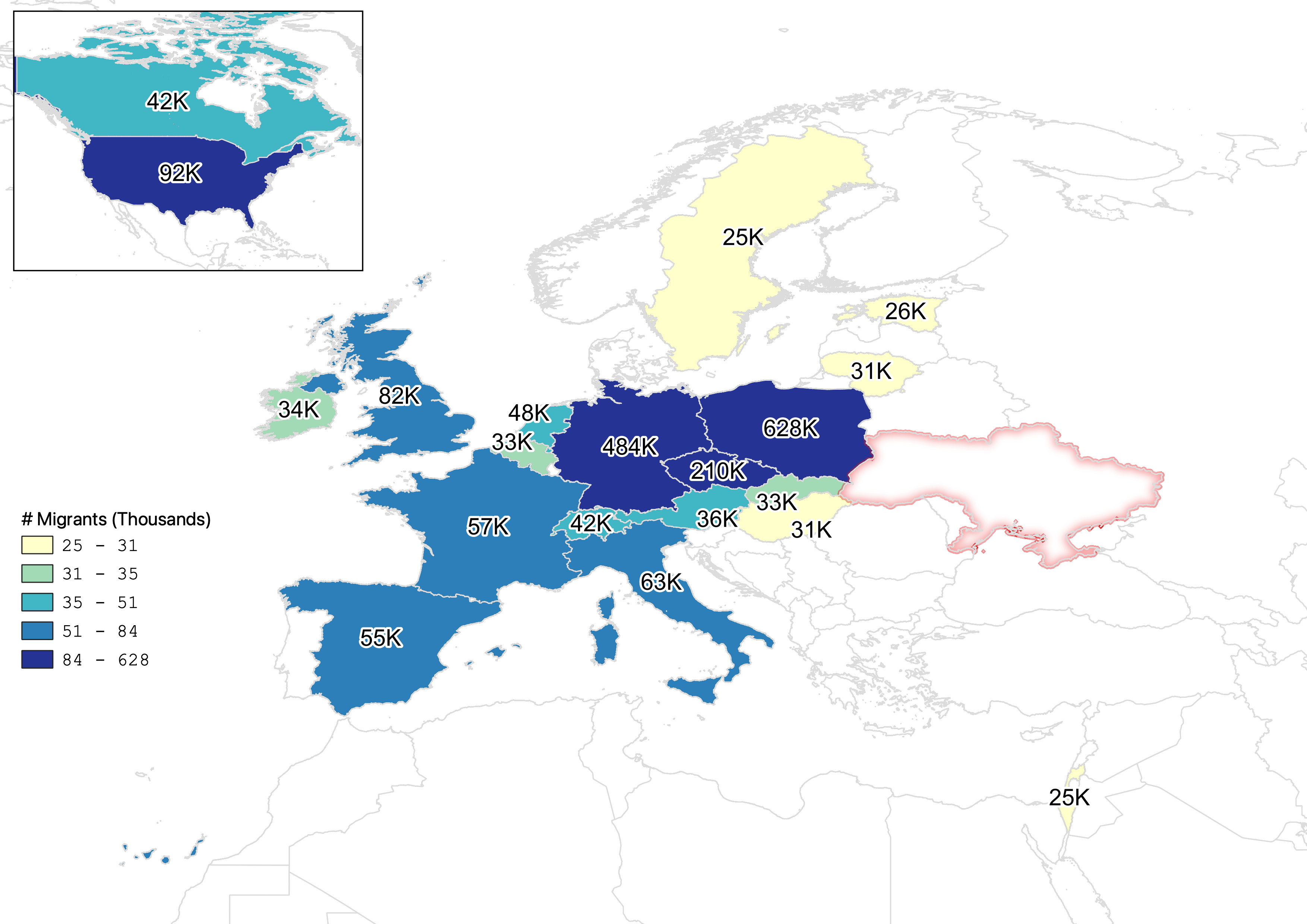} 
    \end{subfigure}
    \vspace{0.5cm}
    \begin{subfigure}[t]{0.8\textwidth}
        \caption{}
        \includegraphics[width=1\linewidth]{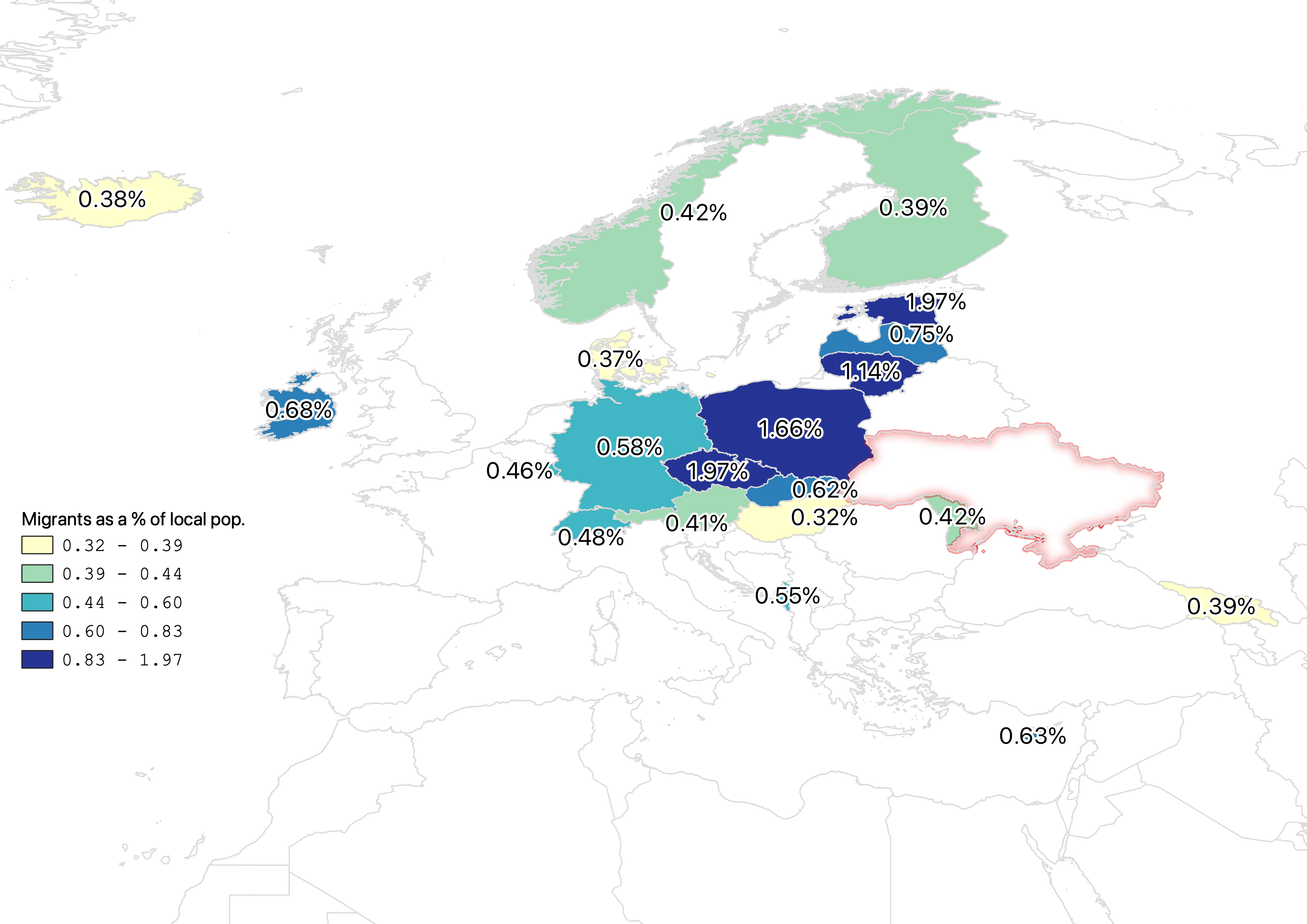} 
    \end{subfigure} 
    \end{center}
    \vspace{-0.6cm}
\caption{Top 20 outflow destinations from Ukraine during the Ukraine war. (A) Total number of migrants from Ukraine in each country; (B) Proportion of migrants from Ukraine over the population in each destination country.  \label{ukraine_map}}
\end{figure}

Fig. \ref{ukraine_map} shows the top 20 destination countries from Ukraine during the Ukraine war in terms of the total number of migrants (Panel A) and the proportion of migrants over the population in the destination country (Panel B). Different from the rank based on the total number of migrants, Lithuania and Estonia have a much higher proportion of migrants over the population -- the migrants from Ukraine between Feb. 2022 to Dec. 2022 account for 1.14\% and 1.97\% of the population in Lithuania and Estonia, respectively.

\begin{figure}
\centering
\includegraphics[width=1\linewidth]{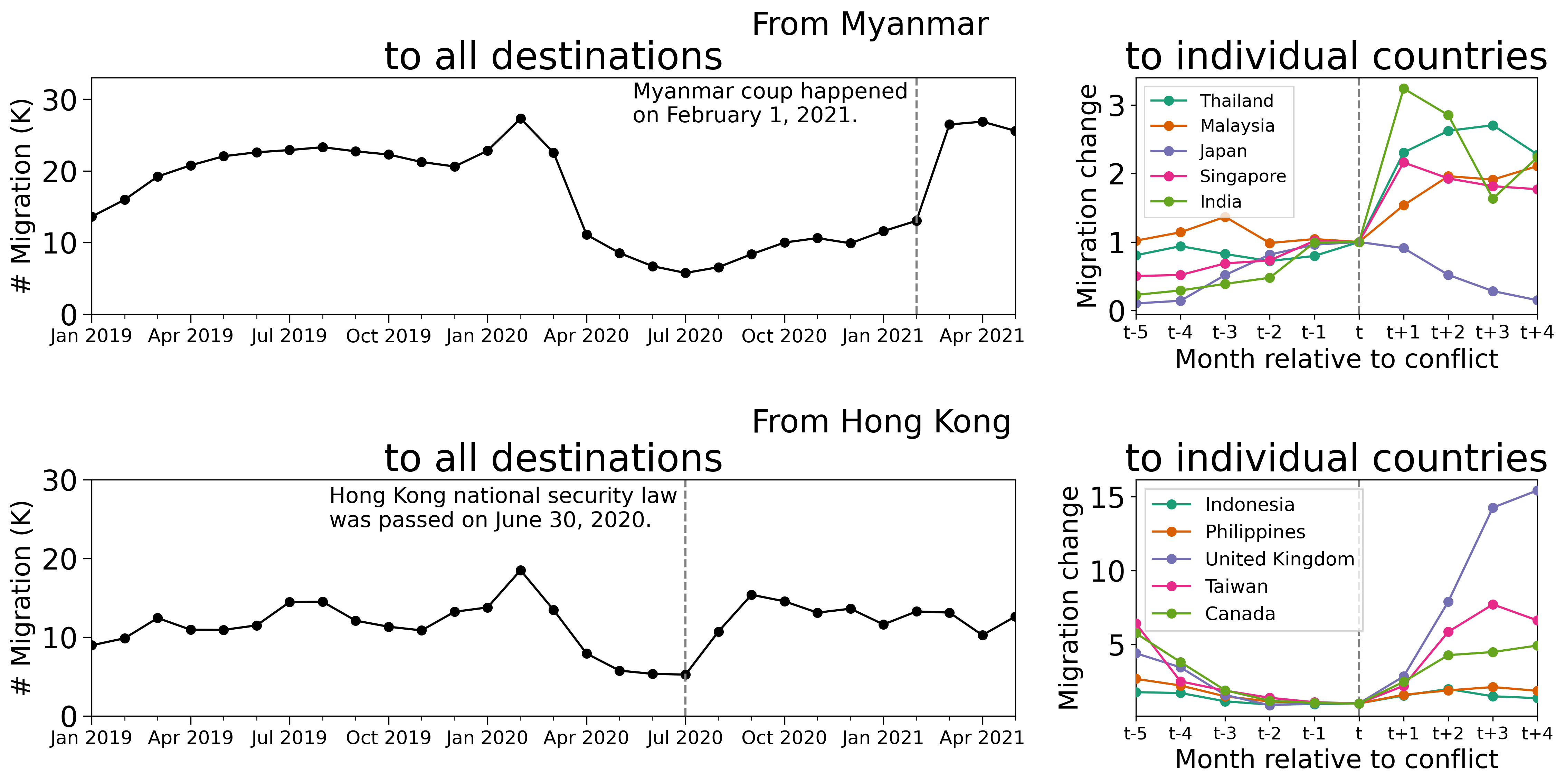}
\caption{Migration flow changes after crisis. Top: Myanmar coup in February 2021; Bottom: Hong Kong national security law in June 2020. The plots in the right column show the migration flow change relative to the month ({\it t}) when the crisis happened. The Y-axis is normalized so the value in month $t$ is 1 for all destinations. The five countries are the top 5 destinations based on the number of migrants in our dataset.  \label{conflict_migration}}
\end{figure}

In Fig. \ref{conflict_migration}, we demonstrate the rapid increase in outward migration flows from Myanmar following the February 2021 coup. Disaggregating these migrants by destination, we observe particularly large increases in India and Thailand. We notice a similar outward migration flow from Hong Kong following the passing of the controversial national security law in June 2020\footnote{See \href{https://www.gov.uk/government/news/hong-kong-bno-visa-uk-government-to-honour-historic-commitment}{here} for the details.}. Most strikingly, outflows from Hong Kong to the United Kingdom increased fifteenfold in the four months following the passage of the law. Many of the departures preceded the United Kingdom's announcement of the British National (Overseas) visa program for Hong Kong residents in January 2021.

\section*{Social networks and migration}

The role of social networks in predicting and facilitating migration has been of immense interest to economists, sociologists, and human geographers \cite{blumenstock_migration_2019,munshi_networks_2003,ryan_changing_2018,ryan_migrants_2011,bertoli_networks_2018}, but has not previously been tested on a global scale. In this section, we produce figures connecting the aggregated migration flows with an index of social connectedness between countries. To measure migration flows, we calculate the migration intensity, which we define as:
\begin{equation*}
    Migration\ intensity_{A,B} = \frac{M_{A \rightarrow B} + M_{B \rightarrow A}}{Pop_A * Pop_B}
\end{equation*}

The Social Connectedness Index, which was introduced in \cite{bailey2018social}, is defined analogously as:
$SCI_{A,B} \propto \frac{Friendships_{A,B}}{FBUsers_{A} \times FBUsers_{B}}$

\begin{figure}
\centering
\includegraphics[width=1\linewidth]{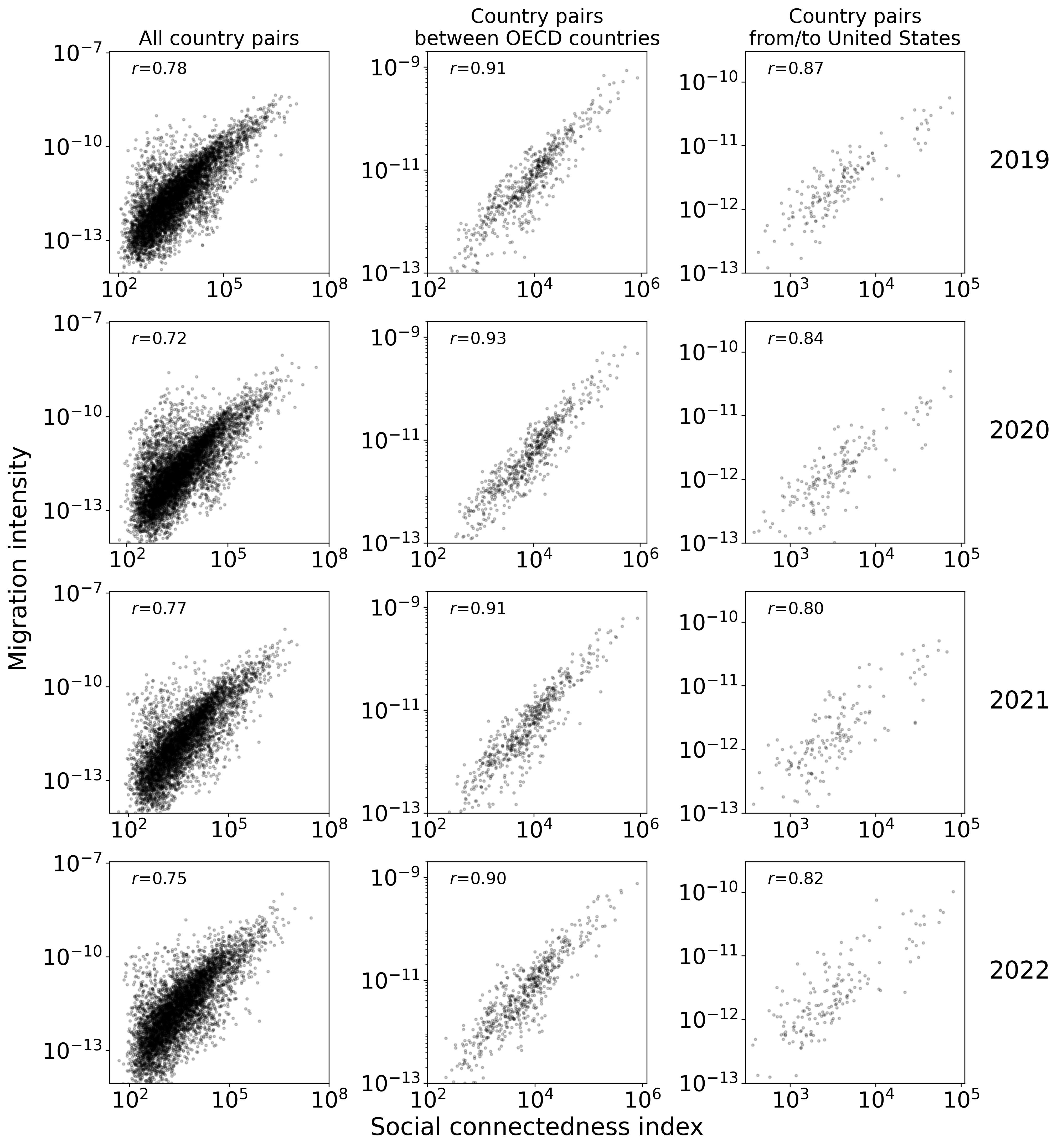}
\caption{Correlation between social connectedness index and migration. The first row is the correlation in 2019; the second row is the correlation in 2020; the third row is the correlation in 2021, and the bottom row is the correlation in 2022. The left column includes all the country pairs, the middle column only includes country pairs between OECD countries, and the right column only includes country pairs that include the United States as the origin or destination. \label{sci_migration}}
\end{figure}

In Fig. \ref{sci_migration}, we present the correlation between these indices on the country pair by year level, using our index of migration to measure flows. We include only those country pairs with at least 20 migrants because 95\% of the noise from differential privacy is in [-6.98, 6.98], which could cause outliers if the number of migrants is very small. We find that the two indices are highly correlated, measuring 0.75 in the full sample in 2022, rising to 0.90 when we limit the sample to only migration between OECD countries.

\begin{table}[]
\centering
\caption{Performance comparison of the segment-based and frequency-based methods}
\label{segment_vs_frequency}
\begin{tabular}{|lll|}
\hline
\multicolumn{1}{|l|}{}                                                                                                                                                                                                                                                                                                  & \multicolumn{1}{l|}{Frequency-based (\%)} & Segment-based  (\%)\\ \hline
\multicolumn{3}{|l|}{\textbf{Modal country}}                                                                                                                                                                                                                                                                                                                                   \\ \hline
\multicolumn{1}{|l|}{\begin{tabular}[c]{@{}l@{}}Share of users whose modal country covers $\geq 90\%$ of \\the year/segment.\end{tabular}}                                                                                                                                                                           & \multicolumn{1}{l|}{70.5}          & 97.8        \\ \hline
\multicolumn{1}{|l|}{\begin{tabular}[c]{@{}l@{}}Share of users for whom the difference between the \\ most-common and second most-common countries is \\ smaller than 20\%.\end{tabular}} & \multicolumn{1}{l|}{6.6}           & 0.004       \\ \hline
\multicolumn{1}{|l|}{Longest segment \textgreater 300 days}                                                                                                                                                                                                                                                             & \multicolumn{1}{l|}{81.3}         & 100         \\ \hline
\multicolumn{3}{|l|}{\textbf{Complexity}}                                                                                                                                                                                                                                                                                                                                      \\ \hline
\multicolumn{1}{|l|}{Complexity \textless 0.01}                                                                                                                                                                                                                                                                         & \multicolumn{1}{l|}{65.7}          & 93.5        \\ \hline
\end{tabular}
\end{table}

\begin{table}[]
\centering
\caption{Comparison of different weighting methods using data on migrants to Sweden. Errors are expressed in thousands of people. The High HDI and Low HDI rows subset to source countries that are, respectively, above and below the global median Human Development Index. In 2022, we exclude migrants from Ukraine from the analysis since Sweden did not include those protected under the Temporary Protection Directive from its immigration figures; see SI Validation Sweden for more details.}
\label{reweight_methods_result}
\begin{tabular}{|r|l|rr|rr|rr|rr|rr|}
\hline
\multicolumn{1}{|l|}{\multirow{2}{*}{\textbf{Year}}} & \multirow{2}{*}{\textbf{HDI type}} & \multicolumn{2}{c|}{\textbf{Raw}}                                     & \multicolumn{2}{c|}{\textbf{Penetration}}                        & \multicolumn{2}{c|}{\textbf{Raking}}                                  & \multicolumn{2}{c|}{\textbf{Coefficient}}                             & \multicolumn{2}{c|}{\textbf{Selection rate}}                          \\ \cline{3-12} 
\multicolumn{1}{|l|}{}                               &                                    & \multicolumn{1}{l|}{\textbf{r}} & \multicolumn{1}{l|}{\textbf{error}} & \multicolumn{1}{l|}{\textbf{r}} & \multicolumn{1}{l|}{\textbf{error}} & \multicolumn{1}{l|}{\textbf{r}} & \multicolumn{1}{l|}{\textbf{error}} & \multicolumn{1}{l|}{\textbf{r}} & \multicolumn{1}{l|}{\textbf{error}} & \multicolumn{1}{l|}{\textbf{r}} & \multicolumn{1}{l|}{\textbf{error}} \\ \hline
\multirow{3}{*}{2019}                                & All                                & \multicolumn{1}{r|}{0.84}       & 55.5                               & \multicolumn{1}{r|}{0.59}       & 83.5                              & \multicolumn{1}{r|}{0.79}       & 62.3                               & \multicolumn{1}{r|}{0.84}       & 42.0                               & \multicolumn{1}{r|}{0.86}       & \textbf{39.0}                      \\ \cline{2-12} 
                                                     & High HDI                           & \multicolumn{1}{r|}{0.92}       & 29.1                               & \multicolumn{1}{r|}{0.56}       & 48.3                              & \multicolumn{1}{r|}{0.75}       & 27.4                               & \multicolumn{1}{r|}{0.92}       & 23.8                               & \multicolumn{1}{r|}{0.93}       & 20.7                               \\ \cline{2-12} 
                                                     & Low HDI                            & \multicolumn{1}{r|}{0.80}       & 26.4                               & \multicolumn{1}{r|}{0.80}       & 35.3                              & \multicolumn{1}{r|}{0.82}       & 34.9                               & \multicolumn{1}{r|}{0.80}       & 18.2                               & \multicolumn{1}{r|}{0.80}       & 18.4                               \\ \hline
\multirow{3}{*}{2020}                                & All                                & \multicolumn{1}{r|}{0.93}       & 39.9                               & \multicolumn{1}{r|}{0.62}       & 50.0                              & \multicolumn{1}{r|}{0.59}       & 71.6                               & \multicolumn{1}{r|}{0.93}       & \textbf{22.2}                               & \multicolumn{1}{r|}{0.94}       & 23.7                      \\ \cline{2-12} 
                                                     & High HDI                           & \multicolumn{1}{r|}{0.94}       & 25.9                               & \multicolumn{1}{r|}{0.58}       & 28.8                              & \multicolumn{1}{r|}{0.51}       & 45.1                               & \multicolumn{1}{r|}{0.94}       & 14.0                               & \multicolumn{1}{r|}{0.95}       & 15.7                               \\ \cline{2-12} 
                                                     & Low HDI                            & \multicolumn{1}{r|}{0.92}       & 14.0                               & \multicolumn{1}{r|}{0.76}       & 21.2                              & \multicolumn{1}{r|}{0.84}       & 26.5                               & \multicolumn{1}{r|}{0.92}       & 8.2                                & \multicolumn{1}{r|}{0.93}       & 7.9                               \\ \hline
\multirow{3}{*}{2021}                                & All                                & \multicolumn{1}{r|}{0.95}       & 44.8                               & \multicolumn{1}{r|}{0.65}       & 49.8                              & \multicolumn{1}{r|}{0.71}       & 51.3                               & \multicolumn{1}{r|}{0.95}       & 23.5                               & \multicolumn{1}{r|}{0.96}       & \textbf{20.8}                      \\ \cline{2-12} 
                                                     & High HDI                           & \multicolumn{1}{r|}{0.95}       & 28.0                               & \multicolumn{1}{r|}{0.58}       & 30.4                              & \multicolumn{1}{r|}{0.62}       & 26.7                               & \multicolumn{1}{r|}{0.95}       & 13.5                               & \multicolumn{1}{r|}{0.95}       & 13.1                               \\ \cline{2-12} 
                                                     & Low HDI                            & \multicolumn{1}{r|}{0.97}       & 16.8                               & \multicolumn{1}{r|}{0.82}       & 19.4                              & \multicolumn{1}{r|}{0.86}       & 24.6                               & \multicolumn{1}{r|}{0.97}       & 10.0                               & \multicolumn{1}{r|}{0.98}       & 7.8                               \\ \hline
\multirow{3}{*}{2022}                                & All                                & \multicolumn{1}{r|}{0.96}       & 50.4                               & \multicolumn{1}{r|}{0.55}       & 75.3                              & \multicolumn{1}{r|}{0.63}       & 62.2                               & \multicolumn{1}{r|}{0.96}       & 28.0                               & \multicolumn{1}{r|}{0.96}       & \textbf{25.7}                      \\ \cline{2-12} 
                                                     & High HDI                           & \multicolumn{1}{r|}{0.95}       & 34.6                               & \multicolumn{1}{r|}{0.45}       & 54.7                              & \multicolumn{1}{r|}{0.52}       & 43.1                               & \multicolumn{1}{r|}{0.95}       & 19.0                               & \multicolumn{1}{r|}{0.95}       & 17.2                               \\ \cline{2-12} 
                                                     & Low HDI                            & \multicolumn{1}{r|}{0.97}       & 15.9                               & \multicolumn{1}{r|}{0.91}       & 20.6                              & \multicolumn{1}{r|}{0.86}       & 19.1                               & \multicolumn{1}{r|}{0.97}       & 9.0                                & \multicolumn{1}{r|}{0.97}       & 8.5                              \\ \hline
\end{tabular}
\end{table}

\begin{table}[]
\centering
\caption{Estimated number of migrants to Sweden. Figures are in the scale of 1000. In 2022, we exclude migrants from Ukraine from the analysis since Sweden did not include those protected under the Temporary Protection Directive from its immigration figures; see SI Validation Sweden for more details.}
\label{reweighting_methods_migrants_num}
\begin{tabular}{|c|l|c|c|c|c|c|c|}
\hline
\textbf{Year} & \multicolumn{1}{c|}{\textbf{HDI Type}} & \textbf{\begin{tabular}[c]{@{}c@{}}Swedish\\ Government\end{tabular}} & \textbf{Raw} & \textbf{Penetration} & \textbf{Raking} & \textbf{Coefficient} & \textbf{Selection} \\ \hline
              & All                                    & 108    &   54  &   165 &  144  &  116  &  112           \\ \cline{2-8}
2019 & High HDI                                  & 67   &   39  &   105 &  84   &  84   &  79           \\ \cline{2-8}
              & Low HDI                                    & 41 &   15  &   60  &  60   &  32   &  33           \\ \hline
              & All                                    & 77 &   38  &   103 &  135  &  77   &  84           \\ \cline{2-8}
2020 & High HDI                                   & 54  &   29  &   66  &  89   &  58   &  61           \\ \cline{2-8}
              & Low HDI                                    & 23 &   9   &   37  &  46   &  19   &  22           \\ \hline
              & All                                        & 86  &   41  &   110 &  114  &  76   &  85           \\ \cline{2-8}
2021          & High HDI                                   & 58  &   30  &   69  &  65   &  55   &  61           \\ \cline{2-8}
              & Low HDI                                    & 28  &   11  &   40  &  49   &  21   &  24           \\ \hline
              & All                                        & 97  &  48  &  142  &  104  &   106  &  99           \\ \cline{2-8}
2022          & High HDI                                   & 67  &  34  &  96   &  61   &   75   &  69           \\ \cline{2-8}
              & Low HDI                                    & 30  &  14  &  46   &  44   &   32   &  31           \\ \hline
\end{tabular}
\end{table}

\begin{table}[]
\centering
\caption{Performance comparison among different weighting methods based on the data from Eurostat.  Errors are in thousands of people.}
\label{reweight_methods_result_eurostat}
\begin{tabular}{|r|l|rr|rr|rr|rr|rr|}
\hline
\multicolumn{1}{|l|}{\multirow{2}{*}{\textbf{Year}}} & \multirow{2}{*}{\textbf{HDI type}} & \multicolumn{2}{c|}{\textbf{Raw}}                                     & \multicolumn{2}{c|}{\textbf{Penetration}}                        & \multicolumn{2}{c|}{\textbf{Raking}}                                  & \multicolumn{2}{c|}{\textbf{Coefficient}}                             & \multicolumn{2}{c|}{\textbf{Selection rate}}                          \\ \cline{3-12} 
\multicolumn{1}{|l|}{}                               &                                    & \multicolumn{1}{l|}{\textbf{r}} & \multicolumn{1}{l|}{\textbf{error}} & \multicolumn{1}{l|}{\textbf{r}} & \multicolumn{1}{l|}{\textbf{error}} & \multicolumn{1}{l|}{\textbf{r}} & \multicolumn{1}{l|}{\textbf{error}} & \multicolumn{1}{l|}{\textbf{r}} & \multicolumn{1}{l|}{\textbf{error}} & \multicolumn{1}{l|}{\textbf{r}} & \multicolumn{1}{l|}{\textbf{error}} \\ \hline
\multirow{3}{*}{2019}                                & All                                & \multicolumn{1}{r|}{0.92}       & 1269.3                             & \multicolumn{1}{r|}{0.89}       & 1274.9                             & \multicolumn{1}{r|}{0.90}       & 1239.7                             & \multicolumn{1}{r|}{0.92}       & 1107.1                              & \multicolumn{1}{r|}{0.93}       & \textbf{1012.7}                     \\ \cline{2-12} 
                                                     & High HDI                           & \multicolumn{1}{r|}{0.92}       & 862.5                              & \multicolumn{1}{r|}{0.86}       & 924.4                              & \multicolumn{1}{r|}{0.91}       & 806.1                              & \multicolumn{1}{r|}{0.92}       & 905.7                               & \multicolumn{1}{r|}{0.93}       & 818.0                               \\ \cline{2-12} 
                                                     & Low HDI                            & \multicolumn{1}{r|}{0.99}       & 406.8                              & \multicolumn{1}{r|}{0.93}       & 350.4                              & \multicolumn{1}{r|}{0.91}       & 433.5                              & \multicolumn{1}{r|}{0.99}       & 201.3                               & \multicolumn{1}{r|}{0.99}       & 194.7                               \\ \hline
\multirow{3}{*}{2020}                                & All                                & \multicolumn{1}{r|}{0.90}       & 891.9                              & \multicolumn{1}{r|}{0.84}       & 815.0                              & \multicolumn{1}{r|}{0.88}       & 838.9                              & \multicolumn{1}{r|}{0.91}       & 642.6                               & \multicolumn{1}{r|}{0.91}       & \textbf{637.3}                      \\ \cline{2-12} 
                                                     & High HDI                           & \multicolumn{1}{r|}{0.89}       & 680.2                              & \multicolumn{1}{r|}{0.81}       & 641.8                              & \multicolumn{1}{r|}{0.85}       & 646.1                              & \multicolumn{1}{r|}{0.89}       & 511.9                               & \multicolumn{1}{r|}{0.90}       & 511.7                               \\ \cline{2-12} 
                                                     & Low HDI                            & \multicolumn{1}{r|}{0.94}       & 211.6                              & \multicolumn{1}{r|}{0.95}       & 173.2                              & \multicolumn{1}{r|}{0.96}       & 192.8                              & \multicolumn{1}{r|}{0.94}       & 130.7                               & \multicolumn{1}{r|}{0.95}       & 125.6                     \\ \hline
\multirow{3}{*}{2021}                                & All                                & \multicolumn{1}{r|}{0.87}       & 981.2                              & \multicolumn{1}{r|}{0.80}       & 904.5                              & \multicolumn{1}{r|}{0.84}       & 855.5                              & \multicolumn{1}{r|}{0.87}       & 729.8                               & \multicolumn{1}{r|}{0.87}       & \textbf{716.0}                      \\ \cline{2-12} 
                                                     & High HDI                           & \multicolumn{1}{r|}{0.84}       & 746.9                              & \multicolumn{1}{r|}{0.78}       & 715.8                              & \multicolumn{1}{r|}{0.81}       & 669.0                              & \multicolumn{1}{r|}{0.85}       & 576.5                               & \multicolumn{1}{r|}{0.84}       & 580.3                               \\ \cline{2-12} 
                                                     & Low HDI                            & \multicolumn{1}{r|}{0.96}       & 234.3                              & \multicolumn{1}{r|}{0.93}       & 188.6                              & \multicolumn{1}{r|}{0.94}       & 186.5                              & \multicolumn{1}{r|}{0.96}       & 153.3                               & \multicolumn{1}{r|}{0.96}       & 135.7                     \\ \hline
\multirow{3}{*}{2022}                                & All                                & \multicolumn{1}{r|}{0.94}       & 1317.2                             & \multicolumn{1}{r|}{0.74}       & 1585.1                             & \multicolumn{1}{r|}{0.91}       & 1235.1                             & \multicolumn{1}{r|}{0.94}       & 1116.0                              & \multicolumn{1}{r|}{0.94}       & \textbf{1032.3}                      \\ \cline{2-12} 
                                                     & High HDI                           & \multicolumn{1}{r|}{0.94}       & 970.8                              & \multicolumn{1}{r|}{0.71}       & 1295.7                             & \multicolumn{1}{r|}{0.90}       & 972.8                              & \multicolumn{1}{r|}{0.94}       & 860.5                               & \multicolumn{1}{r|}{0.94}       & 786.4                               \\ \cline{2-12} 
                                                     & Low HDI                            & \multicolumn{1}{r|}{0.96}       & 346.4                              & \multicolumn{1}{r|}{0.92}       & 289.4                              & \multicolumn{1}{r|}{0.94}       & 262.3                              & \multicolumn{1}{r|}{0.96}       & 255.5                               & \multicolumn{1}{r|}{0.96}       & 246.0                     \\ \hline
\end{tabular}
\end{table}

\clearpage
\begin{longtable}{l|rrrrrr}
\caption{Correlations for 2019 \label{validation_corr_2019}}
\\ \toprule
\multicolumn{1}{l}{} & \multicolumn{2}{c}{\textbf{New Zealand}} & \multicolumn{2}{c}{\textbf{Sweden}} & \multicolumn{2}{c}{\textbf{Eurostat}} \\ 
\cmidrule(lr){2-3} \cmidrule(lr){4-5} \cmidrule(lr){6-7}
\multicolumn{1}{l}{} & \textbf{N} & \textbf{Correlation} & \textbf{N} & \textbf{Correlation} & \textbf{N} & \textbf{Correlation} \\ 
\midrule
Migrants                & 166   &  0.98  &  176  & 0.87  &  3293  &   0.94 \\
Log(Migrants)           & 166   &  0.89  &  168  & 0.91  &  2603  &   0.88 \\
Proportion of Migrants  & 166   &  0.98  &  176  & 0.87  &  3293  &   0.89 \\
Total Outbound          &   0   &   NA   &    0  &  NA   &   179  &   0.96 \\
Total Inbound           &   1   &   NA   &    1  &  NA   &    24  &   0.99 \\
Net Migration           &   0   &   NA   &    0  &  NA   &    24  &   0.99 \\
\bottomrule
\end{longtable}

\clearpage
\begin{longtable}{l|rrrrrr}
\caption{Correlations for 2020 \label{validation_corr_2020}}
\\ \toprule
\multicolumn{1}{l}{} & \multicolumn{2}{c}{\textbf{New Zealand}} & \multicolumn{2}{c}{\textbf{Sweden}} & \multicolumn{2}{c}{\textbf{Eurostat}} \\ 
\cmidrule(lr){2-3} \cmidrule(lr){4-5} \cmidrule(lr){6-7}
\multicolumn{1}{l}{} & \textbf{N} & \textbf{Correlation} & \textbf{N} & \textbf{Correlation} & \textbf{N} & \textbf{Correlation} \\ 
\midrule
Migrants                & 158   &  0.99  &  176  & 0.95  &  2923  &   0.91 \\ 
Log(Migrants)           & 158   &  0.87  &  166  & 0.90  &  2269  &   0.87 \\ 
Proportion of Migrants  & 158   &  0.99  &  176  & 0.95  &  2923  &   0.78 \\ 
Total Outbound          &   0   &   NA   &    0  &  NA   &   179  &   0.96 \\
Total Inbound           &   1   &   NA   &    1  &  NA   &    18  &   0.96 \\
Net Migration           &   0   &   NA   &    0  &  NA   &    18  &   0.94 \\
\bottomrule
\end{longtable}

\clearpage
\begin{longtable}{l|rrrrrr}
\caption{Correlations for 2021  \label{validation_corr_2021}}
\\ \toprule
\multicolumn{1}{l}{} & \multicolumn{2}{c}{\textbf{New Zealand}} & \multicolumn{2}{c}{\textbf{Sweden}} & \multicolumn{2}{c}{\textbf{Eurostat}} \\ 
\cmidrule(lr){2-3} \cmidrule(lr){4-5} \cmidrule(lr){6-7}
\multicolumn{1}{l}{} & \textbf{N} & \textbf{Correlation} & \textbf{N} & \textbf{Correlation} & \textbf{N} & \textbf{Correlation} \\ 
\midrule
Migrants                & 146   &  0.99  &  176  & 0.96  &  2566  &   0.87 \\
Log(Migrants)           & 146   &  0.84  &  169  & 0.89  &  1982  &   0.88 \\
Proportion of Migrants  & 146   &  0.99  &  176  & 0.96  &  2566  &   0.70 \\
Total Outbound          &   0   &   NA   &    0  &  NA   &   179  &   0.95 \\
Total Inbound           &   1   &   NA   &    1  &  NA   &    16  &   0.98 \\
Net Migration           &   0   &   NA   &    0  &  NA   &    16  &   0.97 \\
\bottomrule
\end{longtable}

\clearpage
\begin{longtable}{l|rrrrrr}
\caption{Correlations for 2022. We exclude migrants from Ukraine to Sweden from these analyses since Sweden did not include those protected under the Temporary Protection Directive from its immigration figures; see SI Validation Sweden for more details.  \label{validation_corr_2022}}
\\ \toprule
\multicolumn{1}{l}{} & \multicolumn{2}{c}{\textbf{New Zealand}} & \multicolumn{2}{c}{\textbf{Sweden}} & \multicolumn{2}{c}{\textbf{Eurostat}} \\ 
\cmidrule(lr){2-3} \cmidrule(lr){4-5} \cmidrule(lr){6-7}
\multicolumn{1}{l}{} & \textbf{N} & \textbf{Correlation} & \textbf{N} & \textbf{Correlation} & \textbf{N} & \textbf{Correlation} \\ 
\midrule
Migrants                &   150   &   0.99   &  177  & 0.97  &  2900  &   0.92 \\
Log(Migrants)           &   150   &   0.86   &  170  & 0.91  &  2349  &   0.88 \\
Proportion of Migrants  &   150   &   0.99   &  177  & 0.97  &     2900  &   0.82 \\
Total Outbound          &     0   &    NA   &    0  &  NA   &   178  &   0.98 \\
Total Inbound           &     1   &    NA   &    1  &  NA   &    19  &   0.98 \\
Net Migration           &     0   &    NA   &    0  &  NA   &    19  &   0.97 \\
\bottomrule
\end{longtable}

\clearpage
\begin{longtable}{rllllllll}
\caption{Inflow and outflow of each country} \label{country_summary} 
\\
\toprule
\multicolumn{1}{l}{} & \multicolumn{2}{c}{\textbf{2019}} & \multicolumn{2}{c}{\textbf{2020}} & \multicolumn{2}{c}{\textbf{2021}} & \multicolumn{2}{c}{\textbf{2022}} \\ 
\cmidrule(lr){2-3} \cmidrule(lr){4-5} \cmidrule(lr){6-7} \cmidrule(lr){8-9}
\multicolumn{1}{l}{\textbf{iso2}} & \textbf{Outflow} & \textbf{Inflow} & \textbf{Outflow} & \textbf{Inflow} & \textbf{Outflow} & \textbf{Inflow} & \textbf{Outflow} & \textbf{Inflow} \\ 
\midrule
\endfirsthead
\caption[]{Inflow and outflow of each country} \\
\toprule
\multicolumn{1}{l}{} & \multicolumn{2}{c}{\textbf{2019}} & \multicolumn{2}{c}{\textbf{2020}} & \multicolumn{2}{c}{\textbf{2021}} & \multicolumn{2}{c}{\textbf{2022}} \\ 
\cmidrule(lr){2-3} \cmidrule(lr){4-5} \cmidrule(lr){6-7} \cmidrule(lr){8-9}
\multicolumn{1}{l}{\textbf{iso2}} & \textbf{Outflow} & \textbf{Inflow} & \textbf{Outflow} & \textbf{Inflow} & \textbf{Outflow} & \textbf{Inflow} & \textbf{Outflow} & \textbf{Inflow} \\ 
\midrule
\endhead
\midrule
\multicolumn{9}{r}{Continued on next page} \\
\midrule
\endfoot
\bottomrule
\endlastfoot
AD & 7,301 & 8,390 & 7,414 & 7,908 & 7,446 & 6,923 & 7,142 & 9,685 \\
AE & 769,808 & 1,248,145 & 803,535 & 643,752 & 749,831 & 1,144,597 & 633,752 & 2,151,310 \\
AF & 103,530 & 89,590 & 91,669 & 66,438 & 189,945 & 60,701 & 166,065 & 81,698 \\
AL & 85,476 & 33,450 & 55,911 & 34,778 & 76,694 & 30,131 & 113,508 & 28,662 \\
AM & 42,025 & 52,272 & 25,998 & 59,099 & 65,878 & 43,459 & 36,459 & 103,325 \\
AO & 67,818 & 38,101 & 46,377 & 24,005 & 41,674 & 30,931 & 56,405 & 39,884 \\
AR & 233,137 & 181,051 & 147,823 & 112,947 & 181,623 & 87,456 & 215,615 & 161,802 \\
AT & 75,894 & 103,587 & 78,738 & 76,056 & 66,872 & 82,422 & 61,468 & 154,988 \\
AU & 276,380 & 587,275 & 238,571 & 278,379 & 177,536 & 170,281 & 187,387 & 693,583 \\
AZ & 47,205 & 43,218 & 52,941 & 35,082 & 42,696 & 39,946 & 35,634 & 54,940 \\
BA & 60,246 & 26,869 & 45,815 & 30,241 & 46,173 & 24,059 & 58,076 & 23,016 \\
BB & 9,359 & 8,430 & 9,098 & 8,560 & 9,464 & 7,905 & 9,370 & 8,677 \\
BD & 720,011 & 708,969 & 415,712 & 568,457 & 676,047 & 620,369 & 1,391,318 & 699,291 \\
BE & 154,433 & 156,145 & 120,559 & 104,747 & 87,150 & 114,017 & 86,528 & 178,368 \\
BF & 30,641 & 41,271 & 27,967 & 39,397 & 41,849 & 44,398 & 56,131 & 65,231 \\
BG & 76,590 & 64,826 & 66,512 & 72,705 & 63,037 & 70,065 & 69,845 & 91,452 \\
BH & 121,757 & 99,356 & 110,101 & 52,981 & 107,922 & 80,020 & 91,447 & 154,349 \\
BI & 29,736 & 21,234 & 23,039 & 22,637 & 24,080 & 27,047 & 36,260 & 26,908 \\
BJ & 40,837 & 33,645 & 35,546 & 39,874 & 41,272 & 33,124 & 44,742 & 51,085 \\
BN & 26,077 & 28,325 & 21,014 & 16,634 & 20,570 & 10,483 & 30,105 & 23,008 \\
BO & 94,729 & 75,852 & 83,439 & 56,575 & 97,349 & 67,379 & 142,032 & 60,698 \\
BR & 487,368 & 298,911 & 222,480 & 247,998 & 275,492 & 226,958 & 547,945 & 255,427 \\
BS & 12,798 & 12,414 & 10,844 & 9,273 & 14,689 & 8,259 & 14,192 & 13,480 \\
BT & 20,812 & 13,880 & 17,531 & 15,699 & 13,233 & 9,345 & 23,932 & 10,949 \\
BW & 14,382 & 16,052 & 12,882 & 14,520 & 13,495 & 12,921 & 18,379 & 13,856 \\
BY & 35,465 & 22,849 & 37,073 & 22,561 & 55,299 & 17,779 & 78,686 & 31,644 \\
BZ & 10,786 & 9,825 & 10,275 & 9,020 & 11,515 & 8,204 & 12,071 & 8,981 \\
CA & 231,918 & 749,815 & 204,070 & 351,121 & 196,009 & 533,336 & 210,034 & 1,113,926 \\
CD & 75,747 & 81,033 & 54,918 & 55,883 & 64,357 & 57,956 & 86,923 & 60,998 \\
CF & 18,643 & 19,394 & 18,840 & 14,573 & 15,930 & 11,280 & 13,119 & 13,067 \\
CG & 30,854 & 21,794 & 23,936 & 16,532 & 27,358 & 20,452 & 31,037 & 22,611 \\
CH & 102,644 & 146,245 & 93,016 & 110,440 & 88,937 & 117,810 & 87,892 & 202,667 \\
CI & 92,683 & 68,406 & 80,684 & 53,943 & 92,956 & 79,933 & 120,966 & 99,858 \\
CL & 126,265 & 294,280 & 105,161 & 116,168 & 137,304 & 284,440 & 169,610 & 383,607 \\
CM & 79,813 & 40,463 & 46,436 & 36,322 & 65,176 & 38,529 & 89,662 & 42,586 \\
CO & 476,970 & 991,213 & 312,750 & 442,357 & 517,252 & 550,820 & 1,119,344 & 366,085 \\
CR & 50,693 & 58,578 & 47,433 & 37,343 & 55,813 & 55,487 & 70,084 & 69,725 \\
CV & 17,506 & 9,257 & 13,578 & 8,467 & 14,342 & 8,266 & 25,302 & 8,317 \\
CY & 33,205 & 60,474 & 32,713 & 42,950 & 34,584 & 52,933 & 37,808 & 91,806 \\
CZ & 58,262 & 83,238 & 60,055 & 64,729 & 47,025 & 79,446 & 50,780 & 270,575 \\
DE & 527,011 & 756,908 & 436,989 & 556,905 & 460,019 & 599,642 & 463,342 & 1,345,919 \\
DJ & 15,252 & 11,134 & 13,959 & 11,550 & 21,493 & 8,404 & 29,317 & 7,476 \\
DK & 38,775 & 58,907 & 33,420 & 50,384 & 31,940 & 53,177 & 31,932 & 83,045 \\
DO & 119,376 & 93,491 & 76,814 & 84,215 & 105,066 & 105,580 & 169,978 & 110,224 \\
DZ & 140,366 & 85,849 & 124,884 & 78,264 & 121,219 & 60,543 & 170,152 & 67,476 \\
EC & 159,954 & 202,062 & 128,684 & 91,476 & 235,880 & 113,147 & 287,348 & 103,600 \\
EE & 13,461 & 21,004 & 12,763 & 17,463 & 11,664 & 18,432 & 13,465 & 44,908 \\
EG & 604,339 & 505,476 & 343,670 & 586,969 & 481,975 & 538,491 & 914,587 & 466,232 \\
ER & 20,487 & 6,809 & 10,279 & 6,755 & 7,687 & 6,521 & 7,251 & 6,513 \\
ES & 305,333 & 768,788 & 286,001 & 423,347 & 294,047 & 452,547 & 280,207 & 1,026,335 \\
ET & 67,401 & 133,596 & 56,158 & 95,763 & 76,229 & 120,487 & 102,422 & 130,403 \\
FI & 26,387 & 38,004 & 21,782 & 32,973 & 21,208 & 32,458 & 21,227 & 72,424 \\
FJ & 17,469 & 11,140 & 13,396 & 12,126 & 12,026 & 8,774 & 24,394 & 10,389 \\
FM & 8,230 & 9,657 & 8,606 & 7,280 & 8,786 & 6,061 & 8,844 & 6,709 \\
FR & 364,592 & 594,973 & 280,533 & 456,177 & 266,791 & 465,212 & 302,363 & 700,157 \\
GA & 34,086 & 21,891 & 24,751 & 18,234 & 31,164 & 19,454 & 34,247 & 21,254 \\
GB & 670,647 & 856,989 & 629,902 & 658,333 & 470,148 & 845,371 & 442,728 & 1,336,265 \\
GD & 8,790 & 7,034 & 9,025 & 7,298 & 7,424 & 7,162 & 7,433 & 7,652 \\
GE & 65,832 & 54,661 & 41,642 & 52,556 & 62,183 & 57,104 & 95,827 & 142,900 \\
GH & 87,619 & 84,712 & 58,654 & 76,993 & 85,994 & 86,889 & 127,255 & 97,907 \\
GM & 18,504 & 22,747 & 17,399 & 16,810 & 21,008 & 20,642 & 25,450 & 22,918 \\
GN & 62,463 & 54,171 & 46,123 & 41,460 & 66,017 & 70,647 & 85,691 & 77,770 \\
GQ & 11,128 & 16,634 & 13,523 & 11,028 & 14,384 & 10,483 & 15,482 & 11,216 \\
GR & 108,528 & 105,284 & 95,941 & 84,398 & 119,589 & 66,556 & 128,331 & 87,391 \\
GT & 162,067 & 60,549 & 81,091 & 52,734 & 196,080 & 57,821 & 282,674 & 59,230 \\
GW & 14,917 & 9,507 & 13,082 & 12,302 & 19,486 & 16,716 & 22,865 & 19,837 \\
GY & 15,633 & 15,840 & 12,149 & 12,252 & 15,195 & 14,080 & 17,386 & 12,721 \\
HK & 141,175 & 137,023 & 137,462 & 82,873 & 175,957 & 80,291 & 227,873 & 107,926 \\
HN & 173,028 & 37,383 & 59,554 & 35,671 & 175,243 & 37,907 & 208,427 & 38,202 \\
HR & 41,568 & 34,962 & 35,532 & 30,488 & 32,450 & 30,355 & 31,388 & 54,355 \\
HT & 82,898 & 27,898 & 64,849 & 26,556 & 95,868 & 36,763 & 107,673 & 45,577 \\
HU & 56,915 & 84,142 & 48,943 & 75,922 & 44,967 & 66,566 & 57,382 & 93,763 \\
ID & 596,395 & 748,250 & 251,159 & 650,273 & 206,979 & 619,149 & 533,239 & 574,464 \\
IE & 77,893 & 120,989 & 67,511 & 80,212 & 64,217 & 81,718 & 73,317 & 190,421 \\
IL & 79,346 & 87,826 & 57,908 & 53,830 & 64,564 & 59,696 & 65,442 & 127,439 \\
IN & 2,520,754 & 1,544,878 & 1,329,500 & 1,654,110 & 1,638,248 & 1,684,086 & 3,429,268 & 1,371,867 \\
IQ & 112,660 & 165,750 & 121,287 & 123,981 & 107,685 & 135,884 & 111,173 & 167,685 \\
IS & 9,879 & 13,509 & 9,591 & 12,235 & 8,753 & 12,312 & 8,373 & 17,258 \\
IT & 405,085 & 369,670 & 300,536 & 315,529 & 274,637 & 347,466 & 279,804 & 555,478 \\
JM & 38,397 & 17,443 & 25,464 & 16,686 & 40,516 & 13,721 & 55,940 & 14,104 \\
JO & 202,509 & 101,970 & 151,627 & 74,178 & 152,641 & 121,104 & 150,489 & 111,333 \\
JP & 438,531 & 568,919 & 349,403 & 309,346 & 356,541 & 246,772 & 464,053 & 622,060 \\
KE & 100,905 & 67,724 & 72,441 & 58,121 & 111,218 & 79,081 & 169,873 & 85,134 \\
KG & 39,752 & 39,059 & 35,969 & 45,305 & 62,253 & 66,340 & 28,856 & 89,651 \\
KH & 93,454 & 173,925 & 82,943 & 194,430 & 78,477 & 173,421 & 196,466 & 218,971 \\
KI & 6,670 & 9,143 & 8,727 & 7,348 & 7,002 & 6,906 & 7,290 & 7,072 \\
KM & 12,782 & 10,415 & 10,863 & 9,143 & 11,266 & 9,912 & 13,428 & 10,609 \\
KR & 345,074 & 353,549 & 278,257 & 209,655 & 251,523 & 146,349 & 272,984 & 319,973 \\
KW & 312,143 & 340,821 & 360,242 & 116,791 & 308,802 & 113,502 & 286,569 & 533,330 \\
KZ & 55,445 & 32,515 & 51,387 & 32,202 & 48,786 & 34,870 & 42,802 & 57,501 \\
LA & 62,089 & 65,198 & 54,607 & 104,655 & 83,579 & 82,190 & 216,912 & 81,522 \\
LB & 204,157 & 84,951 & 234,975 & 52,910 & 221,039 & 67,597 & 187,912 & 97,156 \\
LC & 8,017 & 6,664 & 8,243 & 7,369 & 7,900 & 7,091 & 8,117 & 6,774 \\
LK & 122,941 & 108,177 & 66,810 & 81,179 & 83,407 & 100,003 & 275,661 & 78,997 \\
LR & 18,944 & 15,289 & 16,918 & 12,682 & 22,409 & 15,512 & 23,552 & 18,270 \\
LS & 20,638 & 17,119 & 22,808 & 16,465 & 45,749 & 15,615 & 57,473 & 24,316 \\
LT & 31,245 & 37,509 & 24,149 & 40,801 & 19,688 & 40,118 & 23,330 & 69,217 \\
LU & 17,241 & 25,942 & 17,444 & 20,145 & 15,777 & 20,921 & 16,379 & 28,332 \\
LV & 21,469 & 22,560 & 19,860 & 21,307 & 17,715 & 17,245 & 21,066 & 34,189 \\
LY & 79,425 & 110,921 & 79,393 & 89,595 & 95,073 & 168,174 & 129,612 & 177,302 \\
MA & 239,073 & 135,912 & 185,550 & 130,713 & 207,872 & 121,452 & 278,507 & 121,802 \\
MD & 66,040 & 38,669 & 54,829 & 51,937 & 67,960 & 45,912 & 89,714 & 55,291 \\
ME & 14,351 & 13,752 & 14,615 & 13,474 & 12,806 & 11,695 & 12,817 & 28,309 \\
MG & 21,056 & 16,257 & 17,198 & 13,543 & 16,114 & 11,796 & 24,865 & 15,121 \\
MK & 35,171 & 17,390 & 27,351 & 18,332 & 28,926 & 15,696 & 41,734 & 14,797 \\
ML & 49,382 & 60,528 & 45,713 & 51,641 & 52,466 & 65,939 & 67,182 & 70,034 \\
MM & 247,327 & 292,821 & 150,215 & 354,655 & 259,100 & 121,694 & 954,028 & 151,995 \\
MN & 39,405 & 28,897 & 22,975 & 30,543 & 16,988 & 35,027 & 45,643 & 27,197 \\
MO & 24,082 & 27,491 & 24,859 & 16,614 & 20,945 & 9,843 & 27,570 & 9,265 \\
MR & 28,853 & 25,881 & 26,014 & 21,840 & 28,993 & 25,645 & 34,264 & 27,899 \\
MT & 16,529 & 33,435 & 18,589 & 22,666 & 18,497 & 19,710 & 17,857 & 47,200 \\
MU & 22,716 & 25,218 & 18,170 & 16,377 & 21,610 & 12,223 & 28,568 & 21,960 \\
MV & 25,973 & 30,891 & 37,748 & 15,234 & 22,550 & 20,145 & 20,642 & 38,384 \\
MW & 36,791 & 29,157 & 23,924 & 25,068 & 25,942 & 32,807 & 43,313 & 33,674 \\
MX & 535,548 & 411,515 & 404,686 & 353,089 & 704,209 & 323,741 & 1,164,919 & 330,249 \\
MY & 926,923 & 591,865 & 832,539 & 299,361 & 683,455 & 154,379 & 659,573 & 802,993 \\
MZ & 40,836 & 32,534 & 37,925 & 26,657 & 48,312 & 29,798 & 59,238 & 37,773 \\
NA & 14,057 & 13,134 & 12,700 & 11,146 & 11,837 & 10,463 & 13,875 & 10,897 \\
NE & 19,426 & 26,083 & 19,030 & 26,443 & 23,324 & 33,046 & 25,670 & 41,490 \\
NG & 202,936 & 163,859 & 140,807 & 152,620 & 203,680 & 149,229 & 329,187 & 204,944 \\
NI & 93,734 & 41,890 & 42,787 & 48,802 & 97,335 & 50,992 & 261,853 & 47,478 \\
NL & 123,194 & 190,045 & 116,640 & 141,624 & 106,307 & 162,183 & 109,465 & 271,156 \\
NO & 42,603 & 56,454 & 34,938 & 47,086 & 33,575 & 45,569 & 30,832 & 84,883 \\
NP & 428,407 & 499,772 & 234,822 & 482,275 & 366,172 & 410,855 & 911,782 & 409,404 \\
NZ & 85,457 & 149,261 & 60,521 & 100,429 & 68,423 & 56,846 & 87,715 & 102,305 \\
OM & 274,388 & 215,900 & 252,905 & 98,113 & 302,590 & 141,350 & 215,167 & 354,919 \\
PA & 42,882 & 44,852 & 41,986 & 25,528 & 48,968 & 29,552 & 59,879 & 36,537 \\
PE & 250,451 & 352,229 & 168,587 & 135,342 & 270,782 & 197,129 & 507,368 & 169,187 \\
PG & 13,631 & 12,933 & 14,860 & 10,085 & 11,495 & 9,328 & 13,512 & 11,329 \\
PH & 1,113,419 & 960,591 & 457,370 & 752,808 & 632,137 & 639,143 & 1,119,019 & 794,163 \\
PK & 704,808 & 667,201 & 414,645 & 616,570 & 392,041 & 623,130 & 1,413,852 & 485,986 \\
PL & 198,095 & 286,470 & 188,080 & 268,028 & 146,325 & 337,158 & 211,197 & 868,433 \\
PT & 96,512 & 274,785 & 91,092 & 160,892 & 95,237 & 166,035 & 99,885 & 413,718 \\
PY & 64,333 & 64,973 & 52,160 & 47,704 & 40,667 & 52,365 & 98,079 & 40,771 \\
QA & 253,140 & 351,880 & 206,886 & 137,173 & 189,537 & 331,632 & 175,483 & 549,691 \\
RO & 249,475 & 232,758 & 192,241 & 253,355 & 178,711 & 231,180 & 226,445 & 229,828 \\
RS & 61,651 & 48,025 & 47,983 & 48,752 & 47,008 & 42,233 & 56,930 & 58,146 \\
RU & 329,790 & 233,024 & 306,391 & 145,638 & 335,660 & 301,118 & 947,587 & 82,625 \\
RW & 32,520 & 27,748 & 27,939 & 21,730 & 23,723 & 20,930 & 28,377 & 21,490 \\
SA & 2,032,176 & 1,599,367 & 1,666,477 & 840,627 & 1,882,151 & 1,207,862 & 1,715,220 & 2,798,137 \\
SB & 9,325 & 7,611 & 9,348 & 7,714 & 8,396 & 7,519 & 9,320 & 7,868 \\
SD & 206,353 & 131,297 & 175,304 & 101,805 & 297,307 & 132,590 & 337,278 & 166,354 \\
SE & 72,617 & 113,830 & 60,441 & 86,496 & 60,611 & 85,989 & 65,710 & 126,235 \\
SG & 204,891 & 271,327 & 205,474 & 250,163 & 175,723 & 138,652 & 257,990 & 343,202 \\
SI & 17,517 & 27,267 & 17,537 & 21,653 & 15,623 & 20,846 & 15,204 & 30,068 \\
SK & 37,813 & 50,959 & 33,983 & 46,036 & 34,970 & 33,985 & 37,733 & 64,949 \\
SL & 17,927 & 15,443 & 15,940 & 14,322 & 24,110 & 16,610 & 30,365 & 18,165 \\
SN & 89,153 & 89,900 & 70,740 & 61,911 & 92,053 & 73,935 & 105,179 & 84,726 \\
SR & 12,929 & 11,682 & 12,790 & 9,609 & 13,591 & 8,499 & 14,368 & 9,975 \\
SS & 28,860 & 50,026 & 31,488 & 44,168 & 27,686 & 57,511 & 34,086 & 55,336 \\
ST & 9,898 & 7,167 & 9,423 & 7,150 & 9,868 & 6,929 & 14,898 & 6,917 \\
SV & 102,040 & 33,920 & 43,728 & 31,088 & 90,663 & 33,287 & 135,121 & 29,706 \\
SY & 140,730 & 311,007 & 166,373 & 175,655 & 235,297 & 149,338 & 355,909 & 139,160 \\
SZ & 12,421 & 13,474 & 12,730 & 14,210 & 15,023 & 12,345 & 17,884 & 13,636 \\
TD & 19,422 & 36,916 & 19,123 & 26,775 & 22,791 & 30,204 & 27,492 & 38,324 \\
TG & 29,408 & 26,391 & 23,270 & 26,560 & 26,981 & 30,103 & 32,773 & 44,024 \\
TH & 571,133 & 416,558 & 636,965 & 352,336 & 378,601 & 363,565 & 505,074 & 1,114,982 \\
TJ & 24,931 & 30,072 & 17,493 & 28,598 & 67,344 & 38,539 & 33,723 & 66,031 \\
TL & 14,567 & 26,265 & 9,405 & 19,996 & 8,000 & 28,818 & 14,681 & 31,199 \\
TM & 8,728 & 8,735 & 9,103 & 7,906 & 7,849 & 7,528 & 8,818 & 6,859 \\
TN & 100,856 & 58,415 & 86,368 & 54,297 & 102,159 & 50,185 & 176,621 & 45,221 \\
TO & 9,199 & 7,928 & 9,645 & 7,726 & 9,638 & 7,648 & 10,194 & 7,884 \\
TR & 461,574 & 313,810 & 279,334 & 323,111 & 307,931 & 354,659 & 521,337 & 359,872 \\
TT & 17,613 & 29,051 & 13,876 & 20,202 & 19,327 & 14,971 & 22,602 & 15,685 \\
TW & 305,644 & 306,171 & 198,971 & 222,989 & 179,397 & 148,069 & 266,064 & 274,046 \\
TZ & 41,725 & 39,506 & 38,748 & 36,113 & 36,030 & 36,625 & 38,295 & 37,931 \\
UA & 205,591 & 131,064 & 164,592 & 170,622 & 277,814 & 96,964 & 2,402,184 & 66,573 \\
UG & 106,646 & 62,482 & 78,136 & 57,890 & 146,520 & 54,802 & 143,658 & 68,934 \\
US & 1,242,505 & 2,283,893 & 980,507 & 1,264,227 & 806,207 & 2,592,038 & 841,150 & 4,109,309 \\
UY & 29,405 & 28,563 & 21,256 & 22,669 & 28,117 & 19,700 & 29,083 & 22,552 \\
UZ & 37,663 & 73,287 & 28,278 & 65,645 & 68,982 & 88,047 & 49,060 & 127,466 \\
VC & 7,688 & 7,008 & 7,319 & 7,461 & 7,075 & 6,514 & 7,541 & 6,683 \\
VE & 1,701,436 & 221,880 & 502,467 & 234,669 & 809,974 & 343,755 & 777,440 & 560,362 \\
VN & 658,292 & 523,981 & 400,985 & 297,295 & 334,627 & 320,131 & 647,023 & 539,049 \\
VU & 8,160 & 6,952 & 9,521 & 8,123 & 10,613 & 7,626 & 9,532 & 8,634 \\
WS & 7,511 & 8,160 & 9,818 & 8,220 & 9,496 & 7,201 & 11,879 & 8,580 \\
XK & 35,189 & 19,342 & 31,446 & 14,550 & 34,511 & 14,235 & 49,903 & 14,124 \\
YE & 86,237 & 124,178 & 58,452 & 88,148 & 120,317 & 101,771 & 172,619 & 94,908 \\
ZA & 200,864 & 194,751 & 152,393 & 149,340 & 177,802 & 181,685 & 276,584 & 220,776 \\
ZM & 22,016 & 30,454 & 22,893 & 24,468 & 22,263 & 23,560 & 27,844 & 27,207 \\
ZW & 71,874 & 39,165 & 53,204 & 38,031 & 51,055 & 43,564 & 62,530 & 70,722 \\
\end{longtable}

\FloatBarrier

\bibliography{si}